\documentclass[
    pre,
    twocolumn,
    twoside,
    byrevtex,
    superscriptaddress,
    floatfix,
    nofootinbib,
    longbibliography
]{revtex4-1}

\usepackage[realmainfile]{currfile}

\input{\currfilebase.settings}

\usepackage{booktabs}

\usepackage{url}
\usepackage{hyperref}
\hypersetup{
    colorlinks=true,
    linkcolor=grey,
    filecolor=grey,
    urlcolor=grey,
    citecolor=todoblue,
    breaklinks=true,
}

\usepackage{xcolor}

\definecolor{olivegreen}{rgb}{0.33333,.41961,0.18431}
\definecolor{forestgreen}{rgb}{0.13333,.5451,0.13333}
\definecolor{lightgrey}{rgb}{0.7,0.7,0.7}
\definecolor{verylightgrey}{rgb}{0.90,0.90,0.90}
\definecolor{grey}{rgb}{0.5,0.5,0.5}

\definecolor{headerblue}{HTML}{33367E}
\definecolor{unitednationsblue}{HTML}{4D88FF}

\definecolor{charcoal}{HTML}{36454F}
\definecolor{cinerous}{HTML}{98817B}
\definecolor{feldgrau}{HTML}{4D5D53}
\definecolor{glaucous}{HTML}{6082B6}
\definecolor{arsenic}{HTML}{3B444B}
\definecolor{xanadu}{HTML}{738678}

\definecolor{firebrick}{HTML}{B22222}
\definecolor{orangered}{HTML}{FF4500}
\definecolor{tomato}{HTML}{FF6347}

\definecolor{purpletaupe}{HTML}{3B444B}

\definecolor{todoblue}{RGB}{0, 91, 187}

\usepackage{graphicx,epsfig,verbatim,enumerate}
\usepackage{amssymb,amsmath,textcomp}
\usepackage{ifthen}

\usepackage{tabularx}
\usepackage{multirow, makecell}
\newcolumntype{C}{>{\centering\arraybackslash}X}
\newcolumntype{L}{>{\raggedright\arraybackslash}X}
\newcolumntype{R}{>{\raggedleft\arraybackslash}X}
\usepackage{floatrow}

\usepackage{mathtools}

\newboolean{twocolswitch}

\newcommand{\sindex}[1]{}
\newcommand{\nindex}[1]{}

\newcommand{\etal}{\textit{et al.}}
\newcommand{\www}[1]{\url{#1}}

\newcommand{\Req}[1]{Eq.~(\ref{#1})}

\usepackage{lettrine}

\newcommand{\lexiconsymbol}{\mathcal{D}}
\newcommand{\alltweetsym}{\textnormal{AT}}
\newcommand{\organictweetsym}{\textnormal{OT}}
\newcommand{\retweetsym}{\textnormal{RT}}

\newcommand{\lexiconAT}{\lexiconsymbol_{t,\ell;n}}
\newcommand{\lexiconOT}{\lexiconsymbol^{\textnormal{(\organictweetsym)}}_{t,\ell;n}}
\newcommand{\lexiconRT}{\lexiconsymbol^{\textnormal{(\retweetsym)}}_{t,\ell;n}}

\newcommand{\RTrate}{R}
\newcommand{\relativeRTrate}{R^{\textnormal{rel}}}
\newcommand{\MSFE}{\textnormal{MSFE}}
\setboolean{twocolswitch}{true}

\begin{document}

\title{\protect
  Storywrangler: A massive exploratorium for \\
sociolinguistic, cultural, socioeconomic, 
and political timelines using Twitter




}

\author{
\firstname{Thayer}
\surname{Alshaabi}
}
\email{thayer.alshaabi@uvm.edu}
\affiliation{
  Vermont Complex Systems Center,
  Computational Story Lab,
  University of Vermont,
  Burlington, VT 05405.
  }
 \affiliation{
  Department of Computer Science,
  University of Vermont,
  Burlington, VT 05405.
  }

\author{
  \firstname{Jane L.}
  \surname{Adams}
}
\thanks{Equal contribution}
\affiliation{
  Vermont Complex Systems Center,
  Computational Story Lab,
  University of Vermont,
  Burlington, VT 05405.
  }

\author{
  \firstname{Michael V.}
  \surname{Arnold}
}
\thanks{Equal contribution}
\affiliation{
  Vermont Complex Systems Center,
  Computational Story Lab,
  University of Vermont,
  Burlington, VT 05405.
  }

\author{
  \firstname{Joshua R.}
  \surname{Minot}
}
\thanks{Equal contribution}
\affiliation{
  Vermont Complex Systems Center,
  Computational Story Lab,
  University of Vermont,
  Burlington, VT 05405.
  }

\author{
    \firstname{David R.}
    \surname{Dewhurst}
}
\affiliation{
  Vermont Complex Systems Center,
  Computational Story Lab,
  University of Vermont,
  Burlington, VT 05405.
  }
\affiliation{
  Charles River Analytics, 
  Cambridge, MA 02138.
  }

\author{
  \firstname{Andrew J.}
  \surname{Reagan}
}
\affiliation{
  MassMutual Data Science, 
  Amherst, MA 01002.
  }

\author{
  \firstname{Christopher M.}
  \surname{Danforth}
}
\affiliation{
  Vermont Complex Systems Center,
  Computational Story Lab,
  University of Vermont,
  Burlington, VT 05405.
  }
\affiliation{
  Department of Mathematics \& Statistics,
  University of Vermont,
  Burlington, VT 05405.
  }
 \affiliation{
  Department of Computer Science,
  University of Vermont,
  Burlington, VT 05405.
  }

\author{
  \firstname{Peter Sheridan}
  \surname{Dodds}
}
\email{peter.dodds@uvm.edu}
\affiliation{
  Vermont Complex Systems Center,
  Computational Story Lab,
  University of Vermont,
  Burlington, VT 05405.
  }
 \affiliation{
  Department of Computer Science,
  University of Vermont,
  Burlington, VT 05405.
  }
  \affiliation{
  Department of Mathematics \& Statistics,
  University of Vermont,
  Burlington, VT 05405.
  }

\date{\today}

\begin{abstract}
  \protect
  In real-time, Twitter strongly imprints world events, popular culture, and the day-to-day, recording an ever growing compendium of language change.
Vitally, and absent from many standard corpora such as books and news archives, Twitter also encodes popularity and spreading through retweets.
Here, we describe Storywrangler, an ongoing curation of over 100 billion tweets containing 1 trillion 1-grams from 2008 to 2021.
For each day, we break tweets into 1-, 2-, and 3-grams across 100+ languages, generating frequencies for words, hashtags, handles, numerals, symbols, and emojis.
We make the dataset available through an interactive time series viewer, and as downloadable time series and daily distributions.
Although Storywrangler leverages Twitter data, 
our method of tracking dynamic changes in $n$-grams 
can be extended to any temporally evolving corpus.
Illustrating the instrument's potential, we present example use cases including social amplification,
the sociotechnical dynamics of famous individuals,
box office success, and social unrest. 
\end{abstract}

\maketitle

\section{Introduction}
\label{sec:storywrangler.introduction}

Our collective memory lies in our
recordings---in our
written texts, artworks, photographs, audio, and video---and
in our retellings and reinterpretations of that which becomes history.
The relatively recent digitization of historical texts,
from books~\cite{michel2011a,pechenick2015a,christenson2011a,gerlach2020a}
to
news~\cite{nytimescorpus2008a,beeferman2019a,hollink2016a,hong2020a}
to
folklore~\cite{mieder2004a,abello2012a,tangherlini2013a,vuong2018a}
to
governmental records~\cite{woolley2008a}, 
has enabled compelling computational analyses
across many fields~\cite{abello2012a,primack2009a,allen2020a}.
But books, news, and other formal records only constitute a specific type of 
text---carefully edited to deliver a deliberate message to a target audience. 
Large-scale constructions of historical corpora also often fail to encode a fundamental characteristic: 
Popularity (i.e., social amplification).
How many people have read a text?
How many have retold a news story to others?


For text-based corpora, 
we are confronted with the challenge of sorting through different aspects of popularity of 
$n$-grams---sequences of $n$ `words' in a text that are formed 
by contiguous characters, numerals, symbols, emojis, etc.
An $n$-gram may or may not be part of a text's lexicon, 
as the vocabulary of a text gives a base sense of what that text may span meaning-wise~\cite{peirce1906a}.
For texts, 
it is well established that $n$-gram frequency-of-usage (or Zipf) distributions 
are heavy-tailed~\cite{zipf1949a}.
Problematically, 
this essential character of natural language is readily misinterpreted as indicating cultural popularity.
For a prominent example, the Google Books $n$-gram corpus~\cite{michel2011a}, 
which in part provides inspiration for our work here, presents year-scale, $n$-gram frequency time series where each book, in principle, counts only once~\cite{pechenick2015a}.
All cultural fame is stripped away.
The words of George Orwell's 1984 or Rick Riordan's Percy Jackson books, 
indisputably read and re-read by many people around the world, 
count as equal to the words in the least read books published in the same years.
And yet, 
time series provided by the Google Books $n$-gram viewer have regularly been erroneously conflated 
with the changing interests of readers 
(e.g., the apparent decline of sacred words~\cite{bohannon2010a,pechenick2015a,koplenig2017a,pechenick2017a,merritt2018a}).
Further compounded with an increase of scientific literature throughout the 20th Century, 
the corpus remains a deeply problematic database for investigations of sociolinguistic and cultural trends.
It is also very difficult to measure cultural popularity.
For a given book, 
we would want to know sales of the book over time, 
how many times the book has been actually read, 
and to what degree a book becomes part of broader culture.
Large-scale corpora capturing various aspects of popularity exist~\cite{allen2020a}, 
but are hard to compile as the relevant data is either prohibitively expensive or closed 
(e.g., Facebook), and, even when accessible, 
may not be consistently recorded over time (e.g., Billboard's Hot 100).


Now, 
well into the age of the internet, 
our recordings are vast, 
inherently digital,
and capable of being created and shared in the moment.
People, 
news media, governmental bodies, corporations, bots, and many other entities 
all contribute constantly to giant social media platforms.
When open, 
these services provide an opportunity for us to attempt to track myriad statements, reactions, 
and stories of large populations in real-time.
Social media data allows us to explore day-to-day conversations by millions of ordinary people and celebrities 
at a scale that is scarcely conventionalized and recorded. 
And crucially, 
when sharing and commenting mechanisms are native to a social media platform, 
we can quantify popularity of a trending topic and social amplification of a contemporary cultural phenomenon.

In this paper, 
we present Storywrangler---a natural language processing framework that extracts, 
ranks, and organizes $n$-gram time series for social media.
Storywrangler provides an analytical lens to examine discourse on social media, 
carrying both the voices of famous individuals---political figures and celebrities---and 
the expressions of the many.
With a complex, 
ever-expanding fabric of time-stamped messages, 
Storywrangler allows us to capture storylines in over 150 languages in real time.

For a primary social media source, 
we use Twitter for several reasons, while acknowledging its limitations.
Our method of extracting and tracking dynamic changes of $n$-grams can in principle be 
extended to any social media platform  
(e.g., Facebook, Reddit, Instagram, Parler, 4Chan, Weibo).

Twitter acts as a distributed sociotechnical sensor system~\cite{hong2011a,younus2011a}. 
Using Storywrangler, 
we can trace major news events and 
stories---from serious matters such as natural disasters~\cite{sakaki2010a,pickard2011a,gao2011a,lampos2010a,culotta2010a} 
and political events~\cite{steinert2015a} to entertainment such as sports, music, and movies.
Storywrangler also gives us insights into 
discourse around these topics and myriad others
including,
violence,
racism,
inequality,
employment,
pop culture (e.g., fandom), 
fashion trends, 
health,
metaphors, 
emerging memes,
and 
the quotidian.



We can track and explore discussions 
surrounding political and cultural movements that 
are born and nurtured in real-time over social media 
with profound ramifications for society  
(e.g., \#MeToo, \#BlackLivesMatter, \#QAnon).
Modern social movements of all kinds, 
may develop a strong imprint on social media, over years in some cases, before becoming widely known and discussed.

Twitter and social media in general 
differ profoundly from traditional news and print media in various dimensions.
Although amplification is deeply uneven, 
vast numbers of people may now express themselves 
to a global audience on any subject they choose 
(within limits of a service, and not without potential consequences).
Unlike journalists, columnists, or book authors, 
people can instantly record and share messages in 
milliseconds.
Importantly from a measurement perspective, 
this is a far finer temporal resolution than would be reasonably needed 
to explore sociocultural phenomena or reconstruct major events.
The eye witness base for major events is now no longer limited to those physically present because of growing, 
decentralized live-streaming through various social media platforms.
Social media thus enables, and not without peril, 
a kind of mass distributed journalism.




A crucial feature of Storywrangler is the explicit encoding of $n$-gram popularity,
 which is enabled by Twiter's social amplification mechanisms: Retweets and quote tweets. 
For each day and across languages, we create Zipf distributions for: 
(1) $n$-grams from originally authored messages (\organictweetsym), 
excluding all retweeted material (\retweetsym); and 
(2) $n$-grams from all Twitter messages (\alltweetsym).
For each day, we then have three key levels of popularity: 
$n$-gram lexicon, 
$n$-gram usage in organic tweets (originally authored tweets), 
and the rate at which a given $n$-gram is socially amplified 
(i.e., retweeted) on the platform.
Our data curation using Storywrangler yields a rich dataset, 
providing an interdisciplinarily resource for researchers 
to explore transitions in social amplification by reconstructing $n$-gram Zipf distributions 
with a tunable fraction of retweets.


We structure our paper as follows.
In Sec.~\ref{sec:storywrangler.data}, 
we describe in brief our instrument, dataset, and the Storywrangler site
which provides day-scale $n$-gram time series datasets
for $n$=1, 2, and 3,
both as time series and as daily Zipf distributions.
In Sec.~\ref{sec:storywrangler.examples},
we showcase a group of example analyses, arranged by increasing complication:
Simple $n$-gram rank time series
(Sec.~\ref{subsec:storywrangler.timelines});
qualitative comparison to other prominent social signals of Google Trends and cable news
(Sec.~\ref{subsec:storywrangler.comparison});
Contagiograms, time series showing social amplification
(Sec.~\ref{subsec:storywrangler.contagiograms});
analysis for identifying and exploring narratively trending storylines 
(Sec.~\ref{subsec:storywrangler.trending});
and an example set of case studies bridging $n$-gram time series with
disparate data sources to study famous individuals, box office success, and social unrest
(Sec.~\ref{subsec:storywrangler.casestudies}).
In our concluding remarks in
Sec.~\ref{sec:storywrangler.concludingremarks},
we outline some potential future developments for Storywrangler.

\begin{figure*}[tp!]
  \includegraphics[width=\textwidth]{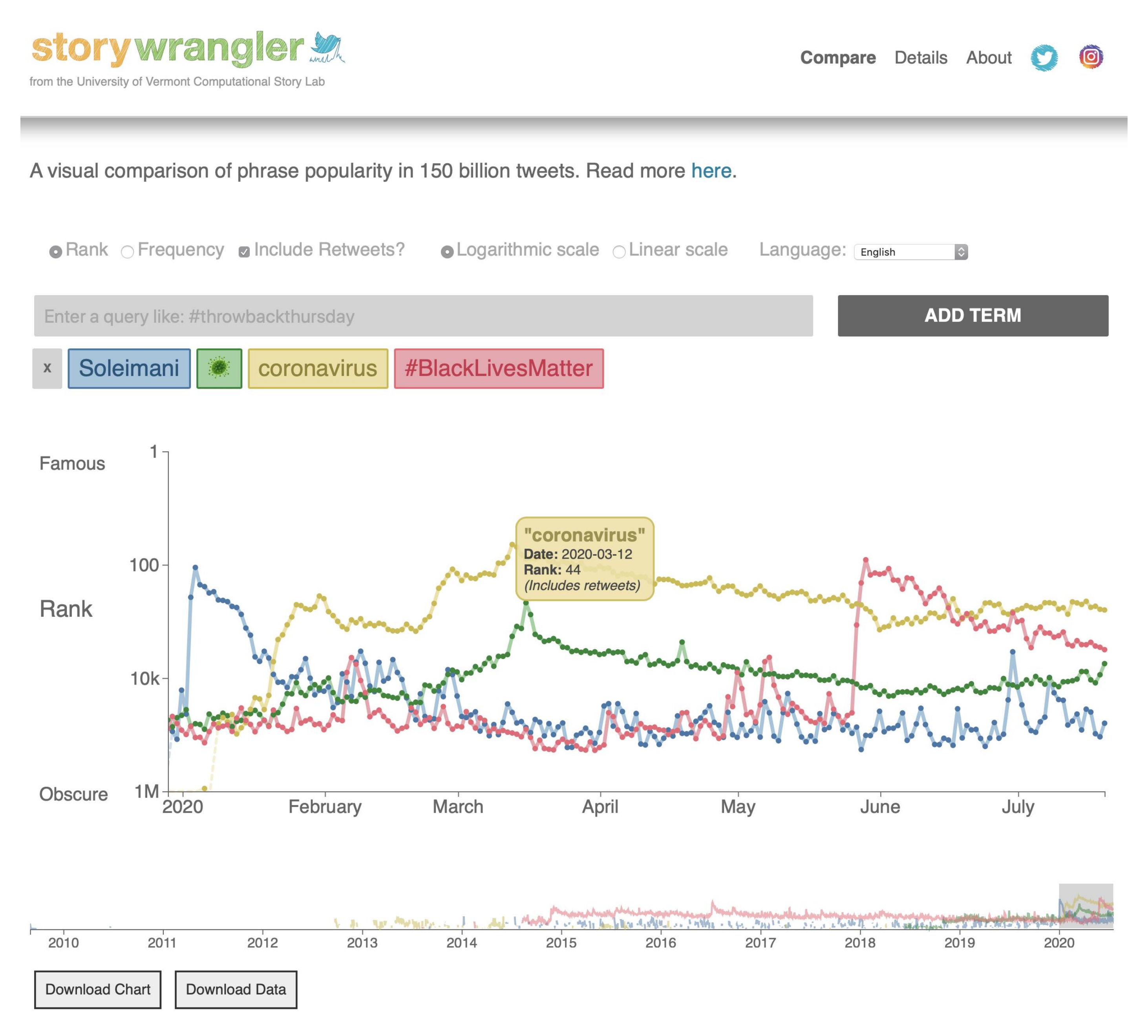}
  \caption{
    \textbf{Interactive online viewer.}
    Screenshot of the Storywrangler site showing example
    Twitter $n$-gram time series
    for the first half of 2020.
    The series reflect three global events:
    The assassination of Iranian general Qasem Soleimani by the United States
    on 2020-01-03,
    the COVID-19 pandemic 
    (the virus emoji
    \protect
    \includegraphics[height=1em]{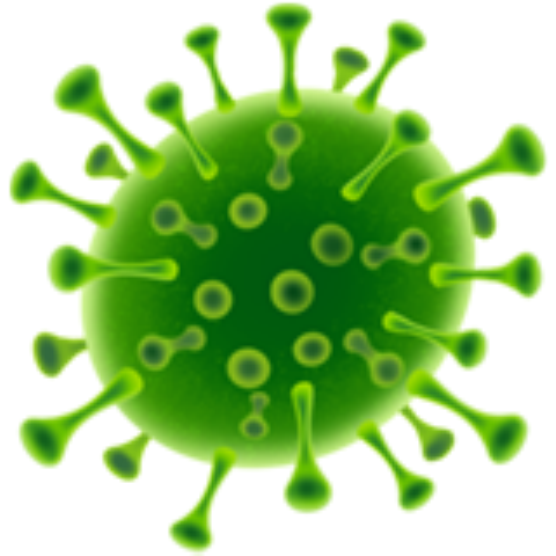}
    and `coronavirus'),
    and the Black Lives Matter protests following the murder of George Floyd
    by Minneapolis police (`\#BlackLivesMatter').
    The $n$-gram Storywrangler dataset for Twitter records the
    full ecology of text elements,
    including punctuation, hashtags, handles, and emojis.
    The default view is for $n$-gram (Zipfian) rank at the day scale (Eastern Time),
    a logarithmic y-axis,
    and for retweets to be included.
    These settings can be respectively switched to normalized frequency, linear scale,
    and organic tweets (\organictweetsym) only.
    The displayed time range can be adjusted with the selector at the bottom,
    and all data is downloadable.
  }
  \label{fig:storywrangler.viewer}
\end{figure*}

\section{Data and Methods}
\label{sec:storywrangler.data}

\subsection{Overview of Storywrangler}
\label{subsec:storywrangler.instrument}

We draw on a storehouse of messages comprising roughly 10\% of all tweets 
collected from 2008-09-09 onwards, 
and covering 150+ languages.
In previous work~\cite{alshaabi2020growing}, 
we described how we re-identified the languages of all tweets in our collection using 
FastText-LID\footnote{\url{https://fasttext.cc/docs/en/language-identification.html}}~\cite{joulin2016a,bojanowskia}, 
uncovering a general increase in retweeting across Twitter over time.
A uniform language re-identification was needed 
as Twitter's own real-time identification algorithm was introduced in late 2012
and then adjusted over time, 
resulting in temporal inconsistencies 
for long-term streaming collection of tweets~\cite{dodds2020long}.
While we can occasionally observe subtle cues of regional dialects and slang, 
especially on a non-mainstream media platform like Twitter, 
we still classify them based on their native languages.
Date and language are the only metadata we incorporate into our database.
For user privacy in particular, 
we discard all other information associated with a tweet.

For each day $t$ (Eastern Time encoding) and for each language $\ell$, 
we categorize tweets into two classes: 
Organic tweets (\organictweetsym), 
and retweets (\retweetsym). 
To quantify the relative effect of social amplification, 
we group originally authored posts---including the comments found in quote tweets 
but not the retweeted content they refer to---into what we call organic tweets. 
We break each tweet into 1-grams, 2-grams, and 3-grams. 
Although we can identify tweets written in continuous-script-based languages 
(e.g., Japanese, Chinese, and Thai), 
our current implementation does not support breaking them into $n$-grams.

We accommodate all Unicode characters, 
including emojis, 
contending with punctuation as fully as possible 
(see Appendix~\ref{appx:storywrangler.dataset} for further details).
For our application, 
we designed a custom $n$-gram tokenizer to preserve 
handles, hashtags, date/time strings, and links 
(similar to the tweet tokenizer in the NLTK library~\cite{loper2002a}).
Although some older text tokenization toolkits followed different criteria, 
our protocol is consistent with modern computational linguistics 
for social media data~\cite{hong2020a,bevensee2020a}.

We derive three essential measures for each $n$-gram: 
raw frequency (or count), 
normalized frequency (interpretable as probability), 
and rank, generating the corresponding Zipf distributions~\cite{zipf1949a}.
We perform this process for all tweets (\alltweetsym), 
organic tweets (\organictweetsym), 
and (implicitly) retweets (\retweetsym).
We then record $n$-grams along with ranks, raw frequencies, normalized frequencies 
for all tweets and organic tweets in a single file, 
with the default ordering according to $n$-gram prevalence in all tweets.

\subsection{Notation and measures}
\label{subsec:storywrangler.notation}

We write an $n$-gram by $\tau$ and
a day's lexicon for language $\ell$---the set of distinct $n$-grams found in all tweets (AT)
for a given date $t$---by $\lexiconAT$.
We write $n$-gram raw frequency as $f_{\tau,t,\ell}$,
and compute its usage rate in all tweets written in language $\ell$
as
\begin{equation}
  p_{\tau,t,\ell}
  =
  \frac{f_{\tau,t,\ell}}
       {\sum_{\tau' \in \lexiconAT} f_{\tau',t,\ell}}.
  \label{eq:storywrangler.freq}
\end{equation}
We further define the set of unique language $\ell$ $n$-grams found in organic tweets
as
$\lexiconOT$,
and the set of unique $n$-grams found in
retweets as
$\lexiconRT$
(hence 
$
\lexiconAT
=
\lexiconOT
\cup
\lexiconRT
$).
The corresponding normalized frequencies for these
two subsets of $n$-grams are then:
\begin{equation} 
  p_{\tau,t,\ell}^{(\organictweetsym)}
  =
  \frac{f_{\tau,t,\ell}^{(\organictweetsym)}}{
    \sum_{\tau'
      \in
      \lexiconOT}
      f_{\tau',t,\ell}^{(\organictweetsym)}}, \ \textnormal{and}
  \label{eq:storywrangler.f_ot}
\end{equation}
\begin{equation} 
  p_{\tau,t,\ell}^{(\retweetsym)}
  =
  \frac{f_{\tau,t,\ell}^{(\retweetsym)}}{
    \sum_{\tau'
      \in
      \lexiconRT}
      f_{\tau',t,\ell}^{(\retweetsym)}}.
  \label{eq:storywrangler.f_rt}
\end{equation}

We rank $n$-grams by raw frequency of usage using fractional ranks for ties.
The corresponding notation is:
\begin{equation}
  r_{\tau,t,\ell},
  \
  r_{\tau,t,\ell}^{(\organictweetsym)},
  \
  \textnormal{and}
  \
  r_{\tau,t,\ell}^{(\retweetsym)}.
  \label{eq:storywrangler.ranks}
\end{equation}

\begin{figure*}[tp!]
  \centering	
  \includegraphics[width=.9\textwidth]{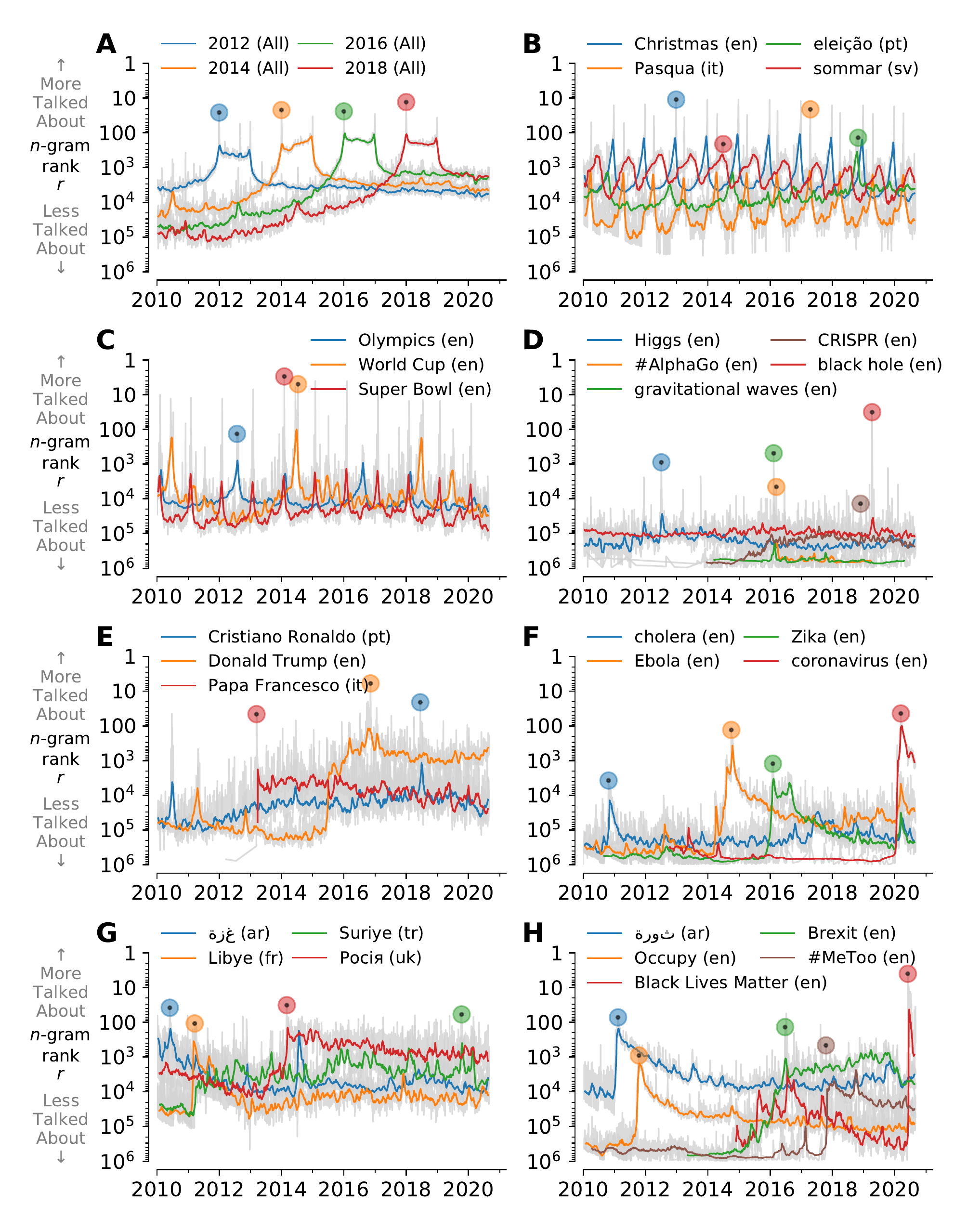}
    \caption{
      \label{fig:storywrangler.chart}
        \textbf{Thematically connected $n$-gram time series.}
        For each $n$-gram,
        we display daily rank in gray overlaid by a centered monthly rolling average (colored lines),
        and highlight the $n$-gram's overall highest rank with a solid disk.
        \textbf{A.}
        Anticipation and memory of calendar years for all of Twitter.
        \textbf{B.}
        Annual and periodic events: 
        Christmas in English (blue), 
        Easter in Italian (orange), 
        election in Portuguese (green), 
        and summer in Swedish (red).
        \textbf{C.}
        Attention around international sports in English: 
        Olympics (blue),
        FIFA world cup (orange), 
        and Super Bowl (red).
        \textbf{D.}
        Major scientific discoveries and technological innovations in English.
        \textbf{E.}
        Three famous individuals in relevant languages:
        Ronaldo (Portuguese), 
        Trump (English), and 
        Pope Francis (Italian).
        \textbf{F.}
        Major infectious disease outbreaks.
        \textbf{G.}
        Conflicts:
        Gaza in Arabic (blue),
        Libya in French (orange),
        Syria in Turkish (green),
        and Russia in Ukrainian (red).
        \textbf{H.}
        Protest and movements:
        Arab Spring (Arabic word for `revolution', blue),
        Occupy movement (English, orange),
        Brexit campaign (English, green),
        \#MeToo movement (English, brown), 
        and Black Lives Matter protests (English, red).
    }
\end{figure*}

\subsection{User interface}
\label{subsec:storywrangler.site}

We make interactive times series based on our $n$-gram dataset viewable
at \href{storywrangling.org}{storywrangling.org}.
In Fig.~\ref{fig:storywrangler.viewer},
we show a screenshot of the site displaying rank time series for the first half of 2020
for
`Soleimani',
the virus emoji
\includegraphics[height=1em]{figures/microbe_1f9a0.pdf},
`coronavirus',
and
`\#BlackLivesMatter'.
Ranks and normalized frequencies for $n$-grams are relative to $n$-grams
with the same $n$, and in the online version
we show time series on separate axes below the main comparison plot.

For each time series, hovering over any data point will pop up
an information box.
Clicking on a data point will take the user to Twitter's search results
for the $n$-gram for the span of three days centered on the given date.
All time series are shareable and downloadable through the site,
as are daily Zipf distributions for the top million ranked $n$-grams in each language.
Retweets may be included (the default) or excluded,
and 
the language, vertical scale, and time frame may all be selected.

\section{Results and discussion}
\label{sec:storywrangler.examples}

\subsection{Basic rank time series}
\label{subsec:storywrangler.timelines}

In Fig.~\ref{fig:storywrangler.chart}, we show
rank time series for eight sets of $n$-grams from all tweets
(i.e., including retweets).
The $n$-gram groups move from simple to increasingly complex in theme,
span a number of languages,
and display a wide range of sociotechnical dynamics.
Because of an approximate obeyance of Zipf's law 
($f \sim r^{-\theta}$),
we observe that normalized frequency of usage time series 
match rank time series in basic form.
We use rank as the default view for its straightforwardness.

Starting with time and calendars, Fig.~\ref{fig:storywrangler.chart}A gives a sense of how
years are mentioned on Twitter.
The dynamics show an anticipatory 
growth, plateau, and then rapid decay, with each
year's start and finish marked by a spike.

Figs.~\ref{fig:storywrangler.chart}B and C
show calendrically anchored rank time series for
seasonal, religious, political, and sporting events that recur at the scale of years
in various languages.
Periodic signatures at the day, week, and year scale are prominent
on Twitter, reflecting the dynamics of the Earth, moon, and sun.
Easter (shown in Italian) 
in particular combines cycles of all three.
Major sporting events produce time series with strong anticipation,
and can reach great heights of attention as exemplified
by a peak rank of $r=3$ for
`Super Bowl'
on 2014-02-02.

We move 
to scientific announcements in Fig.~\ref{fig:storywrangler.chart}D
with the 2012 discovery of the Higgs boson particle (blue),
detection of gravitational waves (green),
and the first imaging of a black hole (red).
For innovations, 
we show the time series of `\#AlphaGo'---the first
artificial intelligence program
to beat the human Go champion (orange),
along with the development of CRISPR technology for editing genomes (brown).
We see that time series for scientific advances
generally show shock-like responses with
little anticipation or memory~\cite{dewhurst2020shocklet}.
CRISPR is an exception for these few examples as
through 2015, it moves
to a higher, enduring state of being referenced.

Fame is the state of being talked about and
famous individuals are well reflected on Twitter~\cite{dodds2019fame}.
In Fig.~\ref{fig:storywrangler.chart}E,
we show 
time series for the Portuguese football player Cristiano Ronaldo,
the 45th US president Donald Trump,
and Pope Francis (Papa Francesco in Italian).
All three show enduring fame,
following sudden rises
for both Trump and Pope Francis.
On November 9, 2016, 
the day after the US election, 
`Donald Trump'
rose to rank $r=6$ among all English 2-grams.

In Fig.~\ref{fig:storywrangler.chart}F,
we show example major infectious disease outbreaks
over the last decade.
Time series for pandemics are shocks followed by long relaxations,
resurging both when the disease returns in prevalance and
also in the context of new pandemics.
Cholera, ebola, and zika all experienced elevated discussion
within the context of the COVID-19 pandemic.

In Fig.~\ref{fig:storywrangler.chart}G,
we show $n$-gram signals of regional unrest and fighting.
The word for Gaza in Arabic 
tracks events of the the ongoing Israeli-Palestinian conflict.
The time series for 
`Libye' points to Op\'{e}ration Harmattan,
the 2011 French and NATO military intervention in Libya.
Similarly, 
the time series for `Syria' in Turkish
indicates the dynamics of the ongoing Syrian civil war on the region,
and the build up and intervention of the Russian military in Ukraine
is mirrored by the use of the Ukrainian word for `Russia'.

In Fig.~\ref{fig:storywrangler.chart}H,
we highlight protests and movements.
Both the time series 
for `revolution' in Arabic
and `Occupy' in English
show strong shocks followed by slow relaxations over the following years.
The social justice movements represented by
\#MeToo and 
`Black Lives Matter' appear abruptly,
and their time series show slow decays punctuated by shocks returning them
to higher ranks.
Black Lives Matter resurged after the murder of George Floyd,
with a highest one day rank
of $r=4$
occurring on 2020-06-02.
By contrast, the time series of `Brexit',
the portmanteau for the movement to withdraw
the United Kingdom from the European Union, 
builds from around the start of 2015
to the referendum in 2016, 
and then continues to climb during the years
of complicated negotiations to follow.

\begin{figure*}[tp!]
  \centering	
  \floatbox[{\capbeside\thisfloatsetup{capbesideposition={right,center},capbesidewidth=.24\textwidth}}]{figure}[0.95\FBwidth]{
    \caption{
      \label{fig:storywrangler.comparison}
        \textbf{Comparison between Twitter, Google Trends, and Cable News.}
        All time series are rescaled between 0 (low interest) to 100 (peak interest)
        to represent rate of usage relative to the highest point for the given time window. 
        For each $n$-gram (case insensitive),
        we display weekly interest over time using Storywrangler for all tweets (AT, black),
        and originally authored tweets (OT, blue),   
        comparing that with Google Trends~\cite{choi2012a} (orange),
        and Cable TV News~\cite{hong2020a} (green).
        \textbf{A.} 
        Similar social attention to working from home across media sources amid the COVID-19 pandemic.  
        \textbf{B.} 
        Discourse of unemployment continues to fluctuate on Twitter in contrast to other mainstream media sources.  
        \textbf{C.} 
        Usage of the bigram `fake news'. 
        \textbf{D.} 
        Discussion of the `QAnon' conspiracy theory. 
        \textbf{E.} 
        Mentions of `Antifa` on digital media.
        \textbf{F.} 
        Social attention of fans in various arenas as part of the ever changing pop culture. 
        \textbf{G.} 
        A growing social movement organized by longtime fans of Britney Spears regarding her conservatorship. 
        \textbf{H.} 
        Official Twitter handle for the South Korean K-pop band: TWICE. 
        \textbf{I.} 
        Conversations surrounding fitness trends, which occasionally pop up in news via commercial advertisement. 
        \textbf{J.} 
        Volatility of collective attention to fashion trends.
    }
  }{\includegraphics[width=0.75\textwidth]{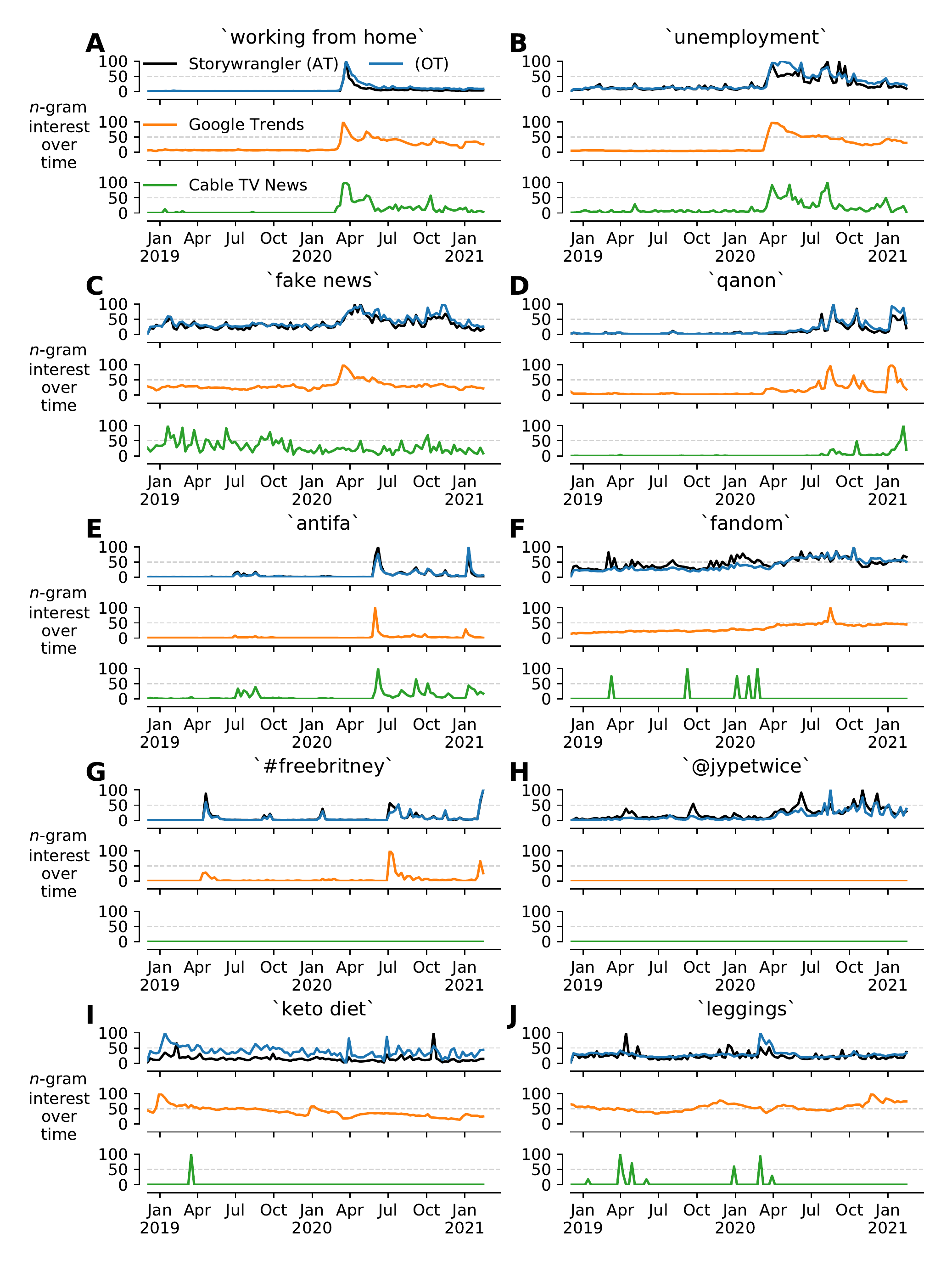}}
\end{figure*}

\subsection{Comparison to other signals}
\label{subsec:storywrangler.comparison}

To highlight key differences Storywrangler offers in contrast to other data sources, 
we display a few example comparisons in Fig.~\ref{fig:storywrangler.comparison}.
In particular, 
we compare usage rate for a set of $n$-grams using Storywrangler, 
Google Trends~\cite{choi2012a}, 
and the Stanford cable TV news analyzer~\cite{hong2020a}.

Each data source has its own unique collection 
scheme that is most appropriate to that venue. 
Google Trends provides search interest scaled relative 
to a given region and 
time.\footnote{\url{https://trends.google.com/trends/?geo=US}}
While Storywrangler is based on daily $n$-gram Zipf distributions, 
the Stanford cable TV news analyzer collects transcripts 
from most cable news outlets and breaks them into $n$-grams, 
recording screen time (seconds per day) for each term~\cite{hong2020a}. 

For the purpose of comparing $n$-gram usage across several disparate data sources, 
we take the weekly rate of usage for each term (case insensitive), 
and normalize each time series between 0 and 100 
relative to the highest observed point within the given time window.
A score of 100 represents the highest observed interest in the given term over time, 
while a value of 0 reflects low interest of that term and/or insufficient data.
We display weekly interest over time for a set of 10 terms 
using Storywrangler for all tweets 
(AT, black), 
and originally authored tweets 
(OT, blue), 
Google trends (orange), 
and cable news (green). 

In Fig.~\ref{fig:storywrangler.comparison}A, 
we show how usage of the trigram `working from home' 
peaks during March 2020 amid the COVID-19 pandemic. 
Although the term may be used in different contexts in each respective media source, 
we observe similar attention signals across all three data sources. 

Similarly, Fig.~\ref{fig:storywrangler.comparison}B 
reveals increased mentions of 
`unemployment' 
on all media platforms during the US national lockdown in April 2020. 
Individuals searching for unemployment claim forms could be
responsible for the Google trends spike, 
while 
news and social media usage of the term resulted from 
coverage of the economic crisis induced by the pandemic.
The time series for 
`unemployment' 
continues to fluctuate on Twitter, 
with distinct patterns across all tweets and originally authored tweets.

In Fig.~\ref{fig:storywrangler.comparison}C, 
we see the bigram `fake news' roiling across social media and news outlets, 
reflecting the state of political discourse in 2020. 
Indeed, 
this period saw the most sustained usage of the term 
since its initial spike following the 2016 US election. 
The term was prominently searched for on Google in March 2020 
during the early stages of the Coronavirus pandemic, 
but no corresponding spike is seen in cable news.

In Fig.~\ref{fig:storywrangler.comparison}D, 
the time series reveal attention to `QAnon' conspiracy theory 
on social media and Google Trends starting in mid 2020.
Using Storywrangler, 
we note a spike of `qanon' 
following Trump's remarks regarding violent far-right groups 
during the first presidential debate on 
September 29, 2020.\footnote{\url{https://www.nytimes.com/2020/09/30/us/politics/debate-takeaways.html}}
We see another spike of interest in October 2020 
in response to the news about a kidnapping plot 
of the governor of Michigan by 
extremists.\footnote{\url{https://www.nytimes.com/2020/10/08/us/gretchen-whitmer-michigan-militia.html}}
Although the time series using both Storywrangler and Google trends 
show sustained usage of the term in 2020,
news outlets do not exhibit similar patterns until the US Capitol insurrection 
on January 6, 2021.\footnote{\url{https://www.theguardian.com/us-news/2021/jan/09/us-capitol-insurrection-white-supremacist-terror}}

Fig.~\ref{fig:storywrangler.comparison}E shows mentions of
`antifa'---a political movement 
that drew attention in response to police violence during protests of the murder of George Floyd.\footnote{\url{https://www.nytimes.com/article/what-antifa-trump.html}} 
We note that mentions surged again in response to false flag allegations in the wake of the Capitol attack, most prominently on Twitter.

In Fig.~\ref{fig:storywrangler.comparison}F,
we display interest over time of the term 
`fandom'---a unigram that is widely used to refer to a group of people 
that share a common interest in 
creative genres, 
celebrities, 
fashion trends, 
modern tech, 
hobbies, 
etc.
While this cultural phenomenon is rarely ever recorded by traditional news outlets, 
it dates back to the enormous fan base of Sherlock Holmes 
as one of the earliest signs of modern fandom, 
with public campaigners mourning the figurative death of their fictional character in 
1893.\footnote{\url{https://www.bbc.com/culture/article/20160106-how-sherlock-holmes-changed-the-world}}
This cultural characteristic can not be easily captured 
with data sources such as Google books or search data. 
Nonetheless, 
it is intrinsic to non-mainstream media, 
illustrating the collective social attention of fans 
in various arenas as part of the ever changing digital pop culture. 

Fig~\ref{fig:storywrangler.comparison}G 
shows a recent example where longtime fans of the pop music star,
Britney Spears, 
organized and launched a social media support campaign  
in light of the controversy surrounding her conservatorship. 
Although the movement dates back to 2009, 
we see a surge of usage of the hashtag 
`\#FreeBritney' in July 2020,
after an interview with Britney's brother, 
revealing some personal details about her struggles 
and reigniting the movement on social media.
The social movement has recently gained stronger cultural currency 
after the release of a documentary film by the New York Times 
in 2021.\footnote{\url{https://www.nytimes.com/article/framing-britney-spears.html}} 

Moreover, Fig~\ref{fig:storywrangler.comparison}H 
shows interest over time of a popular South Korean pop band, `Twice'. 
Although the official handle of the band on Twitter 
(`\@jypetwice') 
is virtually absent in other data sources, 
fans and followers use handles and hashtags regularly on Twitter 
to promote and share their comments for their musical bands.  

In Figs.~\ref{fig:storywrangler.comparison}I and J, 
we see how communications and marketing campaigns of fitness trends 
such as `keto diet', 
and fashion trends such as leggings, and athleisure 
receive sustained interest on Twitter 
while only occasionally popping up in news 
via commercial advertisements on some cable channels.

 \begin{figure*}[tp!]
  \centering	
  \floatbox[{\capbeside\thisfloatsetup{capbesideposition={right,center},capbesidewidth=.30\textwidth}}]{figure}[0.95\FBwidth]{
    \caption{
      \label{fig:storywrangler.maincontagiograms}  
      \textbf{Contagiograms: Augmented time series charting the social amplification of $n$-grams.}
      In each contagiogram,
      above the basic $n$-gram rank time series,
      the top panel displays the monthly
      relative usage of each $n$-gram,
      $\RTrate_{\tau,t,\ell}$
      (\Req{eq:storywrangler.RTrate}),
      indicating whether they appear
      organically in new tweets (OT, blue), or in retweeted content (RT,
      orange).
      The shaded areas denote months when
      the balance favors spreading, suggestive of story contagion.
      The middle (second) panel then shows retweet usage of an $n$-gram relative
      to the background rate of retweeting,
      $\relativeRTrate_{\tau,t,\ell}$
      (\Req{eq:storywrangler.relativeRTrate}).
      \textbf{A--B.} The seasonal
      cycle of the $1$-gram `spring' in Finnish  is
      different than the annual cycle of the word `Carnaval' in
      Portuguese.
      Spring is often mentioned in organic
      tweets while the balance of the word `Carnaval' favors retweets
      exceeding the social contagion threshold starting from 2017.
      \textbf{C.}
      The time series for `Lionel Messi' in Spanish tweets
      exhibits a similar pattern of
      social amplification as a famous soccer player who is talked
      about regularly.
      \textbf{D.}
      The hashtag \#TGIF (`Thank God It's Friday') shows a strong
      weekly cycle, relatively unamplified on Thursday and Friday.
      \textbf{E.}
      The time series of the $1$-gram
      `virus' in French shows strong relative retweeting
      following global news about the early spread of COVID-19 in 2020.
      \textbf{F.}
      We observe mild spikes
      at the beginning of the German dialog around the withdrawal of
      the UK from the EU shifting to a even balance of the 
      $1$-gram `Brexit'  across organic and retweeted
      content.
    }
  }{\includegraphics[width=0.7\textwidth]{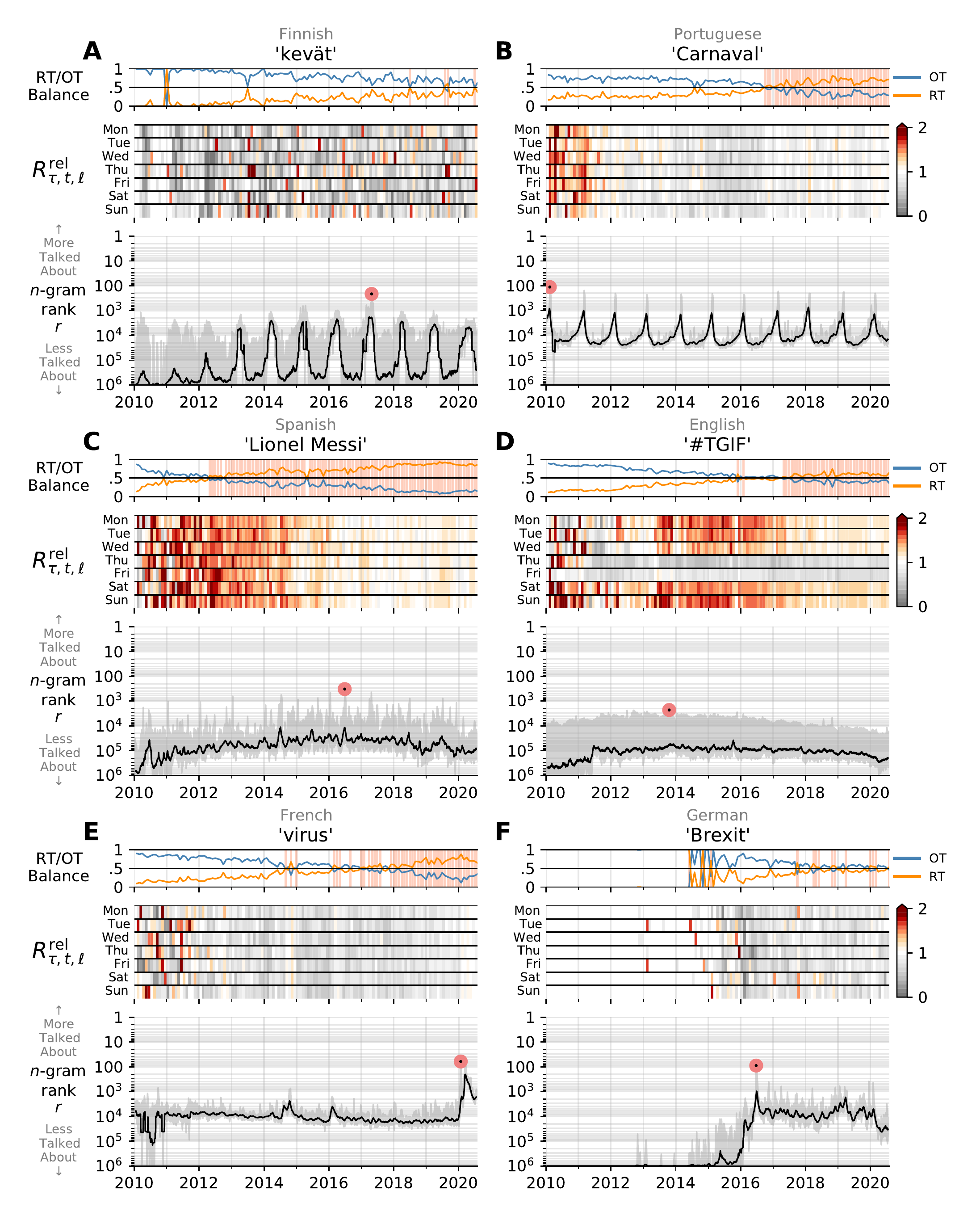}}
\end{figure*}

\subsection{Contagiograms}
\label{subsec:storywrangler.contagiograms}

While rank time series for $n$-grams give us the bare
temporal threads that make up the tapestries of major stories,
our dataset offers more dimensions to explore.
Per our introductory remarks on the limitations of text corpora, the most
important enablement of our database is the ability
to explore story amplification.

In Fig.~\ref{fig:storywrangler.maincontagiograms},
we present a set of six `contagiograms'.
With these expanded time series visualizations,
we convey the degree to which an $n$-gram is retweeted both overall and relative
to the background level of retweeting for a given language.
We show both rates because retweet rates change strongly over
time and variably so across languages~\cite{alshaabi2020growing}.

Each contagiogram has three panels.
The main panel at the bottom
charts, as before, the rank time series for a given $n$-gram.
For contagiograms running over a decade,
we show rank time series in this main panel with  month-scale smoothing (black line),
and add a background shading in gray indicating the highest and lowest rank of each week.

The top two panels of each contagiogram capture the raw and relative social amplification
for each $n$-gram.
First, the top panel displays the raw
\retweetsym/\organictweetsym\
balance,
the monthly relative volumes of each $n$-gram
in retweets (\retweetsym, orange)
and organic tweets (\organictweetsym, blue):
\begin{equation}
  \label{eq:storywrangler.RTrate}
  \RTrate_{\tau,t,\ell}
  =
  f_{\tau,t,\ell}^{(\retweetsym)}
  /
  \left(
  f_{\tau,t,\ell}^{(\retweetsym)}
  +
  f_{\tau,t,\ell}^{(\organictweetsym)}
  \right).
\end{equation}
When the balance of appearances in retweets outweighs those in organic tweets,
$\RTrate_{\tau,t,\ell} > 0.5$,
we view the $n$-gram as nominally being amplified,
and we add a solid background for emphasis.

Second,
in the middle panel of each contagiogram, we display a heatmap of the
values of
the relative amplification rate for $n$-gram $\tau$ in language $\ell$,
$\relativeRTrate_{_{\tau,t,\ell}}$.
Building on from the \retweetsym/\organictweetsym\ balance,
we define $\relativeRTrate_{\tau,t,\ell}$ as:
\begin{equation}
\label{eq:storywrangler.relativeRTrate}
\relativeRTrate_{\tau,t,\ell}
 =
\frac{
  f_{\tau,t,\ell}^{(\retweetsym)}
  /
  \left(
  f_{\tau,t,\ell}^{(\retweetsym)}
  +
  f_{\tau,t,\ell}^{(\organictweetsym)}
  \right)
}{
  \sum_{\tau'}
  f_{\tau',t,\ell}^{(\retweetsym)}
  /
  \sum_{\tau'}
  \left(
  f_{\tau',t,\ell}^{(\retweetsym)}
  +
  f_{\tau',t,\ell}^{(\organictweetsym)}
  \right)
},
\end{equation}
where the denominator gives the overall fraction of $n$-grams that are found in retweets
on day $t$ for language $\ell$.
While still averaging at month scales, 
we now do so based on day of the week.
Shades of red indicate that the relative volume of
$n$-gram $\tau$ is
being socially amplified over the baseline of retweets in language
$\ell$, $\relativeRTrate_{\tau,t,\ell} > 1$,
while gray encodes the opposite, $\relativeRTrate_{\tau,t,\ell} < 1$.

The contagiogram in Fig.~\ref{fig:storywrangler.maincontagiograms}A
for the word for `kev\"{a}t', `spring' in Finnish,
shows an expected annual periodicity.
The word has a general tendency
to appear in organic tweets more than retweets.
But this is true of Finnish words in general,
and we see that from the middle panel that
kevät is in fact relatively, if patchily, amplified
when compared to all Finnish words.
For the anticipatory periodic times series in
Fig.~\ref{fig:storywrangler.maincontagiograms}B,
we track references to
the `Carnival of Madeira' 
festival---held forty days before
Easter in Brazil.
We see `Carnival' has become increasingly amplified over time,
and has been relatively more amplified than Portuguese words
except for 2015 and 2016.

By etymological definition, renowned individuals should
feature strongly in retweets (`renown' derives from `to name again').
Lionel Messi has been one of the most talked about sportspeople on Twitter
over the last decade,
and
Fig.~\ref{fig:storywrangler.maincontagiograms}C
shows his 2-gram is strongly retweeted,
by both raw and relative measures.
(See also Fig.~\ref{fig:storywrangler.contagiograms_samples}F
for the K-pop band BTS's extreme levels of social amplification.)

Some $n$-grams exhibit a consistent weekly amplification signal.
For example, 
`\#TGIF' 
is organically tweeted on Thursdays and Fridays, but
retweeted more often throughout the rest of the week
(Fig.~\ref{fig:storywrangler.maincontagiograms}D). 
At least for those two days, 
individuals expressing relief for the coming weekend overwhelm any
advertising from the eponymous restaurant chain.

Routinely, 
$n$-grams will take off in usage and amplification
due to global events.
In Fig.~\ref{fig:storywrangler.maincontagiograms}E,
we see `virus' in French tweets holding a stable rank throughout
the 2010s before jumping in response to the COVID-19 pandemic,
and showing mildly relatively increased amplification levels.
The word 
`Brexit' 
in German has been prevalent from 2016 on,
balanced in terms of organic tweet and retweet appearances,
and generally relatively more spread than German 1-grams.

The contagiograms
in Fig.~\ref{fig:storywrangler.maincontagiograms}
give just a sample of the rich variety
of social amplification patterns that appear on Twitter.
We include some further examples in the supplementary material
in Figs.~\ref{fig:storywrangler.contagiograms_samples}
and~\ref{fig:storywrangler.contagiograms_langs}.
We provide Python package for generating arbitrary contagiograms
along with further examples
at \url{https://gitlab.com/compstorylab/contagiograms}.
The figure-making scripts interact directly with the Storywrangler database,
and offer a range of configurations.

 \begin{figure*}[tp!]
  \centering	
  \centerline{\includegraphics[width=1.1\textwidth]{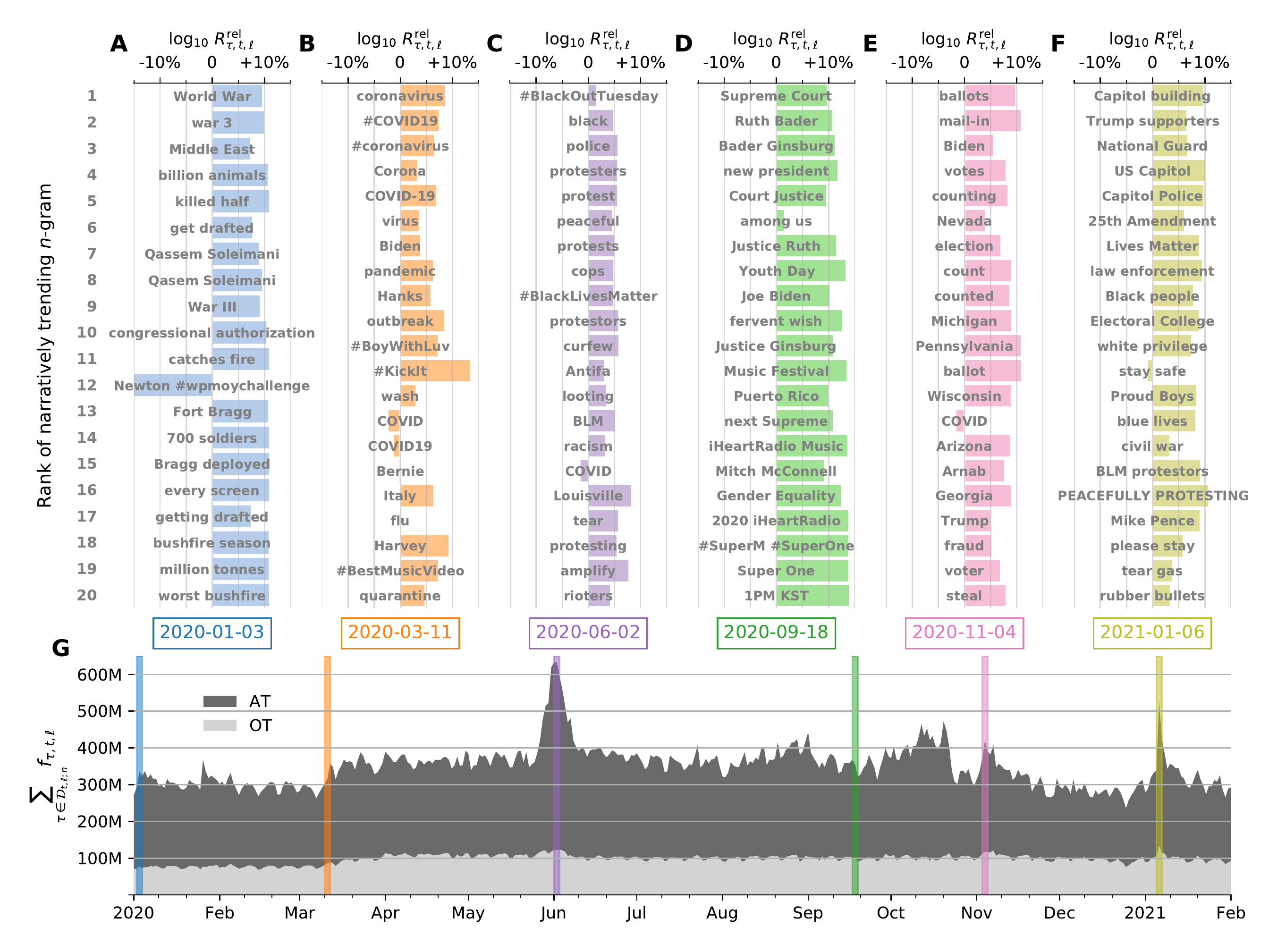}}
    \caption{
      \label{fig:storywrangler.trends}
        \textbf{Narratively trending $n$-grams.}
        We use rank-turbulence divergence (RTD)~\cite{dodds2020allotaxonometry} 
        to find the most narratively trending $n$-grams of each day 
        relative to the year before in English tweets.
        For each day, 
        we display the top 20 $n$-grams sorted by their RTD value on that day. 
        We also display the relative social amplification ratio $\relativeRTrate_{\tau,t,\ell}$
        for each $n$-gram on a logarithmic scale, 
        whereby positive values indicate strong social amplification of that $n$-gram via retweets,
        and negative values imply that the given $n$-gram is often shared in originally authored tweets.   
        \textbf{A.} 
        The assassination of Iranian general Qasem Soleimani by a US drone strike on 2020-01-03 (blue).
        \textbf{B.} 
        WHO declares COVID--19 a global Pandemic on 2020-03-11 (orange).
        \textbf{C.} 
        Mass protests against racism and police brutality on 2020-06-02 (purple).
        \textbf{D.} 
        Death of US Supreme Court justice Ruth Ginsburg from complications of pancreatic cancer on 2020-09-18 (green).
        \textbf{E.} 
        The 2020 US presidential election held on 2020-11-04 (pink).
        \textbf{F.} 
        The deadly insurrection of the US Capitol on 2021-01-06 (yellow).
        \textbf{G.} 
        Daily $n$-gram volume (i.e., number of words) 
        for all tweets (AT, grey), and organic tweets (OT, light-grey).
    }
\end{figure*}

\subsection{Narratively trending storylines}
\label{subsec:storywrangler.trending}

Besides curating daily Zipf distributions, Storywrangler serves 
as an analytical tool to examine and explore the lexicon of emerging storylines in real-time. 
Using rank-turbulence divergence (RTD)~\cite{dodds2020allotaxonometry}, 
we examine the daily rate of usage of each $n$-gram
assessing the subset of $n$-grams that have become most inflated in relative usage.
For each day $t$, 
we compute RTD for each $n$-gram $\tau$ relative to the year before $t'$,
setting the parameter $\alpha$ to 1/4 to examine the lexical turbulence of social media data such that:
\begin{align}
     \delta D^{\textnormal{R}}_{\tau} =
     \bigg| \dfrac{1}{r_{\tau, t, \ell}^{\alpha}} - \dfrac{1}{r_{\tau, t', \ell}^{\alpha}} \bigg|^{1 / (\alpha + 1)}; 
     (\alpha=1/4).
\end{align}
Although our tool uses RTD to determine dramatic shifts in relative usage of $n$-grams, 
other divergence metrics will yield similar lists.

In Fig.~\ref{fig:storywrangler.trends}, 
we show an example analysis of all English tweets for a few days of interest in 2020. 
First, 
we determine the top 20 narratively dominate $n$-grams of each day using RTD, 
leaving aside links, emojis, handles, and stop words but keeping hashtags.  
Second, 
we compute the relative social amplification ratio $\relativeRTrate_{\tau,t,\ell}$ 
to examine whether a given $n$-gram $\tau$ is prevalent in originally authored tweets, 
or socially amplified via retweets on day $t$. 
For ease of plotting, 
we have further chosen to display $\relativeRTrate_{\tau,t,\ell}$ at a logarithmic scale. 
Positive values of $\log_{10} \relativeRTrate_{\tau,t,\ell}$ 
imply strong social amplification of $\tau$, 
whereas negative values show that $\tau$ is 
relatively more predominant in organic tweets.

Fig.~\ref{fig:storywrangler.trends}A 
gives us a sense of the growing discussions and fears of a global warfare 
following the assassination of Iranian general Qasem Soleimani 
by a US drone airstrike on 
2020-01-03.\footnote{\url{https://www.nytimes.com/2020/01/02/world/middleeast/qassem-soleimani-iraq-iran-attack.html}}
While most of the terms are socially amplified, 
we note that the bigram 
`Newton \#wpmoychallenge' 
was trending in organic tweets, 
reflecting the ongoing campaign and nomination of Cam Newton 
for Walter Payton NFL Man of the Year 
Award---an annual reward that is granted for an NFL player 
for their excellence and 
contributions.\footnote{\url{https://www.panthers.com/news/cam-newton-named-walter-payton-nfl-man-of-the-year-nominee}}

In Fig.~\ref{fig:storywrangler.trends}B, 
we see how conversations of the Coronavirus disease 
becomes the most prevailing headline 
on Twitter with the World Health Organization (WHO) 
declaring COVID--19 a global Pandemic 
on 2020-03-11.\footnote{\url{https://www.washingtonpost.com/health/2020/03/11/who-declares-pandemic-coronavirus-disease-covid-19/}}

In light of the social unrest sparked by the murder of George Floyd in Minneapolis, 
we observe the growing rhetoric of the Black Lives Matter movement on Twitter 
driven by an enormous increase of retweets in Fig.~\ref{fig:storywrangler.trends}C. 
The top narratively trending unigram is 
`\#BlackOutTuesday'---a newborn movement that matured overnight on social media, 
leading to major music platforms such as Apple and Spotify 
to shut down their operations 
on 2020-06-02 
in support of the nationwide protests against racism 
and police brutality.\footnote{\url{https://www.nytimes.com/2020/06/02/arts/music/what-blackout-tuesday.html}} 

In Fig.~\ref{fig:storywrangler.trends}D, 
we see the name of the US Supreme Court justice Ruth Bader Ginsburg amplified on Twitter, 
mourning her death from complications of pancreatic cancer 
on 2020-09-18.\footnote{\url{https://www.npr.org/2020/09/18/100306972/justice-ruth-bader-ginsburg-champion-of-gender-equality-dies-at-87}}
We also see names of politicians embodying the heated discourse on Twitter 
preceding the first US presidential debate. 
Emerging pop culture trends can also be observed in the anticipation of the first album 
by a K-pop South Korean band `SuperM', 
entitled `Super One'.\footnote{\url{https://en.wikipedia.org/wiki/Super_One_(album)}}

In Fig.~\ref{fig:storywrangler.trends}E, 
we see names of swing states and political candidates come to the fore during the US presidential election held on 2020-11-04. 
We observe another surge of retweets during 
the storming of the US Capitol by Trump supporters on 2021-01-06. 
Fig.~\ref{fig:storywrangler.trends}F shows the top 20 prevalent bigrams emerging on Twitter 
in response to the deadly insurrection. \\

\bigskip
In Fig.~\ref{fig:storywrangler.trends}G, 
we display the daily daily $n$-gram volume (i.e., number of words)
throughout the year for all tweets (AT, grey), 
and organic tweets (OT, light-grey).

We provide more examples 
in Appendix~\ref{appx:storywrangler.trending_ngrams}
and
Figs.~\ref{fig:storywrangler.trending_1grams}--\ref{fig:storywrangler.trending_2grams}, 
demonstrating the wide variety of sociocultural and sociotechnical phenomena 
that can be identified and examined using Storywrangler.

\begin{figure*}[tp!]
  \centering
  \floatbox[{\capbeside\thisfloatsetup{capbesideposition={right,center},capbesidewidth=.23\textwidth}}]{figure}[0.95\FBwidth]{
    \caption{
      \label{fig:storywrangler.studies}
      \textbf{Three case studies joining Storywrangler with other data sources.}
      \textbf{A.}
      Monthly rolling average of rank $\langle r \rangle$ for the
      top-5 ranked Americans born in the last century in
      each category for a total of 960 individuals found in the
      Pantheon dataset~\cite{yu2016a}.
      \textbf{B.}
      Kernel density estimation for the top rank 
      $r_{\textnormal{min}}$ achieved by 751 personalities in the film and
      theater industry as a function of their age.
      \textbf{C.}
      Rank time series for example movie titles showing anticipation and decay.
      \textbf{D.}
      Contrasting with \textbf{C},
      rank time series for TV series titles.
      \textbf{E--F.}
      Time series and half-life revenue comparison for 636
      movie titles with gross revenue at or above the 95th
      percentile released between 2010-01-01 and
      2017-07-31~\cite{harper2015a}.
      \textbf{G--H.}
      The Storywrangler dataset can also be
      used to potentially predict political and financial turmoil.
      Percent change in the words `rebellion' and `crackdown' 
      in month $m$ are significantly associated with percent change in a
      geopolitical risk index in month $m$+1~\cite{caldara2017a}.
      \textbf{G.} 
      Percent change time series.  
      \textbf{H.} 
      Distributions of coefficients of a fit linear model.
      See Appendix~\ref{appx:storywrangler.pantheon_case_study},
      \ref{appx:storywrangler.movie_case_study},
      and
      \ref{appx:storywrangler.risk_case_study}
      for details of each study.
    }
  }{\includegraphics[width=0.75\textwidth]{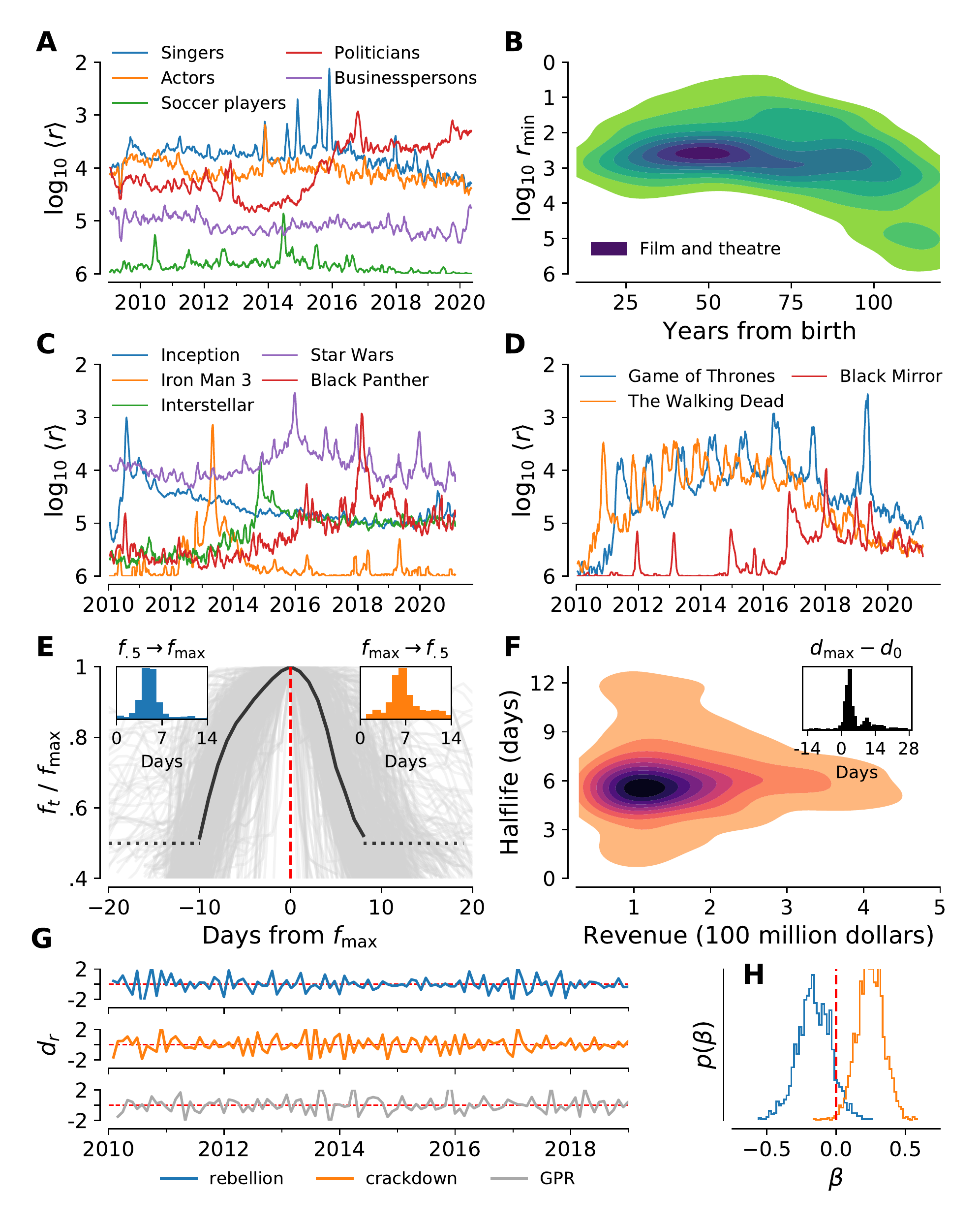}}
\end{figure*}

\subsection{Case studies}
\label{subsec:storywrangler.casestudies}

As a demonstration of our dataset's potential value to a diverse set of disciplines, 
we briefly present three case studies.
We analyze
(1)
The dynamic behavior of
famous individuals' full names and their association with the
individuals' ages;
(2)
The relationship between movie revenue and
anticipatory dynamics in title popularity;
and
(3)
The potential of social unrest related
words to predict future geopolitical risk.

We examine the dialog around celebrities by cross-referencing our
English $2$-grams corpus with names of famous personalities from the
Pantheon dataset~\cite{yu2016a}.  We searched through our
English $n$-grams dataset and selected names that were found in the
top million ranked $2$-grams for at least one day between 2010-01-01 and
2020-06-01.
In Fig.~\ref{fig:storywrangler.studies}A,
we display a monthly rolling average (centered) of the
average rank for the top 5 individuals for each category $\langle
r_{\text{min}(5)} \rangle$
(see also Fig.~\ref{fig:storywrangler.pantheon_age}).
In Fig.~\ref{fig:storywrangler.studies}B,
we display a kernel
density estimation of the top rank achieved by any of these
individuals in each industry as a function of the number of years
since the recorded year of birth.
We note high density of individuals marking their best rankings
between 40 and 60 years of age in the film and theatre industry.
Different dynamics can be observed in
Fig.~\ref{fig:storywrangler.pantheon_age} for other industries.

We next investigate the conversation surrounding major film releases
by tracking $n$-grams that appear in titles for 636 movies with gross
revenue above the 95th percentile during the period ranging from
2010-01-01 to 2017-07-01~\cite{harper2015a}.
We find a median value of 3 days post-release for
peak normalized frequency of usage for movie $n$-grams
(Fig.~\ref{fig:storywrangler.studies}F inset).  Growth of $n$-gram
usage from 50\% ($f_{.5}$) to maximum normalized frequency
($f_{\textnormal{max}}$) has a median value of 5 days across our
titles.
The median value of time to return to $f_{.5}$ from
$f_{\textnormal{max}}$ is 6 days.
Looking at
Fig.~\ref{fig:storywrangler.studies}E we see the median shape of the
spike around movie release dates tends to entail a gradual increase to
peak usage, and a relatively more sudden decrease when returning to
$f_{.5}$.
There is also slightly more spread in the time to return to
$f_{.5}$ than compared with the time to increase from $f_{.5}$ to
$f_{\textnormal{max}}$ (Fig.~\ref{fig:storywrangler.studies}E insets).

In Figs.~\ref{fig:storywrangler.studies}G and H, we show
that changes in word usage can be associated to future changes in
geopolitical risk, which we define here as 
``a decline in real activity,
lower stock returns, 
and movements in capital flows away from emerging
economies'', 
following the US Federal Reserve~\cite{caldara2017a}.
We chose a set of words that
we \textit{a priori} 
believed might be relevant to geopolitical
risk as design variables and a geopolitical index created by the US
Federal Reserve as the response.  
We fit a linear model using the values of the predictors at month $m$ 
to predict the value of the geopolitical risk index at month $m$+1.  
Two of the words,
`rebellion' and `crackdown', are significantly associated with
changes in the geopolitical risk index.

Although global events and breaking news are often recorded across conventional and modern social media platforms, 
Storywrangler uniquely tracks ephemeral day-to-day conversations and sociocultural trends. 
In creating Storywrangler, 
we sought to develop and maintain a large-scale daily record of everyday dialogues 
that is orthogonal to existing data sources, 
but equally vital to identify and study emerging sociotechnical phenomena.
For details about our methodology and further results,
see Appendix~\ref{appx:storywrangler.pantheon_case_study},
\ref{appx:storywrangler.movie_case_study},
and
\ref{appx:storywrangler.risk_case_study}.

\section{Concluding remarks}
\label{sec:storywrangler.concludingremarks}

With this initial effort, we aim to introduce Storywrangler as a platform enabling research in computational social science, data journalism, natural language processing, and the digital humanities.
Along with phrases associated with important events, Storywrangler encodes casual daily conversation in a format unavailable through newspaper articles and books.
While its utility is clear, there are many potential improvements to introduce to Storywrangler.
For high volume languages, we would aim for higher temporal resolution---at
the scale of minutes---and in such an implementation we
would be limited by requiring $n$-gram counts to exceed
some practical minimum.
We would also want to expand the language parsing
to cover continuous-script languages such as Japanese and Chinese.

Another large space of natural improvements would be
to broadly categorize tweets in ways other
than by language identification---while preserving privacy---such as geography, user type (e.g., people, institutions, or automated),
and topic (e.g., tweets containing `fake news').
We note that for Twitter, features like location and user type are more difficult to establish with
as much confidence as for language identification.
Increasingly by design, geographic information is limited on Twitter as are user demographics,
though some aspects may be gleaned 
indirectly~\cite{liu2013a,cohen2013a,preoctiuc2015a,malik2015a,zheng2018a}.
Regardless, 
in this initial curation of Twitter $n$-grams,
we purposefully do not
attempt to incorporate any metadata beyond identified language
into the $n$-gram database.

Topic-based subsets are particularly promising as they would
allow for explorations of language use, ambient framings, narratives,
and conspiracy theories.
Parsing into 2-grams and 3-grams makes possible certain analyses of the temporal evolution of 1-grams adjacent to an anchor 1-gram or 2-gram.
Future development will enable the use of wild cards so that linguists will in principle be able to track patterns of popular
language use in a way that the Google Books $n$-gram corpus is unable to do~\cite{michel2011a,pechenick2015a}.
Similarly, 
journalists and political scientists could chart $n$-grams being used around,
for example, `\#BlackLivesMatter' or `Trump' over time~\cite{dodds2020computational}.

Looking outside of text, a major possible expansion of the instrument
would be to incorporate image and video captions,
a growing component of social media communications
over the last decade.
Moving away from Twitter, we could employ Storywrangler for other
platforms where social amplification is a recorded feature (e.g., Reddit, 4Chan, Weibo, Parler).

There are substantive limitations to Twitter data,
some of which are evident in many large-scale text corpora.
Our $n$-gram dataset
contends with popularity, 
allowing for the examination of story amplification,
and we emphasize the importance of using contagiograms 
as visualization tools
that go beyond presenting simple time series.
Popularity, however, is notoriously difficult to measure.
The main proxy we use for popularity is 
the relative rate of usage of a given $n$-gram across originally authored tweets, 
examining how each term or phrase is socially amplified via retweets. 
While Twitter attempts to measure popularity by counting impressions, is increasingly difficult to capture the number of people exposed to a tweet. 
Twitter's centralized trending feature is yet another dimension 
that alters the popularity of terms on the platform,
personalizing each user timeline and
inherently amplifying algorithmic bias.
We have also observed a growing passive behavior across the platform leading to an increasing preference for retweets over original tweets 
for most languages on Twitter during the past few years~\cite{alshaabi2020growing}.  

Twitter's userbase, while broad, is
clearly not representative of the populace~\cite{mellon2017a},
and is moreover compounded by the mixing of voices from people, organizations, and bots, and
has evolved over time as new users have joined.
Still, modern social media provides an open platform for all people to carry out conversations that matter to their lives.   
Storywrangler serves as instrument to depict discourse on social media 
at a larger scale and finer time resolution than current existing resources. 
Indeed, sociocultural biases that are inherently intrinsic to these platforms 
will be exposed using Storywrangler, 
which can inspire developers to enhance their platforms 
accordingly~\cite{blodgett2016a,koenecke2020a}.  

Remarkably, 
social structures 
(e.g., news and social media platforms) 
form and reshape individual behavior, 
which evidently alters social structures 
in an algorithmic feedback loop fashion~\cite{giddens1984a}.
For instance, 
a trending hashtag can embody a social movement (e.g., \#MeToo), 
such that an $n$-gram may become mutually constituted to a behavioral and sociocultural revolution.
Social and political campaigns can leverage an $n$-gram in their organized marketing strategies, 
seeking sustained collective attention on social media platforms encoded through spikes in $n$-gram usage rates.
There are many examples of this emerging sociotechnical phenomenon on Twitter ranging from 
civil rights (e.g., \#WomensMarch)
to 
gender identity (e.g., \#LGBTQ)
to 
political conspiracy theories (e.g., \#QAnon) 
to 
academy awards promotions (e.g., \#Oscar)
to 
movie advertisement (e.g., \#Avengers), etc.
The Canadian awareness campaign `Bell Let's Talk' 
is another example of an annual awareness campaign 
that subsidises mental health institutions across Canada, 
donating 5 cents for every (re)tweet containing the hashtag `\#BellLetsTalk' 
to reduce stigma surrounding mental illness.
Marketing campaigns have also grasped the periodic feature of key trending $n$-grams 
and adjusted their language accordingly. 
Marketers and bots often exploit this periodicity 
by hijacking popular hashtags to broadcast their propaganda 
(e.g., including \#FF and \#TGIF as trending hashtags for Friday promotions).

In building Storywrangler, we have prioritized privacy by aggregating statistics to day-scale resolution for individual languages, truncating distributions, ignoring geography, and masking all metadata. 
We have also endeavored to make our work as transparent as possible by releasing all code associated with the API. 

Although we frame Storywrangler as a research focused instrument akin to a microscope or telescope for the advancement of science, it does not have built-in ethical guardrails.
There is potential for misinterpretation and mischaracterization of the data, 
whether purposeful or not.
For example, we strongly caution against cherry picking isolated time series
that might suggest a particular story or social trend.
Words and phrases may drift in meaning and other terms take their place.
For example, `coronavirus' gave way to `covid' as the dominant term of reference
on Twitter for the COVID-19 pandemic in the first six months of 2020~\cite{alshaabi2021how}.
To in part properly demonstrate a trend, researchers would need to at least marshal together thematically related $n$-grams,
and do so in a data-driven way, as we have attempted to do for our case studies.
Thoughtful consideration of overall and normalized frequency of usage would also be needed to show whether a topic is changing in real volume.

In building Storywrangler,
our primary goal has been to build an instrument to curate and share a rich, 
language-based ecology of interconnected $n$-gram time series derived from social media.
We see some of the strongest potential for future work in the coupling of Storywrangler with other data streams to enable, for example,
data-driven, computational versions of journalism, linguistics, history, economics, and political science.



\acknowledgments
The authors are grateful for the computing resources provided by the Vermont Advanced Computing Core 
and financial support from the Massachusetts Mutual Life Insurance Company and Google Open Source under the Open-Source Complex Ecosystems And Networks (OCEAN) project.
The authors appreciate discussions and correspondence with Colin Van Oort, 
James Bagrow,
and Randall Harp.
We thank many of our colleagues 
at the Computational Story Lab 
for their feedback on this project.
Computations were performed on the Vermont Advanced Computing Core supported in part by NSF award No. OAC-1827314.

\bibliographystyle{unsrtabbrv}
\bibliography{\filenamebase}

\begin{thebibliography}{10}

\bibitem{michel2011a}
J.~B. Michel, Y.~K. Shen, A.~P. Aiden, A.~Veres, M.~K. Gray, {The Google Books
  Team}, J.~P. Pickett, D.~Hoiberg, D.~Clancy, P.~Norvig, J.~Orwant, S.~Pinker,
  M.~A. Nowak, and E.~A. Lieberman.
\newblock Quantitative analysis of culture using millions of digitized books.
\newblock {\em Science Magazine}, 331:176--182, 2011.

\bibitem{pechenick2015a}
E.~A. Pechenick, C.~M. Danforth, and P.~S. Dodds.
\newblock Characterizing the google books corpus: Strong limits to inferences
  of socio-cultural and linguistic evolution.
\newblock {\em PLOS ONE}, 10(10):1--24, 2015.

\bibitem{christenson2011a}
H.~Christenson.
\newblock Hathitrust: {A} research library at web scale.
\newblock {\em Library Resources \& Technical Services}, 55:93--102, 2011.

\bibitem{gerlach2020a}
M.~Gerlach and F.~Font-Clos.
\newblock A standardized {P}roject {G}utenberg corpus for statistical analysis
  of natural language and quantitative linguistics.
\newblock {\em Entropy}, 22(1):126, 2020.

\bibitem{nytimescorpus2008a}
E.~Sandhaus.
\newblock The {N}ew {Y}ork {T}imes {A}nnotated {C}orpus, 2008.

\bibitem{beeferman2019a}
D.~Beeferman, W.~Brannon, and D.~Roy.
\newblock {RadioTalk}: {A} large-scale corpus of talk radio transcripts.
\newblock In {\em Proceedings of Interspeech 2019}, pages 564--568.
  International Speech Communication Association, 2019.

\bibitem{hollink2016a}
L.~Hollink, A.~Bedjeti, M.~van Harmelen, and D.~Elliott.
\newblock A corpus of images and text in online news.
\newblock In {\em Proceedings of the Tenth International Conference on Language
  Resources and Evaluation ({LREC}'16)}, pages 1377--1382, Portoro{\v{z}},
  Slovenia, 2016. European Language Resources Association (ELRA).

\bibitem{hong2020a}
J.~Hong, W.~Crichton, H.~Zhang, D.~Y. Fu, J.~Ritchie, J.~Barenholtz, B.~Hannel,
  X.~Yao, M.~Murray, G.~Moriba, M.~Agrawala, and K.~Fatahalian.
\newblock Analyzing who and what appears in a decade of {U}{S} cable {T}{V}
  news, 2020.
\newblock Available online at
  \href{https://arxiv.org/abs/2008.06007}{https://arxiv.org/abs/2008.06007}.

\bibitem{mieder2004a}
W.~Mieder.
\newblock {\em Proverbs: A Handbook}.
\newblock Greenwood folklore handbooks. Greenwood Press, 2004.

\bibitem{abello2012a}
J.~Abello, P.~Broadwell, and T.~R. Tangherlini.
\newblock Computational folkloristics.
\newblock {\em Communications of the ACM}, 55(7):60--70, 2012.

\bibitem{tangherlini2013a}
T.~R. Tangherlini and P.~Leonard.
\newblock Trawling in the sea of the {G}reat {U}nread: {S}ub-corpus topic
  modeling and {H}umanities research.
\newblock {\em Poetics}, 41(6):725--749, 2013.

\bibitem{vuong2018a}
Q.-H. Vuong, Q.-K. Bui, V.-P. La, T.-T. Vuong, V.-H.~T. Nguyen, M.-T. Ho,
  H.-K.~T. Nguyen, and M.-T. Ho.
\newblock Cultural additivity: {B}ehavioural insights from the interaction of
  {C}onfucianism, {B}uddhism and {T}aoism in folktales.
\newblock {\em Palgrave Communications}, 4(1):1--15, 2018.

\bibitem{woolley2008a}
J.~T. Woolley and G.~Peters.
\newblock The {A}merican presidency project, 1999.
\newblock Available online at \url{http://www.presidency.ucsb.edu/}.

\bibitem{primack2009a}
R.~B. Primack, H.~Higuchi, and A.~J. Miller-Rushing.
\newblock The impact of climate change on cherry trees and other species in
  {J}apan.
\newblock {\em Biological Conservation}, 142(9):1943--1949, 2009.
\newblock The Conservation and Management of Biodiversity in Japan.

\bibitem{allen2020a}
J.~Allen, B.~Howland, M.~Mobius, D.~Rothschild, and D.~J. Watts.
\newblock Evaluating the fake news problem at the scale of the information
  ecosystem.
\newblock {\em Science Advances}, 6(14), 2020.

\bibitem{peirce1906a}
C.~S. Sanders~Peirce.
\newblock {Prolegomena to an Apology for Pragmaticism.}
\newblock {\em The Monist}, 16(4):492--546, 2015.

\bibitem{zipf1949a}
G.~K. Zipf.
\newblock {\em Human Behaviour and the Principle of Least-Effort}.
\newblock Addison-Wesley, Cambridge, MA, 1949.

\bibitem{bohannon2010a}
J.~Bohannon.
\newblock Google opens books to new cultural studies.
\newblock {\em Science}, 330(6011):1600--1600, 2010.

\bibitem{koplenig2017a}
A.~Koplenig.
\newblock The impact of lacking metadata for the measurement of cultural and
  linguistic change using the {G}oogle {N}gram data sets—reconstructing the
  composition of the {G}erman corpus in times of {WWII}.
\newblock {\em Digital Scholarship in the Humanities}, 32(1):169--188, 2015.

\bibitem{pechenick2017a}
E.~A. Pechenick, C.~M. Danforth, and P.~S. Dodds.
\newblock Is language evolution grinding to a halt? {T}he scaling of lexical
  turbulence in {E}nglish fiction suggests it is not.
\newblock {\em Journal of Computational Science}, 21:24--37, 2017.

\bibitem{merritt2018a}
J.~Merritt and S.~Niequist.
\newblock {\em Learning to Speak {G}od from Scratch: {W}hy Sacred Words Are
  Vanishing--and How We Can Revive Them}.
\newblock Crown Publishing Group, 2018.

\bibitem{hong2011a}
S.~Hong and D.~Nadler.
\newblock Does the early bird move the polls? {T}he use of the social media
  tool `{T}witter' by {U}{S} politicians and its impact on public opinion.
\newblock In {\em Proceedings of the 12th Annual International Digital
  Government Research Conference: Digital Government Innovation in Challenging
  Times}, dg.o '11, page 182–186, New York, NY, USA, 2011. Association for
  Computing Machinery.

\bibitem{younus2011a}
A.~Younus, M.~A. Qureshi, F.~F. Asar, M.~Azam, M.~Saeed, and N.~Touheed.
\newblock What do the average twitterers say: {A} {T}witter model for public
  opinion analysis in the face of major political events.
\newblock In {\em 2011 International Conference on Advances in Social Networks
  Analysis and Mining}, pages 618--623. Institute of Electrical and Electronics
  Engineers, 2011.

\bibitem{sakaki2010a}
T.~Sakaki, M.~Okazaki, and Y.~Matsuo.
\newblock Earthquake shakes {T}witter users: {R}eal-time event detection by
  social sensors.
\newblock In {\em Proceedings of the 19th International Conference on World
  Wide Web}, WWW '10, page 851–860, New York, NY, USA, 2010. Association for
  Computing Machinery.

\bibitem{pickard2011a}
G.~Pickard, W.~Pan, I.~Rahwan, M.~Cebrian, R.~Crane, A.~Madan, and A.~Pentland.
\newblock Time-critical social mobilization.
\newblock {\em Science}, 334(6055):509--512, 2011.

\bibitem{gao2011a}
H.~{Gao}, G.~{Barbier}, and R.~{Goolsby}.
\newblock Harnessing the crowdsourcing power of social media for disaster
  relief.
\newblock {\em IEEE Intelligent Systems}, 26(3):10--14, 2011.

\bibitem{lampos2010a}
V.~Lampos and N.~Cristianini.
\newblock Tracking the flu pandemic by monitoring the social web.
\newblock In {\em 2010 2nd International Workshop on Cognitive Information
  Processing}, pages 411--416. Institute of Electrical and Electronics
  Engineers, 2010.

\bibitem{culotta2010a}
A.~Culotta.
\newblock Towards detecting influenza epidemics by analyzing {T}witter
  messages.
\newblock In {\em Proceedings of the First Workshop on Social Media Analytics},
  SOMA 10, page 115–122, New York, NY, USA, 2010. Association for Computing
  Machinery.

\bibitem{steinert2015a}
Z.~C. Steinert-Threlkeld, D.~Mocanu, A.~Vespignani, and J.~Fowler.
\newblock Online social networks and offline protest.
\newblock {\em EPJ Data Science}, 4(1):19, 2015.

\bibitem{alshaabi2020growing}
T.~Alshaabi, D.~R. Dewhurst, J.~R. Minot, M.~V. Arnold, J.~L. Adams, C.~M.
  Danforth, and P.~S. Dodds.
\newblock The growing amplification of social media: {M}easuring temporal and
  social contagion dynamics for over 150 languages on {T}witter for 2009--2020.
\newblock {\em EPJ Data Science}, 10(15), 2021.

\bibitem{joulin2016a}
A.~Joulin, E.~Grave, P.~Bojanowski, and T.~Mikolov.
\newblock Bag of tricks for efficient text classification.
\newblock In {\em Proceedings of the 15th Conference of the {E}uropean Chapter
  of the Association for Computational Linguistics: Volume 2, Short Papers},
  pages 427--431, Valencia, Spain, 2017. Association for Computational
  Linguistics.

\bibitem{bojanowskia}
P.~Bojanowski, E.~Grave, A.~Joulin, and T.~Mikolov.
\newblock Enriching word vectors with subword information.
\newblock {\em Transactions of the Association for Computational Linguistics},
  5:135--146, 2017.

\bibitem{dodds2020long}
P.~S. Dodds, J.~R. Minot, M.~V. Arnold, T.~Alshaabi, J.~L. Adams, D.~R.
  Dewhurst, A.~J. Reagan, and C.~M. Danforth.
\newblock Long-term word frequency dynamics derived from {T}witter are
  corrupted: {A} bespoke approach to detecting and removing pathologies in
  ensembles of time series, 2020.
\newblock Available online at
  \href{https://arxiv.org/abs/2008.11305}{https://arxiv.org/abs/2008.11305}.

\bibitem{loper2002a}
E.~Loper and S.~Bird.
\newblock {NLTK}: {T}he natural language toolkit.
\newblock In {\em Proceedings of the ACL-02 Workshop on Effective Tools and
  Methodologies for Teaching Natural Language Processing and Computational
  Linguistics - Volume 1}, ETMTNLP '02, page 63–70, USA, 2002. Association
  for Computational Linguistics.

\bibitem{bevensee2020a}
E.~Bevensee, M.~Aliapoulios, Q.~Dougherty, J.~Baumgartner, D.~McCoy, and
  J.~Blackburn.
\newblock {S}{M}{A}{T}: {T}he social media analysis toolkit.
\newblock In {\em Workshop Proceedings of the 14th International AAAI
  Conference on Web and Social Media}, volume~14, 2020.

\bibitem{dewhurst2020shocklet}
D.~R. Dewhurst, T.~Alshaabi, D.~Kiley, M.~V. Arnold, J.~R. Minot, C.~M.
  Danforth, and P.~S. Dodds.
\newblock The shocklet transform: {A} decomposition method for the
  identification of local, mechanism-driven dynamics in sociotechnical time
  series.
\newblock {\em EPJ Data Science}, 9(1):3, 2020.

\bibitem{dodds2019fame}
P.~S. Dodds, J.~R. Minot, M.~V. Arnold, T.~Alshaabi, J.~L. Adams, D.~R.
  Dewhurst, A.~J. Reagan, and C.~M. Danforth.
\newblock Fame and {U}ltrafame: {M}easuring and comparing daily levels of
  `being talked about' for {U}nited {S}tates' presidents, their rivals, {G}od,
  countries, and {K}-pop, 2019.
\newblock Available online at
  \href{https://arxiv.org/abs/1910.00149}{https://arxiv.org/abs/1910.00149}.

\bibitem{choi2012a}
H.~Choi and H.~Varian.
\newblock Predicting the present with google trends.
\newblock {\em Economic Record}, 88(s1):2--9, 2012.

\bibitem{dodds2020allotaxonometry}
P.~S. Dodds, J.~R. Minot, M.~V. Arnold, T.~Alshaabi, J.~L. Adams, D.~R.
  Dewhurst, T.~J. Gray, M.~R. Frank, A.~J. Reagan, and C.~M. Danforth.
\newblock Allotaxonometry and rank-turbulence divergence: {A} universal
  instrument for comparing complex systems, 2020.
\newblock Available online at
  \href{https://arxiv.org/abs/2002.09770}{https://arxiv.org/abs/2002.09770}.

\bibitem{yu2016a}
A.~Z. Yu, S.~Ronen, K.~Hu, T.~Lu, and C.~A. Hidalgo.
\newblock Pantheon 1.0, a manually verified dataset of globally famous
  biographies.
\newblock {\em Scientific Data}, 3(1):1--16, 2016.

\bibitem{harper2015a}
F.~M. Harper and J.~A. Konstan.
\newblock The {MovieLens} datasets: {H}istory and context.
\newblock {\em ACM Transactions on Interactive Intelligent Systems}, 5(4),
  2015.

\bibitem{caldara2017a}
D.~Caldara and M.~Iacoviello.
\newblock Measuring geopolitical risk.
\newblock {\em FRB International Finance Discussion Paper}, (1222), 2018.

\bibitem{liu2013a}
W.~Liu and D.~Ruths.
\newblock What's in a name? {U}sing first names as features for gender
  inference in {T}witter.
\newblock In {\em AAAI Spring Symposium: {A}nalyzing Microtext}, volume
  SS-13-01 of {\em AAAI Technical Report}. AAAI, 2013.

\bibitem{cohen2013a}
R.~Cohen and D.~Ruths.
\newblock Classifying political orientation on {T}witter: It's not easy!
\newblock In {\em Proceedings of the International AAAI Conference on Web and
  Social Media}, volume~7, 2013.

\bibitem{preoctiuc2015a}
D.~Preo{\c{t}}iuc-Pietro, S.~Volkova, V.~Lampos, Y.~Bachrach, and N.~Aletras.
\newblock Studying user income through language, behaviour and affect in social
  media.
\newblock {\em PLOS ONE}, 10(9):1--17, 2015.

\bibitem{malik2015a}
M.~Malik, H.~Lamba, C.~Nakos, and J.~Pfeffer.
\newblock Population bias in geotagged tweets.
\newblock In {\em Proceedings of the International AAAI Conference on Web and
  Social Media}, volume~9, 2015.

\bibitem{zheng2018a}
X.~Zheng, J.~Han, and A.~Sun.
\newblock A survey of location prediction on {T}witter.
\newblock {\em IEEE Transactions on Knowledge and Data Engineering},
  30(9):1652--1671, 2018.

\bibitem{dodds2020computational}
P.~S. Dodds, J.~R. Minot, M.~V. Arnold, T.~Alshaabi, J.~L. Adams, A.~J. Reagan,
  and C.~M. Danforth.
\newblock Computational timeline reconstruction of the stories surrounding
  {T}rump: {S}tory turbulence, narrative control, and collective chronopathy,
  2020.
\newblock Available online at
  \href{https://arxiv.org/abs/2008.07301}{https://arxiv.org/abs/2008.07301}.

\bibitem{mellon2017a}
J.~Mellon and C.~Prosser.
\newblock Twitter and {F}acebook are not representative of the general
  population: {P}olitical attitudes and demographics of {B}ritish social media
  users.
\newblock {\em Research \& Politics}, 4(3):2053168017720008, 2017.

\bibitem{blodgett2016a}
S.~L. Blodgett, L.~Green, and B.~O{'}Connor.
\newblock Demographic dialectal variation in social media: A case study of
  {A}frican-{A}merican {E}nglish.
\newblock In {\em Proceedings of the 2016 Conference on Empirical Methods in
  Natural Language Processing}, pages 1119--1130, Austin, Texas, 2016.
  Association for Computational Linguistics.

\bibitem{koenecke2020a}
A.~Koenecke, A.~Nam, E.~Lake, J.~Nudell, M.~Quartey, Z.~Mengesha, C.~Toups,
  J.~R. Rickford, D.~Jurafsky, and S.~Goel.
\newblock Racial disparities in automated speech recognition.
\newblock {\em Proceedings of the National Academy of Sciences},
  117(14):7684--7689, 2020.

\bibitem{giddens1984a}
A.~Giddens.
\newblock {\em The Constitution of Society: Outline of the Theory of
  Structuration}.
\newblock Outline of the Theory of Structuration. University of California
  Press, 1984.

\bibitem{alshaabi2021how}
T.~Alshaabi, M.~V. Arnold, J.~R. Minot, J.~L. Adams, D.~R. Dewhurst, A.~J.
  Reagan, R.~Muhamad, C.~M. Danforth, and P.~S. Dodds.
\newblock How the world's collective attention is being paid to a pandemic:
  {COVID-19} related n-gram time series for 24 languages on {T}witter.
\newblock {\em PLOS ONE}, 16(1):1--13, 2021.

\bibitem{ke2017a}
Q.~Ke, Y.~Ahn, and C.~R. Sugimoto.
\newblock A systematic identification and analysis of scientists on {T}witter.
\newblock {\em PLOS ONE}, 12(4):1--17, 2017.

\bibitem{simon1955a}
H.~A. Simon.
\newblock On a class of skew distribution functions.
\newblock {\em Biometrika}, 42(3-4):425--440, 1955.

\bibitem{price1976a}
D.~D.~S. Price.
\newblock A general theory of bibliometric and other cumulative advantage
  processes.
\newblock {\em Journal of the American Society for Information Science},
  27(5):292--306, 1976.

\bibitem{hill1975a}
B.~M. Hill.
\newblock A simple general approach to inference about the tail of a
  distribution.
\newblock {\em The Annals of Statistics}, 3(5):1163--1174, 1975.

\bibitem{powers1998a}
D.~M.~W. Powers.
\newblock Applications and explanations of {Z}ipf{'}s law.
\newblock In {\em New Methods in Language Processing and Computational Natural
  Language Learning}, 1998.

\bibitem{piantadosi2014a}
S.~T. Piantadosi.
\newblock Zipf's word frequency law in natural language: {A} critical review
  and future directions.
\newblock {\em Psychonomic Bulletin \& Review}, 21(5):1112--1130, 2014.

\bibitem{bokanyi2019a}
E.~Bokányi, D.~Kondor, and G.~Vattay.
\newblock Scaling in words on twitter.
\newblock {\em Royal Society Open Science}, 6(10):190027, 2019.

\bibitem{pfeffer2018a}
J.~Pfeffer, K.~Mayer, and F.~Morstatter.
\newblock Tampering with {T}witter's sample {API}.
\newblock {\em EPJ Data Science}, 7(1):50, 2018.

\bibitem{williams2015a}
J.~R. Williams, P.~R. Lessard, S.~Desu, E.~M. Clark, J.~P. Bagrow, C.~M.
  Danforth, and P.~S. Dodds.
\newblock Zipf's law holds for phrases, not words.
\newblock {\em Nature Scientific Reports}, 5:12209, 2015.

\bibitem{hoffman2014a}
M.~D. Hoffman and A.~Gelman.
\newblock The {No-U-Turn} sampler: {A}daptively setting path lengths in
  {H}amiltonian {M}onte {C}arlo.
\newblock {\em Journal of Machine Learning Research}, 15(1):1593–1623, 2014.

\bibitem{gelman1992a}
A.~Gelman and D.~B. Rubin.
\newblock Inference from iterative simulation using multiple sequences.
\newblock {\em Statistical Science}, 7(4):457 -- 472, 1992.

\bibitem{chenoweth2014a}
E.~Chenoweth and M.~J. Stephan.
\newblock Drop your weapons: {W}hen and why civil resistance works.
\newblock {\em Foreign Affairs}, 93:94, 2014.

\end{thebibliography}

\clearpage


\newwrite\tempfile
\immediate\openout\tempfile=startsupp.txt
\immediate\write\tempfile{\thepage}
\immediate\closeout\tempfile

\renewcommand{\thesection}{S\arabic{section}}
\renewcommand{\thefigure}{S\arabic{figure}}
\renewcommand{\thetable}{S\arabic{table}}
\setcounter{section}{0}
\setcounter{figure}{0}
\setcounter{table}{0}
\setcounter{footnote}{0}

\appendix

\section{Twitter Dataset}
\label{appx:storywrangler.dataset}

\subsection{Language identification and detection}
\label{subsec:storywrangler.decahose}

\begin{figure*}[tp!] 
\includegraphics[width=\textwidth]{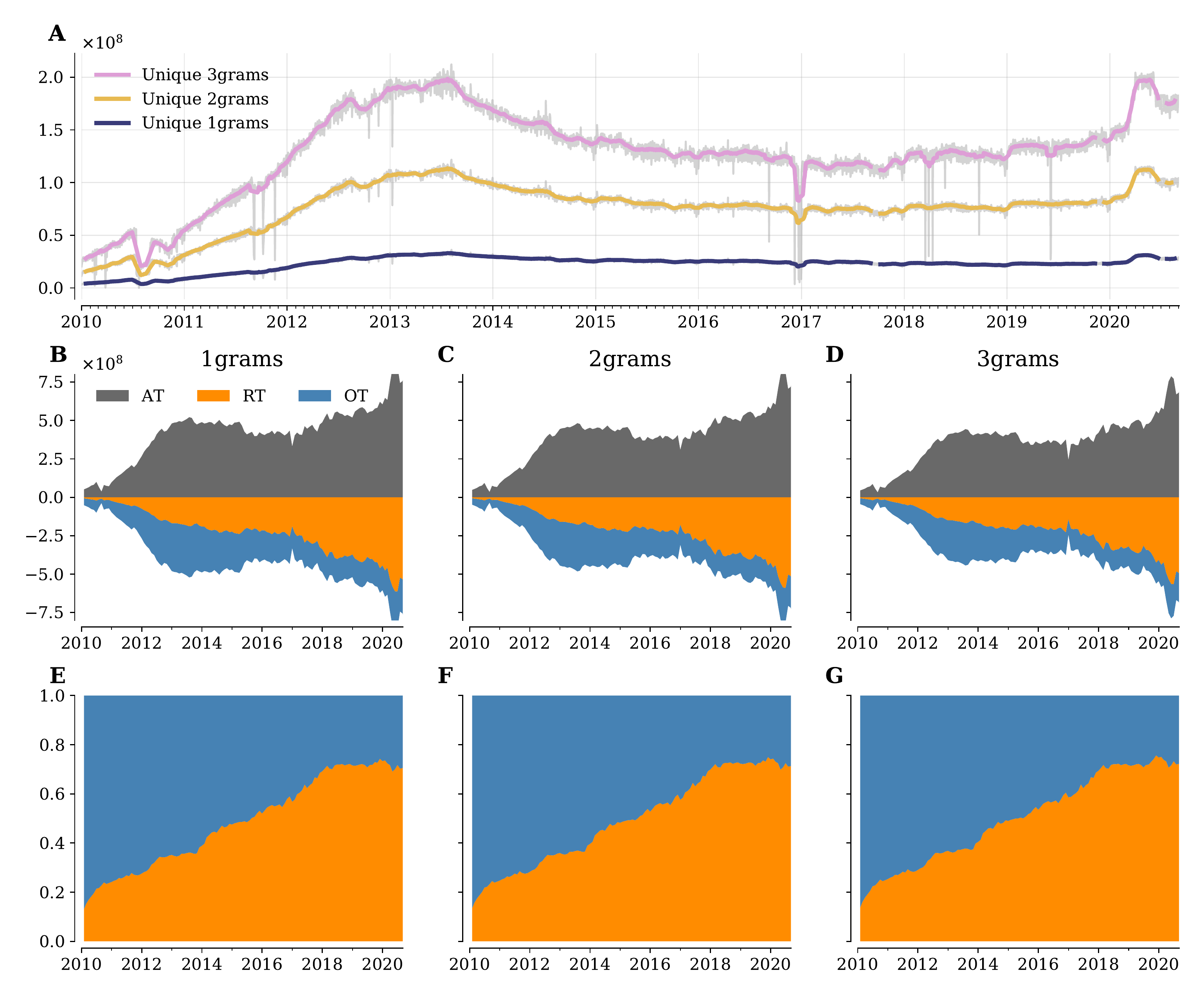}
\caption{
  \textbf{Temporal summary statistics}.
  \textbf{A.}
  The grey bars show the daily unique number of $n$-grams,
  while the lines show a monthly rolling average 
  for $1$-grams (purple), $2$-grams (yellow), and for $3$-grams (pink).
  \textbf{B--D.} 
  The growth of $n$-grams in our dataset by each category 
  where $n$-grams captured from organic tweets (OT) are displayed in blue,
  retweets RT in green, and all tweets combined in grey.  
  \textbf{E--G.} Normalized frequencies to illustrate the growth of retweets over time.  
}   
\label{fig:storywrangler.stats_timeseries} 
\end{figure*}

\begin{figure*}[tp!]  
\includegraphics[width=\linewidth]{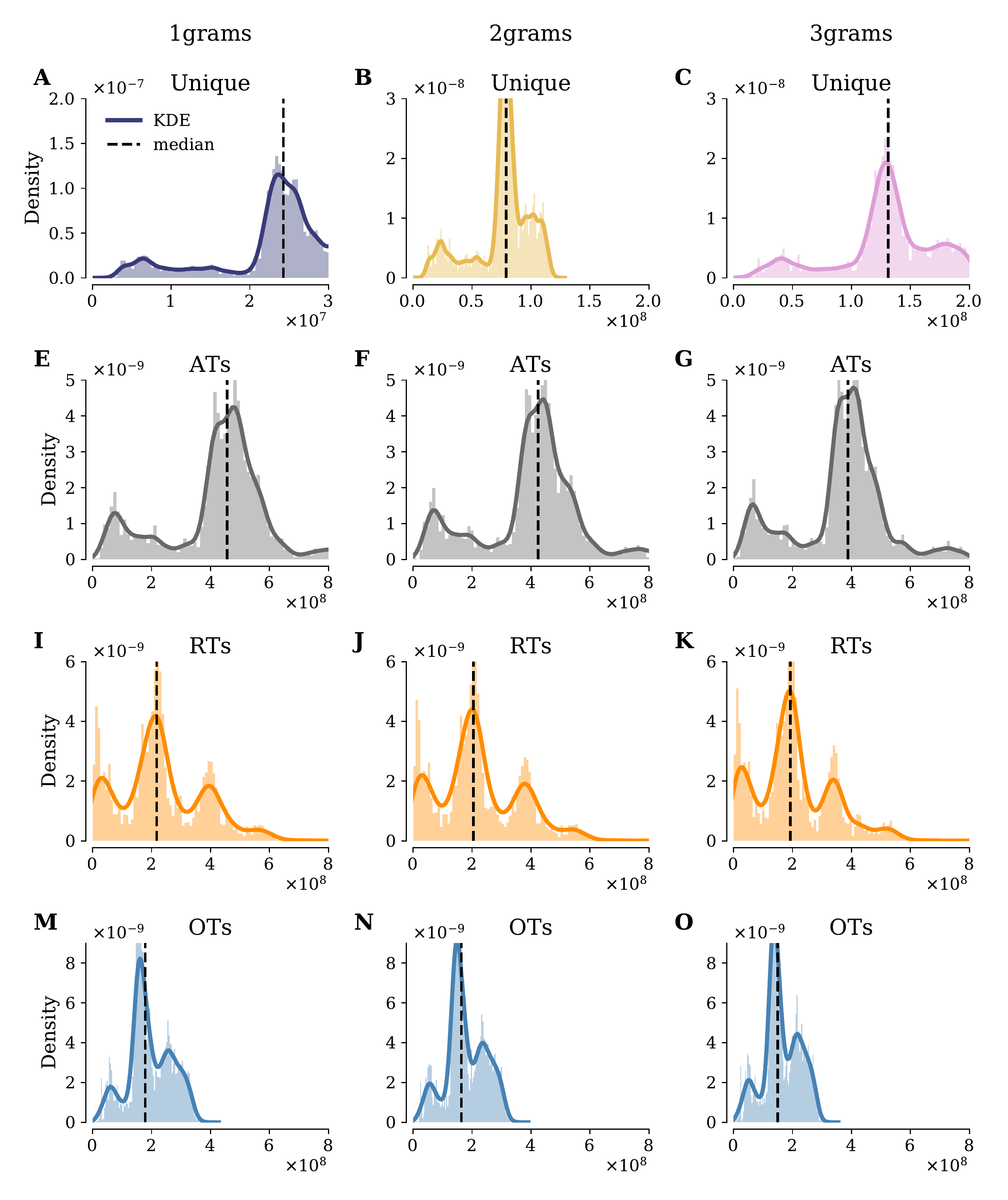}
\caption{
  \textbf{Kernel density estimations.}
  \textbf{A--C.}
  Distributions of unique 
  unigrams, bigrams, and trigrams
  captured daily throughout the last decade.
  \textbf{E--G.}
  Distributions of $n$-grams occurrences in all tweets.
  \textbf{I--K.}
  Distributions of $n$-grams parsed from retweets (RT) only.
  \textbf{M--O.}
  Distributions of $n$-grams parsed from organic tweets (OT) only.
  }
\label{fig:storywrangler.stats_dists} 
\end{figure*}

In previous work~\cite{alshaabi2020growing}, 
we described how we re-identified the languages of all tweets in our collection using FastText~\cite{joulin2016a}. 
A uniform language re-identification was needed 
as Twitter's own real-time identification algorithm was introduced in late 2012
and then adjusted over time, 
resulting in temporal inconsistencies 
for long-term streaming collection of tweets~\cite{dodds2020long}.

While FastText is a language model that can be used for various text mining tasks,
it requires an additional step of producing vector language representations to be used for LID. 
To accomplish that, we use an off-the-shelf language identification 
tool that uses the word embeddings produced by the
model.\footnote{\url{https://fasttext.cc/docs/en/language-identification.html}} 

The word embeddings provided by FastText spans a wide set of languages,
including some regional dialects. 
However, 
language detection of short text remains an outstanding challenge in NLP. 
While we hope to expand our language detection in future work, 
we still classify messages based on the languages identified by FastText-LID.

Importantly, in this work,
we do not intend to reinvent FastText-LID, 
or improve upon existing LID tools. 
FastText-LID is on par with deep learning models 
in terms of accuracy and consistency, 
yet orders of magnitude faster in terms of inference and training time~\cite{joulin2016a}.
They also show that FastText-LID outperforms previously introduced LID tools such as langid.\footnote{\url{https://fasttext.cc/blog/2017/10/02/blog-post.html}}
We use FastText-LID as a light, fast, and reasonably accurate language detection tool 
to overcome the challenge of missing language labels in our Twitter historical feed.

We group tweets by day according to Eastern Time (ET).
Date and language are the only metadata we incorporate into our database.
For user privacy in particular, we discard all other information associated with a tweet.

\subsection{Social amplification and contagion}
\label{subsec:storywrangler.categorization}

Twitter enables social amplification on the platform through the use of retweets and, 
from 2015 on, quote tweets.
Users---including the general public, 
celebrities, scientists, decision-makers, and social 
bots~\cite{ke2017a}---can intervene in the information spread process 
and amplify the volume of any content being shared. 
We categorize tweets into two major classes: 
organic tweets (OT) and retweets (RT). 
Organic tweets represent the set of new information being shared on the platform, 
whereas retweets reflect information being socially amplified on Twitter. 
During our process, consistent with Twitter's original encoding of retweets, 
we enrich the text body of a retweet with 
(RT @userHandle: ...) 
to indicate the original user of the retweeted text. 
Our categorization enables users of the Storywrangler data set to tune the amplification processes 
of the rich-get-richer mechanism~\cite{simon1955a,price1976a} 
by dialing the ratio of retweets added to the $n$-grams corpus.

\subsection{Detailed dataset statistics}
\label{subsec:storywrangler.dataset_stats}

From 2008-09-09 on, 
we have been collecting a random subset of
approximately 10\% of all public messages using Twitter's Decahose
API.\footnote{\url{https://developer.twitter.com/en/docs/twitter-ads-api/campaign-management/api-reference}}  
Every day, half a billion messages are shared
on Twitter in hundreds of languages.  
By the end 2020,
our data collection
comprised around 150 billion messages, 
requiring over 100TB of storage.

It is worth noting again that this is an approximate daily leaderboard of language usage and word popularity.
It is well established that $n$-gram frequency-of-usage (or Zipf) distributions are heavy-tailed~\cite{zipf1949a}. 
Researchers have thoroughly investigated ways to study Zipf distributions and estimate the robustness and stability of their tails~\cite{hill1975a,powers1998a,piantadosi2014a,bokanyi2019a}. Investigators have also examined various aspects of the Twitter’s Sample API~\cite{pfeffer2018a}, and how that may affect the observed daily word distributions~\cite{williams2015a}.

Our Twitter corpus contains an average of 23 million unique $1$-grams every day 
with a maximum of a little over 36 million unique $1$-grams captured on 2013-08-07. 
The numbers of unique bigrams and trigrams strongly outweigh the number of unique unigrams 
because of the combinatorial properties of language. 
On average, we extract around
76 million unique $2$-grams and 128 million unique $3$-grams for each day. 
On 2013-08-07, we recorded a high
of 121 million unique 2-grams, 
and a high of 212 million unique $3$-grams.

We emphasize that these maxima for $n$-grams reflect only our data
set, and not the entirety of Twitter.  
We are unable to make assertions about the size
of Twitter's user base or message volume.  
Indeed, 
because we do not have knowledge of Twitter's overall volume 
(and do not seek to per Twitter's Terms of Service), 
we deliberately focus on ranks and relative
usage rates for $n$-grams away from the tails of their distributions.
Raw frequencies of exceedingly rare words are roughly one-tenth of the true values with regards to all of Twitter, 
however, rankings are likely to be subject to change.

In Tab.~\ref{tab:1grams_stats}, 
we show daily summary statistics for $1$-grams
broken by each category in our data set. 
We demonstrate the same statistical information for $2$-grams and $3$-grams
in Tab.~\ref{tab:2grams_stats} and Tab.~\ref{tab:3grams_stats} respectively. 
We show a time series of the unique number of $n$-grams captured daily in Fig.~\ref{fig:storywrangler.stats_timeseries}
and the statistical distributions of each category in Fig.~\ref{fig:storywrangler.stats_dists}.

\begin{table}[tp!]
\centering
\caption{Average daily summary statistics for $1$-grams.}
\begin{tabularx}{\columnwidth}{l|C|C||C|C}
    & \multicolumn{2}{c||}{\textbf{AT}}         & \multicolumn{2}{c}{\textbf{OT}}         \\ 
    & \textbf{Volume}    & \textbf{Unique}    & \textbf{Volume}    & \textbf{Unique}    \\
\hline
$\mu$                  & 4.25 $\times 10^8$ & 2.27 $\times 10^7$ & 1.93 $\times 10^8$ & 1.68 $\times 10^7$ \\
$25^{\textnormal{th}}$ & 3.87 $\times 10^8$ & 2.20 $\times 10^7$ & 1.52 $\times 10^8$ & 1.43 $\times 10^7$ \\
$50^{\textnormal{th}}$ & 4.56 $\times 10^8$ & 2.42 $\times 10^7$ & 1.78 $\times 10^8$ & 1.74 $\times 10^7$ \\
$75^{\textnormal{th}}$ & 5.16 $\times 10^8$ & 2.67 $\times 10^7$ & 2.53 $\times 10^8$ & 2.05 $\times 10^7$ \\
max                    & 1.13 $\times 10^9$ & 3.61 $\times 10^7$ & 3.83 $\times 10^8$ & 2.90 $\times 10^7$
\end{tabularx}
\label{tab:1grams_stats}
\end{table} 

\begin{table}[htp!]
\centering
\caption{Average daily summary statistics for $2$-grams.}
\begin{tabularx}{\columnwidth}{l|C|C||C|C}
    & \multicolumn{2}{c||}{\textbf{AT}}         & \multicolumn{2}{c}{\textbf{OT}}         \\ 
    & \textbf{Volume}    & \textbf{Unique}    & \textbf{Volume}    & \textbf{Unique}    \\
\hline
$\mu$                  & 3.98 $\times 10^8$ & 7.60 $\times 10^7$ & 1.77 $\times 10^8$ & 5.41 $\times 10^7$ \\
$25^{\textnormal{th}}$ & 3.59 $\times 10^8$ & 7.29 $\times 10^7$ & 1.41 $\times 10^8$ & 4.71 $\times 10^7$ \\
$50^{\textnormal{th}}$ & 4.23 $\times 10^8$ & 7.90 $\times 10^7$ & 1.63 $\times 10^8$ & 5.24 $\times 10^7$ \\
$75^{\textnormal{th}}$ & 4.83 $\times 10^8$ & 8.79 $\times 10^7$ & 2.33 $\times 10^8$ & 6.56 $\times 10^7$ \\
max                    & 1.09 $\times 10^9$ & 1.21 $\times 10^8$ & 3.51 $\times 10^8$ & 9.34 $\times 10^7$
\end{tabularx}
\label{tab:2grams_stats}
\end{table}

\begin{table}[htp!]
\centering
\caption{Average daily summary statistics for $3$-grams.}
\begin{tabularx}{\columnwidth}{l|C|C||C|C}
    & \multicolumn{2}{c||}{\textbf{AT}}         & \multicolumn{2}{c}{\textbf{OT}}         \\ 
    & \textbf{Volume}    & \textbf{Unique}    & \textbf{Volume}    & \textbf{Unique}    \\
\hline
$\mu$                  & 3.66 $\times 10^8$ & 1.28 $\times 10^8$ & 1.62 $\times 10^8$ & 9.04 $\times 10^7$ \\
$25^{\textnormal{th}}$ & 3.30 $\times 10^8$ & 1.17 $\times 10^8$ & 1.30 $\times 10^8$ & 7.66 $\times 10^7$ \\
$50^{\textnormal{th}}$ & 3.88 $\times 10^8$ & 1.31 $\times 10^8$ & 1.50 $\times 10^8$ & 8.65 $\times 10^7$ \\
$75^{\textnormal{th}}$ & 4.42 $\times 10^8$ & 1.52 $\times 10^8$ & 2.13 $\times 10^8$ & 1.12 $\times 10^8$ \\
max                    & 1.03 $\times 10^9$ & 2.12 $\times 10^8$ & 3.19 $\times 10^8$ & 1.61 $\times 10^8$
\end{tabularx}
\label{tab:3grams_stats}
\end{table}

\subsection{Twitter $n$-grams}
\label{subsec:storywrangler.storyons}

For the initial version of Storywrangler, we have extracted 
$n$-grams from tweets where $n \in \{1, 2, 3\}$. 
We record raw $n$-gram frequency (or count)
at the day scale for each language (including unidentified), and for Twitter as a whole.

\begin{figure*}[tp!]
  \centering	
  \floatbox[{\capbeside\thisfloatsetup{capbesideposition={right,center},capbesidewidth=.2\textwidth}}]{figure}[0.95\FBwidth]{
    \caption{
    \label{fig:storywrangler.regex}   
      \textbf{Screenshot of Storywrangler's $n$-gram regular expression pattern recognition.} 
        For our application, 
        we designed a custom $n$-gram tokenizer to accommodate all Unicode characters.
        Our $n$-gram parser is case sensitive.
        We preserve contractions, handles, hashtags, date/time strings, currency, HTML codes, and links 
        (similar to the Tweet Tokenizer in the NLTK library~\protect\cite{loper2002a}).
        We endeavor to combine contractions and acronyms as single objects and parse them out as $1$-grams 
        (e.g., `It's', `well-organized', and `B\&M').
        In addition to text-based $n$-grams, 
        we track all emojis as $1$-grams. 
        While we can identify tweets written in continuous-script-based languages 
        (e.g., Japanese, Chinese, and Thai),
        our current parser does not support breaking them into $n$-grams.
        Although some older text tokenization toolkits followed different criteria, 
        our protocol is consistent with modern computational linguistics 
        for social media data and is widely adopted among researchers~\protect\cite{hong2020a,bevensee2020a}.
    }
  }{\includegraphics[width=0.6\textwidth]{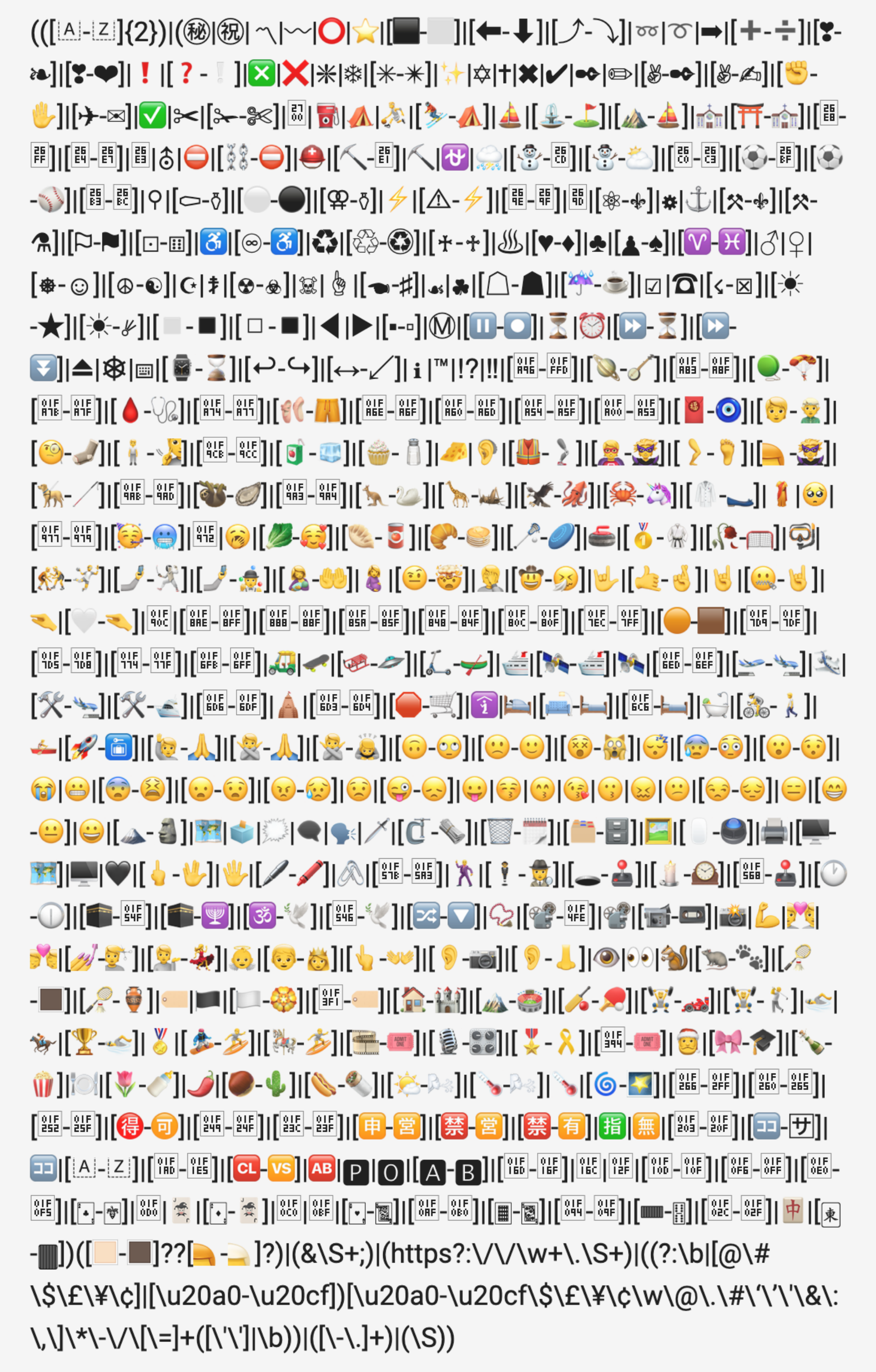}}
\end{figure*}

Although some older text tokenization toolkits followed different criteria 
(e.g., Google books~\cite{michel2011a}), 
we had to design a custom $n$-gram tokenizer to preserve all Unicode characters. 
In particular, 
to accommodate the rich lexicon of social media data, 
we have to preserve handles, hashtags, tickers, and emojis, 
which are irrelevant for books, 
but rather integral to social media platforms such as Twitter. 
We try to maintain the structures in each message as fully as possible, 
giving the user the option of filtering out our daily Zipf distributions in ways that would fulfill their needs. 
We display a screenshot of our regular expression pattern matching in Fig.~\ref{fig:storywrangler.regex}.
Our source code along with our documentation 
is publicly available online on a 
Gitlab repository.\footnote{\url{https://gitlab.com/compstorylab/storywrangler}}

A $1$-gram is a continuous string of characters bounded by either whitespace or punctuation marks. 
For example, the word `the' is one of the prominent $1$-grams in English. 
The $2$-gram `here?' consists of the $1$-grams: `here' and `?'.
Numbers and emojis also count as $1$-grams.
Similarly, a word bounded by two quotes (e.g., ``sentient'') 
would be a $3$-gram, and the expression `see the light' is a $3$-gram, and so forth.

We parse currency (e.g., \$9.99), 
floating numbers (e.g., 3.14), 
and date/time strings (e.g., 2001/9/11, 2018-01-01, 11:59pm) 
all as $1$-grams. 
We curate 
links (e.g., https://www.google.com/), 
handles (e.g., @NASA), 
stocks/tickers (e.g., \$AAPL)
and hashtags (e.g., \#metoo) as $1$-grams.
We endeavor to combine contractions and acronyms as single objects and parse them out as $1$-grams 
(e.g., `It's', `well-organized', and `B\&M').

Emojis are uniquely and interestingly complex entities.\footnote{\url{https://www.unicode.org/Public/emoji/12.0/emoji-data.txt}}
People-centric emojis can be composed of skin-tone modifiers, hair-type modifiers, and family structures.
Emoji flags are two component objects.
The most elaborate emojis are encoded by seven or more unicode elements, rendering them difficult to extract as single entities.
After contending with many emoji-parsing problems, 
we record all emojis as $1$-grams.
We consider repeated emojis with no intervening whitespace---a common feature in tweets---to be a series of $1$-grams.

\begin{figure*}[tp!]  
  \includegraphics[width=\linewidth]{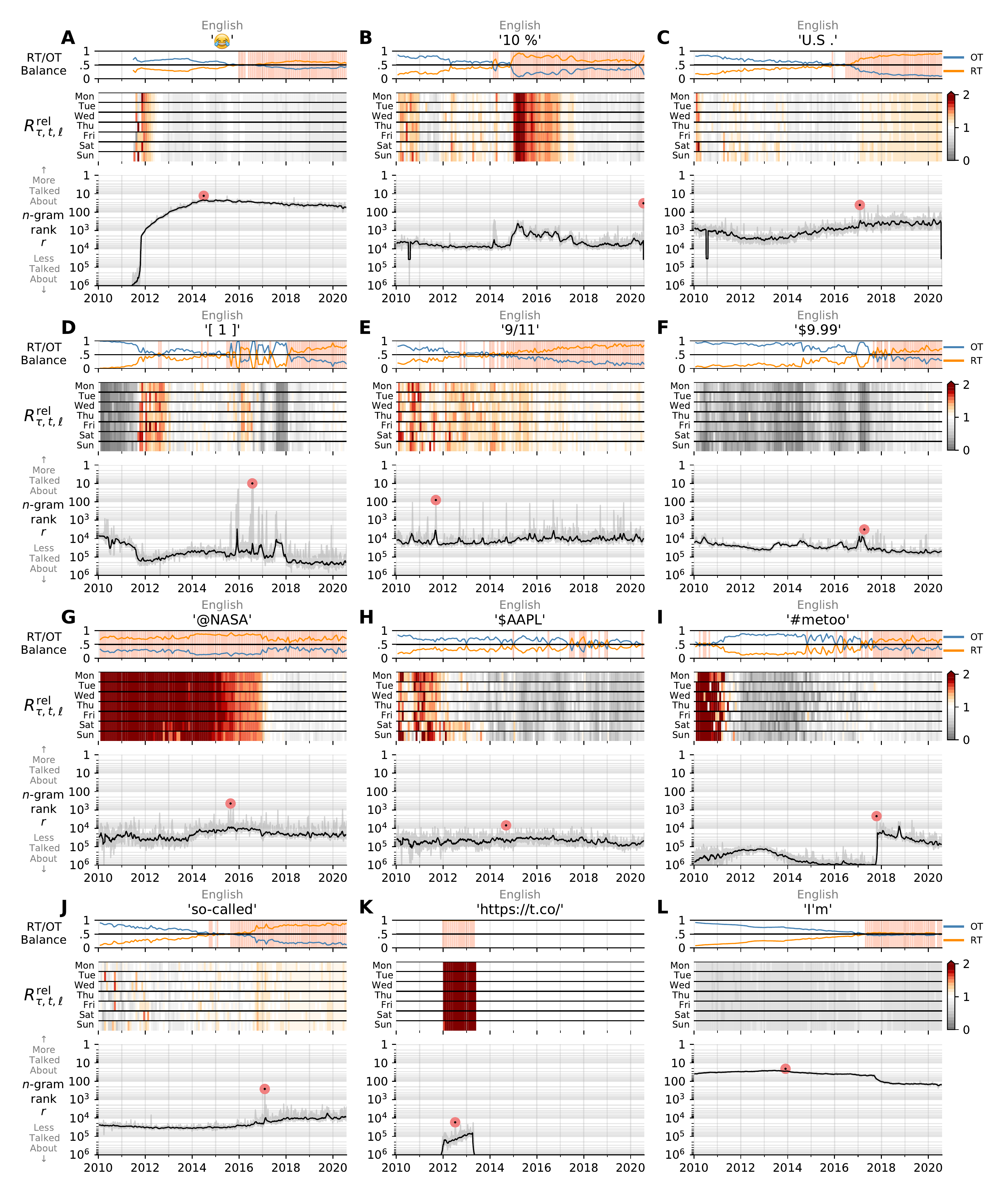}
  \caption{
    \label{fig:storywrangler.contagiograms_samples}
    \textbf{Contagiograms.}
    Example timeseries showing social amplification for Twitter $n$-grams involving emojis, punctuation, numerals, and so on.
  }  
\end{figure*}  

\begin{figure*}[tp!]  
\includegraphics[width=\linewidth]{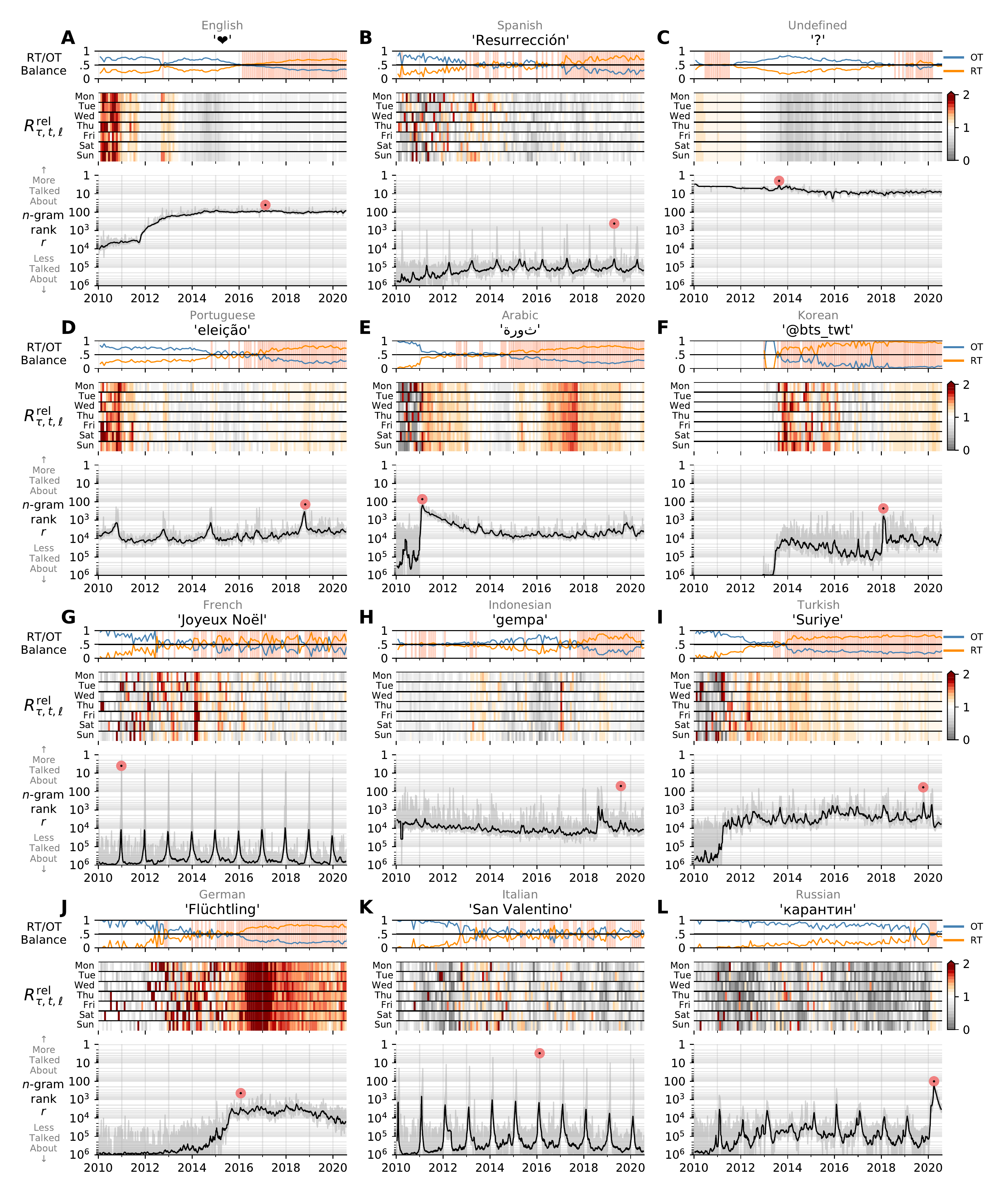}
\caption{
  \label{fig:storywrangler.contagiograms_langs} 
  \textbf{The interplay of social amplification across various languages.}  
  We observe a wide range of sociotechnical dynamics starting with
  $n$-grams that are often mentioned within OTs and RTs equivalently to
  others that spread out after a geopolitical event and more extreme
  regimes whereby some $n$-grams are consistently amplified.
  English translations of $n$-grams:
  \textbf{A.}  Heart emoji,
  \textbf{B.} `Resurrection',
  \textbf{C.}  Question mark,
  \textbf{D.} `election',
  \textbf{E.} `revolution',
  \textbf{F.}  Official handle for the South Korean boy band `BTS',
  \textbf{G.} `Merry Christmas',
  \textbf{H.} `earthquake',
  \textbf{I.} `Syria',
  \textbf{J.} `Refugee',
  \textbf{K.} `Saint Valentine',
  and
  \textbf{L.} `quarantine'.
}
\end{figure*}  

Our protocol is similar to the Tweet Tokenizer 
developed as part of the Natural Language Toolkit (NLTK)~\cite{loper2002a} 
and is adopted in recent applications of modern computational linguistics for social media~\cite{hong2020a,bevensee2020a}. 
In Fig.~\ref{fig:storywrangler.contagiograms_samples}, we show
contagiograms for 12 example $n$-grams that involve punctuation,
numbers, handles, hashtags, and emojis. 
A few more examples across 
various languages can be seen in 
Fig.~\ref{fig:storywrangler.contagiograms_langs}.
We provide a Python package for generating arbitrary contagiograms
along with further examples
at \url{https://gitlab.com/compstorylab/contagiograms}.
The figure-making scripts interact directly with the Storywrangler database,
and offer a range of configurations.

Our $n$-gram parser is case sensitive. 
For example, search queries made on \href{storywrangling.org}{storywrangling.org}
for `New York City' and a search for `new york city' would return different results.
Normalized frequencies and rankings in our daily Zipf distributions are consequently for case-sensitive $n$-grams.

\begin{figure*}[tp!]
\includegraphics[width=\textwidth]{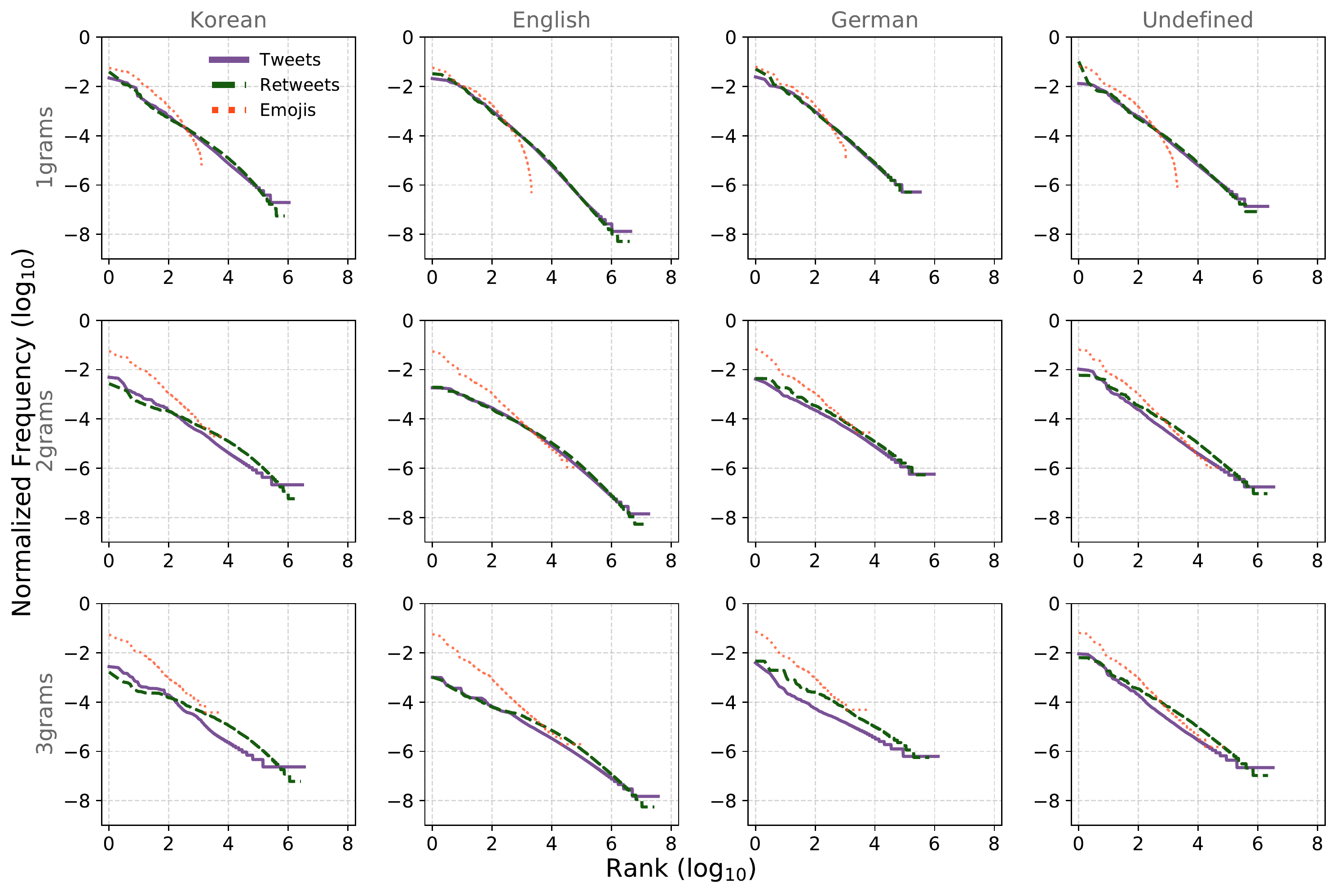}
\caption{ 
\textbf{Zipf distributions for Korean, English, German and undefined language categories for October 16, 2019 on Twitter.}
``Tweets'' refer to organic content, ``retweets'' retweeted content, and ``emojis'' are $n$-grams comprised of strictly emojis (organic and retweets combined). 
}
\label{fig:storywrangler.ccdf}   
\end{figure*}  

Although we can identify tweets written in continuous-script-based languages 
(e.g., Japanese, Chinese, and Thai),
our current parser does not support breaking them into $n$-grams.
We label tweets as Undefined (und) to indicate tweets that we could not classify their language with a confidence score above 25\%. 
The resulting $n$-grams are allocated to an ``Undefined'' 
category as well as to the overall Twitter $n$-gram data set.

\begin{figure*}[tp!]
\includegraphics[width=.9\textwidth]{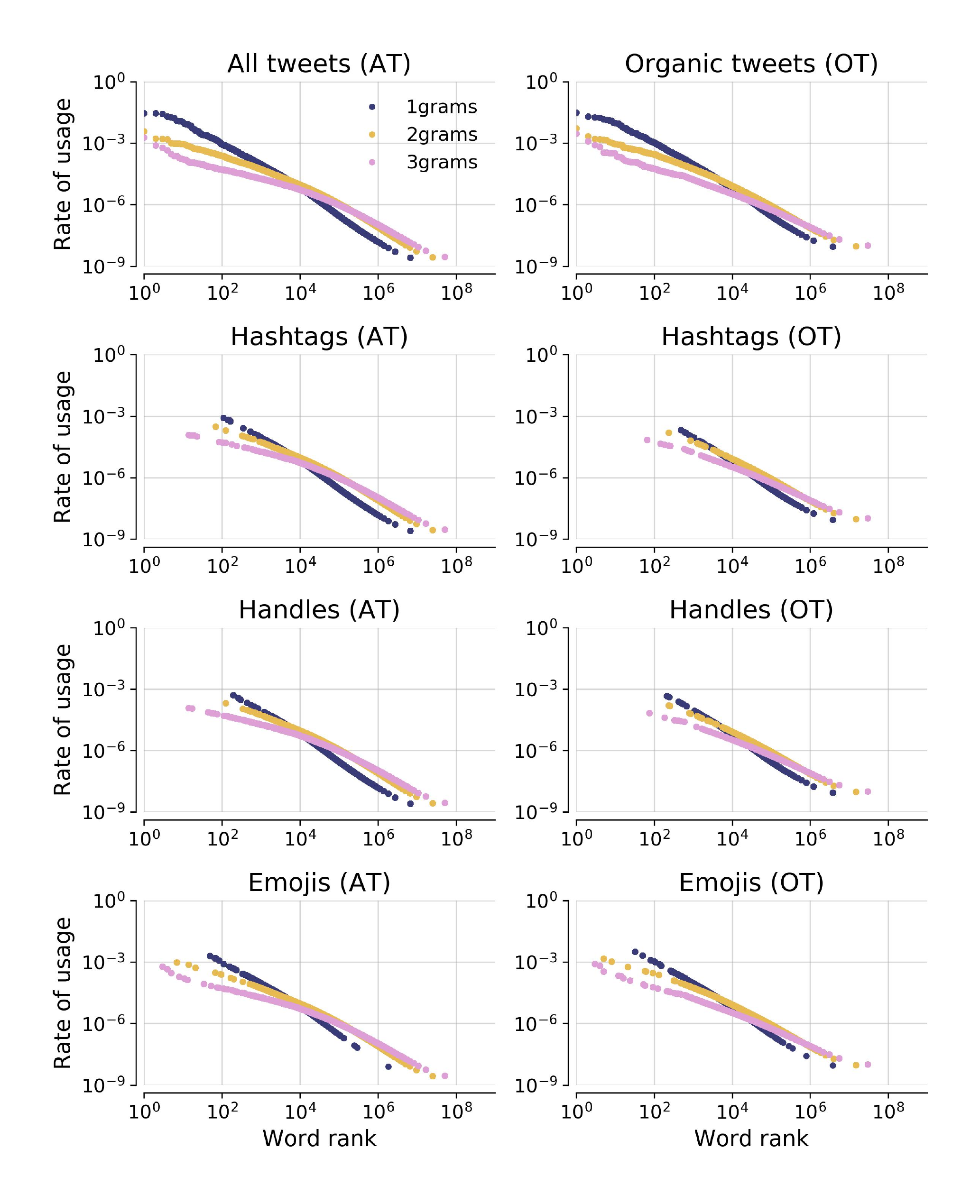}
\caption{ 
\textbf{Daily Zipf distributions for English on May 1st, 2020.} 
We show a weighted 1\% random sample of $1$-grams (blue), $2$-grams (yellow), and $3$-grams (pink) 
in all tweets (AT) and organic tweets (OT) accordingly. 
On the vertical axis, we plot the relative rate of usage of each $n$-gram in our random sample 
whereas the horizontal axis displays the rank of that $n$-gram in the English corpus of that day. 
We first display Zipf distributions for all $n$-grams observed in our sample in the first row.
We also demonstrate the equivalent distributions for 
hashtags (second row),
handles (third row),
and emojis (last row). 
}
\label{fig:storywrangler.en_ngrams_zipf}   
\end{figure*}

To enable access to our dataset, we maintain a MongoDB database of
the top million ranked $n$-grams on each day for each
language. 
We index these collections by date, to allow efficient
queries for all $n$-grams on a given day, as well as by $n$-gram,
which allows for rapid time series queries. 
Data is typically inserted
within two days 
(i.e., counts from Monday will be available by midnight Wednesday).

\subsection{Constructing daily Zipf distributions}
\label{subsec:storywrangler.metadata}

For ease of usability, we maintain two sets of daily measurements for
each $n$-gram in our data set:
raw frequency (count), normalized frequency (probability), and
tied rank with and without retweets.  
We make the default
ordering for the Zipf distribution files according to usage levels of
$n$-grams for all of a given language on Twitter (i.e., including all
retweets and quote tweets).  
Again, all daily distributions are made
according to Eastern Time calendar days.

We denote an $n$-gram by $\tau$ and
a day's lexicon for language $\ell$---the set of distinct $n$-grams found in all tweets (AT)
for a given date $t$---by $\lexiconAT$.
We further define the set of unique language $\ell$ $n$-grams found in organic tweets
as
$\lexiconOT$,
and the set of unique $n$-grams found in
retweets as
$\lexiconRT$, such that
\begin{equation}
\lexiconAT
=
\lexiconOT
\cup
\lexiconRT.
\end{equation}

We compute relative daily rate of usage by dividing total number of
occurrences of a given $n$-gram by the total number of $n$-grams for
that day.  
We write $n$-gram raw frequency as $f_{\tau,t,\ell}$,
and compute its usage rate in all tweets written in language $\ell$
as
\begin{equation}
  p_{\tau,t,\ell}
  =
  \frac{f_{\tau,t,\ell}}
       {\sum_{\tau' \in \lexiconAT} f_{\tau',t,\ell}}.
\end{equation}
The corresponding normalized frequencies for $n$-grams in organic tweets and retweets are then defined as
\begin{equation} 
  p_{\tau,t,\ell}^{(\organictweetsym)}
  =
  \frac{f_{\tau,t,\ell}^{(\organictweetsym)}}{
    \sum_{\tau'
      \in
      \lexiconOT}
      f_{\tau',t,\ell}^{(\organictweetsym)}}, \ \textnormal{and}
\end{equation}
\begin{equation} 
  p_{\tau,t,\ell}^{(\retweetsym)}
  =
  \frac{f_{\tau,t,\ell}^{(\retweetsym)}}{
    \sum_{\tau'
      \in
      \lexiconRT}
      f_{\tau',t,\ell}^{(\retweetsym)}}.
\end{equation}

We then rank all $n$-grams for a given day to create daily
Zipf distribution~\cite{zipf1949a} for all languages in our data set.
If two or more distinct $n$-grams have the same number of instances
(raw frequency), then we adjust their ranks by taking the average
rank (i.e., tied-rank).
The corresponding notation is:
\begin{equation}
  r_{\tau,t,\ell},
  \
  r_{\tau,t,\ell}^{(\organictweetsym)},
  \
  \textnormal{and}
  \
  r_{\tau,t,\ell}^{(\retweetsym)}.
\end{equation}

We do not mix $n$-grams for different values of $n$, and leave this as
an important future upgrade~\cite{williams2015a}.  Users of the viewer
\href{storywrangling.org}{storywrangling.org}, will need to keep this
in mind when considering time series of $n$-grams for different $n$.
In comparing, for example, `NYC' (a $1$-gram) to `New York City' (a
$3$-gram), the shapes of the curves can be meaningfully compared while the ranks
(or raw frequencies) of the $1$-gram and $3$-gram may not be.

We show complementary cumulative distribution functions (CCDFs) of
organic tweets, retweets, and emojis 
for $1$-, $2$-, and $3$-grams in
Fig.~\ref{fig:storywrangler.ccdf} and
Fig.~\ref{fig:storywrangler.en_ngrams_zipf}.

\clearpage
\section{Narratively trending $n$-grams}
\label{appx:storywrangler.trending_ngrams}

\begin{figure*}[tp!]
\includegraphics[width=\textwidth]{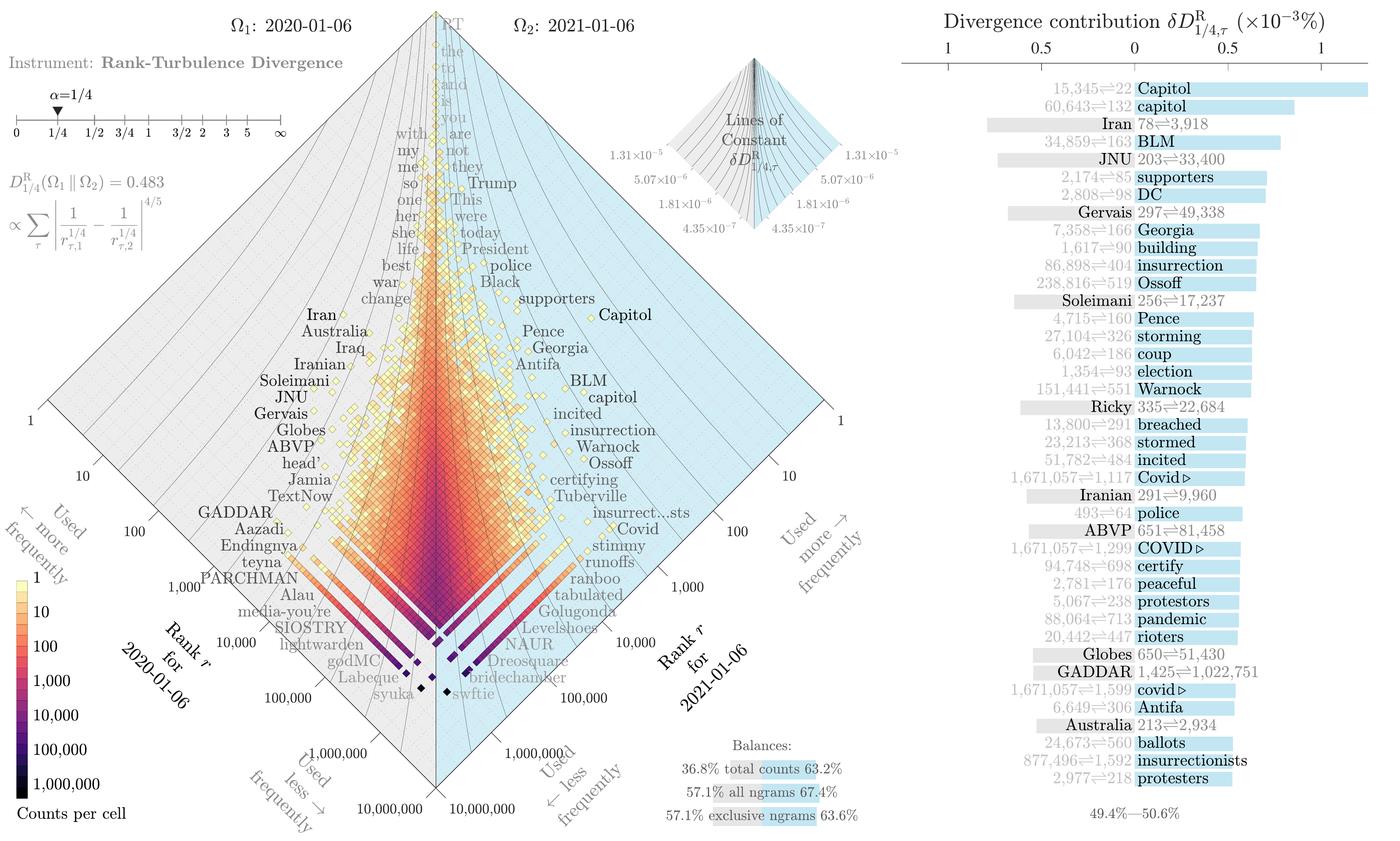}
\caption{   
  \textbf{Allotaxonograph using rank-turbulence divergence for
    English word usage on 2021-01-06 compared to 2020-01-06}. The word `Capitol' was the 22nd most common $1$-gram on January 6, 2021, up from 15,345th most common one year earlier. Similarly, `COVID' was the 1,117th most popular word January 6, 2021, and did not make the top million on January 6, 2020. See Dodds~\etal~\cite{dodds2020allotaxonometry} for further details on the allotaxonometric instrument.
}   
\label{fig:storywrangler.allotaxonograph}   
\end{figure*} 

\begin{figure*}[tp!]
\includegraphics[width=.9\textwidth]{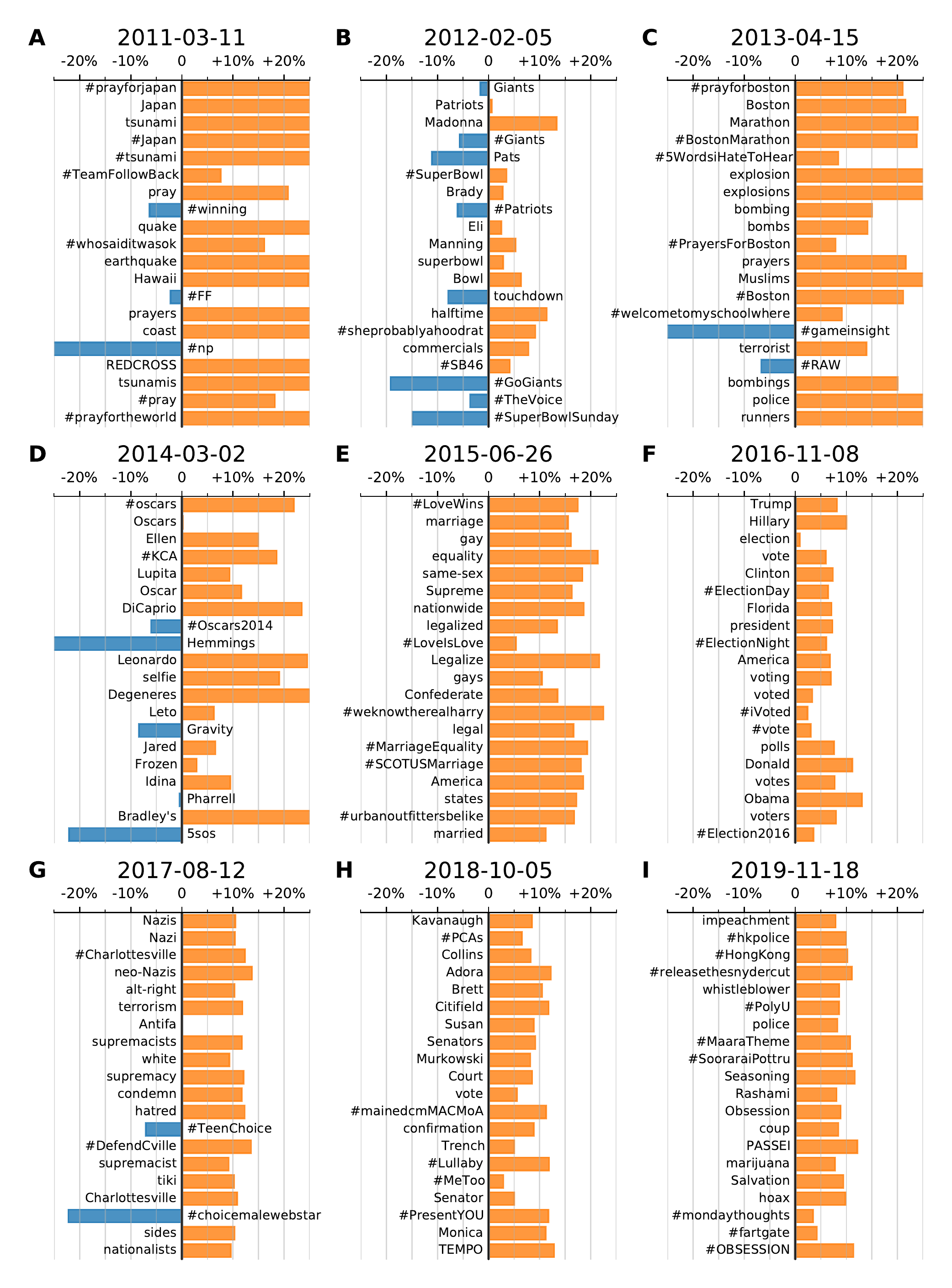}
\caption{   
  \textbf{Narratively trending 1-grams}. 
  Top 20 narratively dominate 1-grams for a few days of interest throughout the last decade
  (sorted by their rank-turbulence divergence contribution).  
  Positive values (orange) indicate strong social amplification via retweets,
  whereas negative values (blue) show terms that are prevalent in originally authored tweets. See Supplementary text for details on each date.
}   
\label{fig:storywrangler.trending_1grams}   
\end{figure*}

\begin{figure*}[tp!]
\includegraphics[width=.9\textwidth]{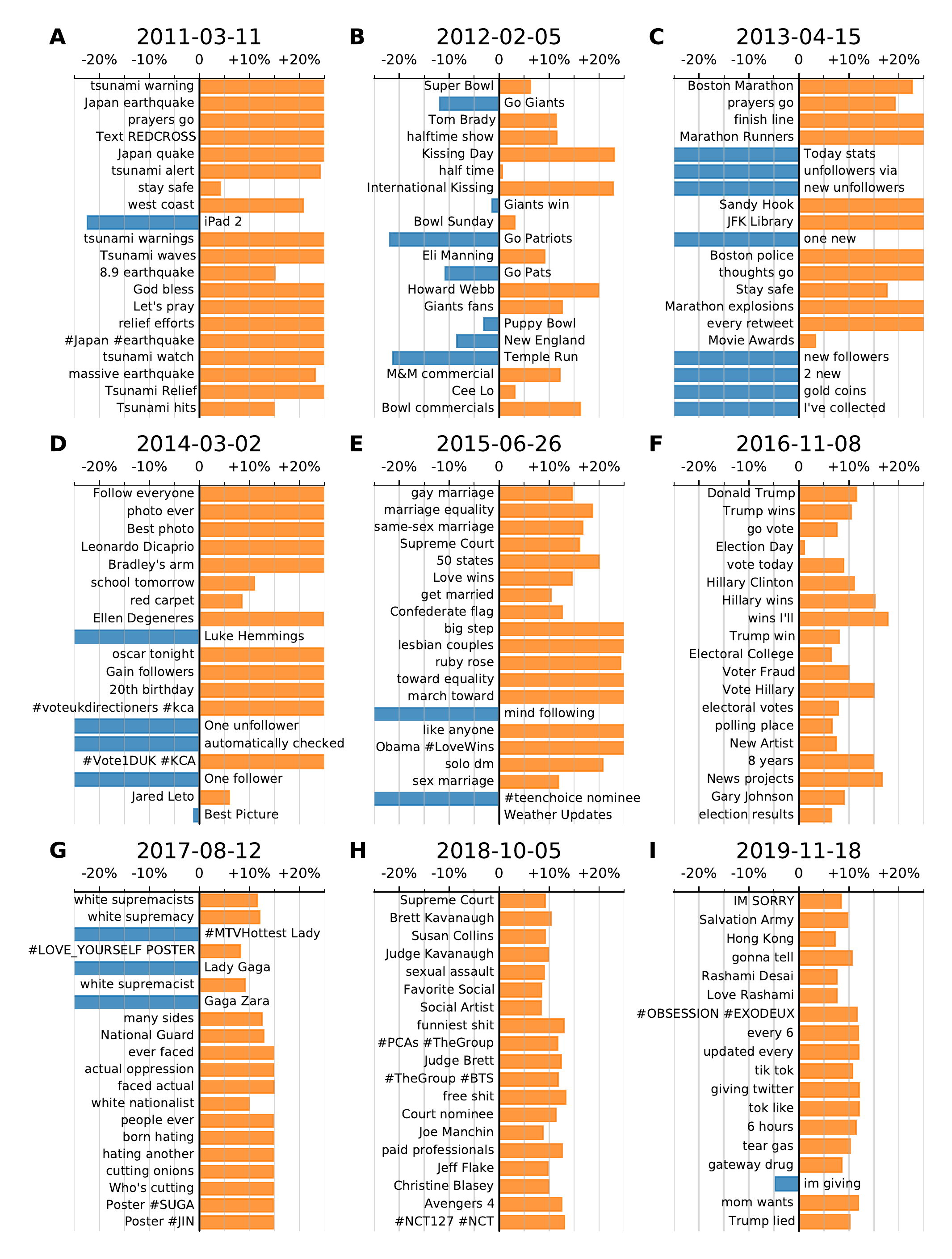}
\caption{   
  \textbf{Narratively trending 2-grams}.
  Top 20 narratively dominate 2-grams for the same days shown in Fig.~\ref{fig:storywrangler.trending_1grams}. See Supplementary text for details on each date.
}   
\label{fig:storywrangler.trending_2grams}   
\end{figure*}

In addition to curating daily Zipf distributions, 
we also examine lexical turbulence and emerging storylines in real-time. 
We do so using rank-turbulence divergence (RTD)~\cite{dodds2020allotaxonometry}, 
an instrument designed to examine the lexical turbulence of two heavy-tailed distributions. 

For each day $t$ and each language $\ell$, 
we take the Zipf distribution $\Omega_{t, \ell}$
and compare it with the corresponding distribution of that day from the same day one year prior $\Omega_{t', \ell}$,
identifying $n$-grams that have become most elevated in relative usage.
We set the parameter $\alpha$ to 1/4, 
which provides a reasonable fit for the lexical turbulence of social media data 
as recommended by Dodds~\etal~\cite{dodds2020allotaxonometry}.
We compute rank divergence contribution such that:
\begin{align}
\begin{split}
     D^{\textnormal{R}}_{\alpha}(\Omega_{t, \ell}||\Omega_{t', \ell})
     &= \sum \delta D^{\textnormal{R}}_{\tau, \ell} \\
     &= \frac{\alpha + 1}{\alpha} 
     \sum_\tau \bigg| \dfrac{1}{r_{\tau, t, \ell}^{\alpha}} - \dfrac{1}{r_{\tau, t', \ell}^{\alpha}} \bigg|^{1 / (\alpha + 1)},
\end{split}
\end{align}
where $r_{\tau, t, \ell}$ is the rank of $n$-gram $\tau$ on day $t$,
and $r_{\tau, t', \ell}$ is the rank of the same $n$-gram on day $t'$ (52 weeks earlier).
Although we use rank-turbulence divergence to determine shifts in relative word rankings, 
other divergence metrics will provide similar lists.

We show an example Allotaxonograph using rank-turbulence divergence 
for English word usage on 2021-01-06 compared to 2020-01-06 
(see Fig.~\ref{fig:storywrangler.allotaxonograph}).
The main plot shows a 2D histogram of $n$-gram ranks on each day. 
Words located near the center vertical line are used equivalently on both days, 
while words on either side highlight higher usage of the given $n$-gram on either date. 

The right side of the plot displays the most narratively dominant $n$-grams
ordered by their rank divergence contribution. 
For ease of plotting, we use the subset of words containing Latin characters only.
Notably, 
words associated with the storming of the US Capitol by Trump supporters 
dominate the contributions from 2021-01-06 such as 
`Capitol', 
`supporters',
`DC',
and
`insurrection'.

In Fig.~\ref{fig:storywrangler.trending_1grams} 
and Fig.~\ref{fig:storywrangler.trending_2grams}, 
we show example analyses of English tweets highlighting 
the top 20 narratively trending 1-grams and 2-grams, respectively. 

First, 
we compute RTD values to find the most narratively dominate $n$-grams 
for a few days of interest throughout the last decade. 
We filter out links, emojis, handles, and stop words, but keep hashtags 
to focus on the relevant linguistic terms on these days.

Second, 
we further compute the relative social amplification ratio $\relativeRTrate_{\tau,t,\ell}$ 
to measure whether a given $n$-gram $\tau$ is socially amplified, 
or rather shared organically in originally authored tweets.

For ease of plotting, 
we display $\relativeRTrate_{\tau,t,\ell}$ on a logarithmic scale. 
Positive values of $\log_{10} \relativeRTrate_{\tau,t,\ell}$ 
imply strong social amplification of $\tau$,
whereas negative values show that $\tau$ is rather predominant in organic tweets.

Figs.~\ref{fig:storywrangler.trending_1grams}A and \ref{fig:storywrangler.trending_2grams}A
show the top terms used to discuss the earthquake off the Pacific coast of Tokyo
leading to a sequence of massive tsunami waves on 
2011-03-11.\footnote{\url{https://www.britannica.com/event/Japan-earthquake-and-tsunami-of-2011}}
Although most terms are socially amplified, 
referring to the catastrophic event in Japan, 
other cultural references can be found in organic tweets 
such as `\#np' (i.e., no problem), and `\#FF' (i.e., follow Friday) 
where users recommend accounts to follow. 

In Figs.~\ref{fig:storywrangler.trending_1grams}B and \ref{fig:storywrangler.trending_2grams}B, 
we see trending terms discussing the Super Bowl held on 2012-02-05. 

Figs.~\ref{fig:storywrangler.trending_1grams}C and \ref{fig:storywrangler.trending_2grams}C 
show salient $n$-grams in response to the terrorist attack 
during the annual Boston Marathon on 
2013-04-15.\footnote{\url{https://www.history.com/topics/21st-century/boston-marathon-bombings}}
We also observe unusual bigrams in organic tweets generated by artificial Twitter bots 
for a new automated service that went viral in 2013, allowing users 
to keep track of their new followers 
and unfollowers by tweeting their stats daily.\footnote{\url{https://who.unfollowed.me/}}

Figs.~\ref{fig:storywrangler.trending_1grams}D and \ref{fig:storywrangler.trending_2grams}D 
show names of celebrities and movie titles that went viral on Twitter 
during the $86^{\textnormal{th}}$ Academy Awards hosted by Ellen DeGeneres on 
2014-03-02.\footnote{\url{https://en.wikipedia.org/wiki/86th_Academy_Awards}}
Most socially amplified terms can be traced back to a single message with more than 3 million retweets. 
We do, however,
see some $n$-grams trending in organic tweets such as `5sos', 
referring to an Australian pop rock band that was slowly taking off that year.

On 2015-06-26,
the US Supreme Court ruled same-sex marriage is a legal right in the US\footnote{\url{https://www.usatoday.com/story/news/politics/2020/06/25/lgbtq-rights-five-years-after-gay-marriage-ruling-battles-continue/3242992001/}}, 
prompting a wave of reactions on social media as seen in 
Figs.~\ref{fig:storywrangler.trending_1grams}E and \ref{fig:storywrangler.trending_2grams}E.

The 2016 US presidential election had a similar chain of reactions on Twitter.
In Figs.~\ref{fig:storywrangler.trending_1grams}F and \ref{fig:storywrangler.trending_2grams}F,
we see names of candidates and politicians across the aisle being amplified collectively on the platform.

As we have seen throughout the paper, cultural movements and social media are profoundly integrated. 
In Figs.~\ref{fig:storywrangler.trending_1grams}G and \ref{fig:storywrangler.trending_2grams}G, 
we see the top 20 narratively trending terms in response to the deadly white supremacist rally 
that took place on 2017-08-12 in Charlottesville, Virginia.\footnote{\url{https://time.com/charlottesville-white-nationalist-rally-clashes/}} 
In particular, we notice several bigrams from Obama's tweet that went viral on that day, 
quoting Nelson Mandela’s remarks on racism. 

In Figs.~\ref{fig:storywrangler.trending_1grams}H and \ref{fig:storywrangler.trending_2grams}H, 
we see some headlines and narratively amplified $n$-grams,
in light of Kavanaugh’s testimony before the Senate Judiciary Committee 
for his nomination to the US Supreme Court 2018-10-05.\footnote{\url{https://www.washingtonpost.com/politics/2018/10/05/brett-kavanaughs-testimony-what-was-misleading-what-was-not/}}

Figs.~\ref{fig:storywrangler.trending_1grams}I and \ref{fig:storywrangler.trending_2grams}I 
display the top $n$-grams on 2019-11-18 
whereby we see several terms and hashtags amplified on Twitter 
triggered by the Siege of the Hong Kong Polytechnic University 
amid the nationwide protests in Hong Kong.\footnote{\url{https://www.theguardian.com/world/2019/nov/18/hong-kong-university-siege-a-visual-guide}} 
Moreover, we also see notable references to the first impeachment of Trump 
that took place between September 2019 and February 2020.

\section{Pantheon case study}
\label{appx:storywrangler.pantheon_case_study}

\begin{figure*}[tp!]
\includegraphics[width=.9\textwidth]{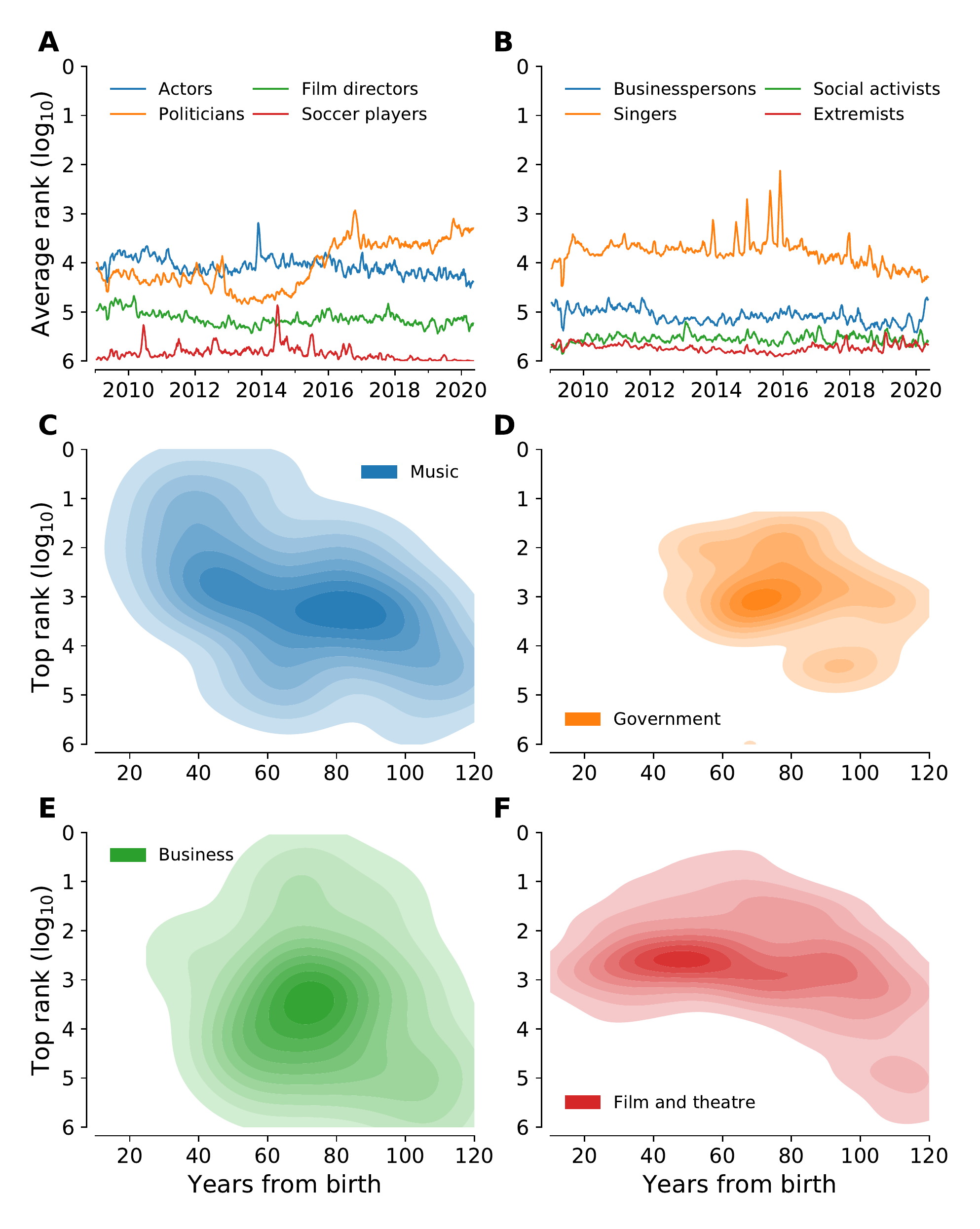}
\caption{   
  \textbf{Rankings of celebrities on Twitter}.
  We take a closer look at rankings of famous figures
  by cross-referencing our English corpus with names of 
  celebrities from the Pantheon dataset~\protect\cite{yu2016a}.
  We use their first and last name to search through our $2$-grams data set.
  We select names of Americans who were born in the last century and can be found 
  in the top million ranked $2$-grams for at least a day between 2010-01-01 and 2020-06-01.
  In panels \textbf{A} and \textbf{B},
  we display a centered monthly rolling average of the average rank for the 
  top 5 individuals for each category $\langle r_{\textnormal{min}(5)} \rangle$.
  We also plot the kernel density estimation of the best rank achieved 
  by another 1162 famous characters in each of the following industries:
  \textbf{C.}
  music,
  \textbf{D.}
  government,
  \textbf{E.}
  business,
  and
  \textbf{F.}
  film. 
}   
\label{fig:storywrangler.pantheon_age}   
\end{figure*}

We examine the dialog around celebrities by cross-referencing our English $2$-grams corpus
with names of famous personalities from the Pantheon data set~\cite{yu2016a}. 
The data set has over 10 thousand biographies. 
We use the place and date of birth to select Americans born in the last century. 

We searched through our English $n$-grams data set and selected names that were found 
in the top million ranked $2$-grams for at least a day between 2010-01-01 and 2020-06-01.
Our list contains 1010 individuals.
We show the average best rank $\bar{r}_{\textnormal{min}}$, 
median best rank $\tilde{r}_{\textnormal{min}}$, 
and best rank $r^*_{\textnormal{min}}$
for all individuals in each occupation in Tab.~\ref{tab:occupations}.
In Figs.~\ref{fig:storywrangler.pantheon_age}A and B,
we display a monthly rolling average (centered) of the average rank for the 
top 5 individuals for each category $\langle r_{\textnormal{min}(5)} \rangle$.   

\begin{table}[htp!]
\centering
\caption{Celebrities by occupation}
\begin{tabularx}{\columnwidth}{l||C|C|C|C}
Occupation & $n$ & $\bar{r}_{\textnormal{min}}$ & $\tilde{r}_{\textnormal{min}}$ & $r^*_{\textnormal{min}}$  \\ 
\hline
\textbf{Actors}  &  674 & 40,632 & 9,255 & 2 \\
\textbf{Singers}  &  162  & 59,713 & 3,479  & 4 \\
\textbf{Politicians}  &  59  & 6,365 & 1,376 & 6 \\
\textbf{Film directors}  &  57 & 75,783 & 10,580 & 13 \\
\textbf{Business-persons}  &  26 & 35,737 & 4195 & 15 \\
\textbf{Soccer players}  &  12  & 20,868  & 8,507 & 25 \\
\textbf{Social activists}  &  10  & 20,302 & 1,781 & 841 \\
\textbf{Extremists}  &  10  & 104,621 & 20,129  & 117 \\
\end{tabularx}
\label{tab:occupations}
\end{table} 

Additionally, we select a total of 1162 celebrities that were also found in the top million ranked $2$-grams for at least a day between 2010-01-01 and 2020-06-01 in a few selected industries (see Tab.~\ref{tab:industries})

\begin{table}[htp!]
\centering
\caption{Celebrities filtered through Pantheon and Twitter rank, by industry}
\begin{tabularx}{\columnwidth}{X||C}
Industry & Individuals  \\ 
\hline
\textbf{Film and theater}  &  751  \\
\textbf{Music}  &  324  \\
\textbf{Government}  &  59  \\
\textbf{Business}  &  28 
\end{tabularx}
\label{tab:industries}
\end{table} 

For each of these individuals, we track their age and top daily rank of their names (first and last). 
In Figs.~\ref{fig:storywrangler.pantheon_age}A, B, C, and D, 
we display kernel density estimation of the top rank achieved by any of these individuals 
in each industry as a function of the number of years since the recorded year of birth (age of the cohort).

\section{Movies case study} 
\label{appx:storywrangler.movie_case_study}

We investigate the conversation surrounding major film releases by
tracking $n$-grams that appear in movie titles.  From the MovieLens
dataset~\cite{harper2015a}, we selected 636 movies with gross
revenue above the 95th percentile during the period ranging from
2010-01 to 2017-07.  We then retrieved the normalized frequency time series for
up to the first 3-grams of a movie's title (e.g., ``Prince of Persia:
The Sands of Time'' would correspond to the 3-gram time series ``Prince
of Persia'').  From there, we look for the maximum daily normalized frequency.
To disambiguate between movies within the same franchise and/or titles
with common $n$-grams, we restrain this search to the release year of
the given movie.  With the peak usage in the year of a movie's
release, we then search backward for the date on which the $n$-gram
usage first breaks 50\% of the peak usage normalized frequency, $f_{.5}$.
Similarly we search forward, from peak usage, for the date on which
the time series first declines below $f_{.5}$.

Peak conversation surrounding major movies tends to occur a few days after the release data of the title. 
We find a median value of 3 days post-release for peak normalized
frequency of usage for movie $n$-grams (Fig.~\ref{fig:storywrangler.studies}F inset).
Growth of $n$-gram usage from 50\% to maximum
normalized frequency has a median value of 5 days across our 636 titles. 
The median value of time to return to 50\% from maximum normalized frequency is 6 days. 
Looking at Fig.~\ref{fig:storywrangler.studies}E we see the median shape of the spike around movie release dates tend to entail a gradual increase to peak usage, and a more sudden decrease when returning to 50\% of maximum normalized frequency. 
There is also slightly more spread in the time to return to 50\% normalized frequency of usage than compared with the time to increase from 50\% to maximum usage (Fig.~\ref{fig:storywrangler.studies}E insets).

\section{Geopolitical risk case study}
\label{appx:storywrangler.risk_case_study}

\begin{figure*}[tp!]
  \centering	
  \floatbox[{\capbeside\thisfloatsetup{capbesideposition={right,center},capbesidewidth=.21\textwidth}}]{figure}[0.95\FBwidth]{
    \caption{
    \label{fig:storywrangler.geopolitical_risk_linear_model}   
    \textbf{Empirical distributions of the $\beta$ coefficient of each word}. 
    Percent change in word popularity is significantly associated with
    percent change of future geopolitical risk
    (GPR) index level for a few words out of a panel of eight words: 
    ``revolution'', ``rebellion'', ``uprising'', ``coup'', 
    ``overthrow'', ``unrest'', ``crackdown'', and ``protests''. 
    We assess significance of model coefficients using centered $80\%$ credible intervals (CI). 
    The sign of the coefficient differs between the words, 
    with positive associations shown in orange and negative associations shown in blue. 
    Using Storywrangler, 
    we note the words ``crackdown'', and ``uprising'' are positively associated with GPR, 
    whereas ``rebellion'' is negatively associated.
    We see some overlap between Storywrangler and other data streams. 
    Percent change of the word ``rebellion'' is also negatively associated with GPR,
    using Google trends search data~\protect\cite{choi2012a}, but not statistically significant.
    By contrast, mentions of the word ``coup'' in cable news is positively associated with GPR 
    using the Stanford cable TV news analyzer~\protect\cite{hong2020a}.
    }
  }{\includegraphics[width=0.8\textwidth]{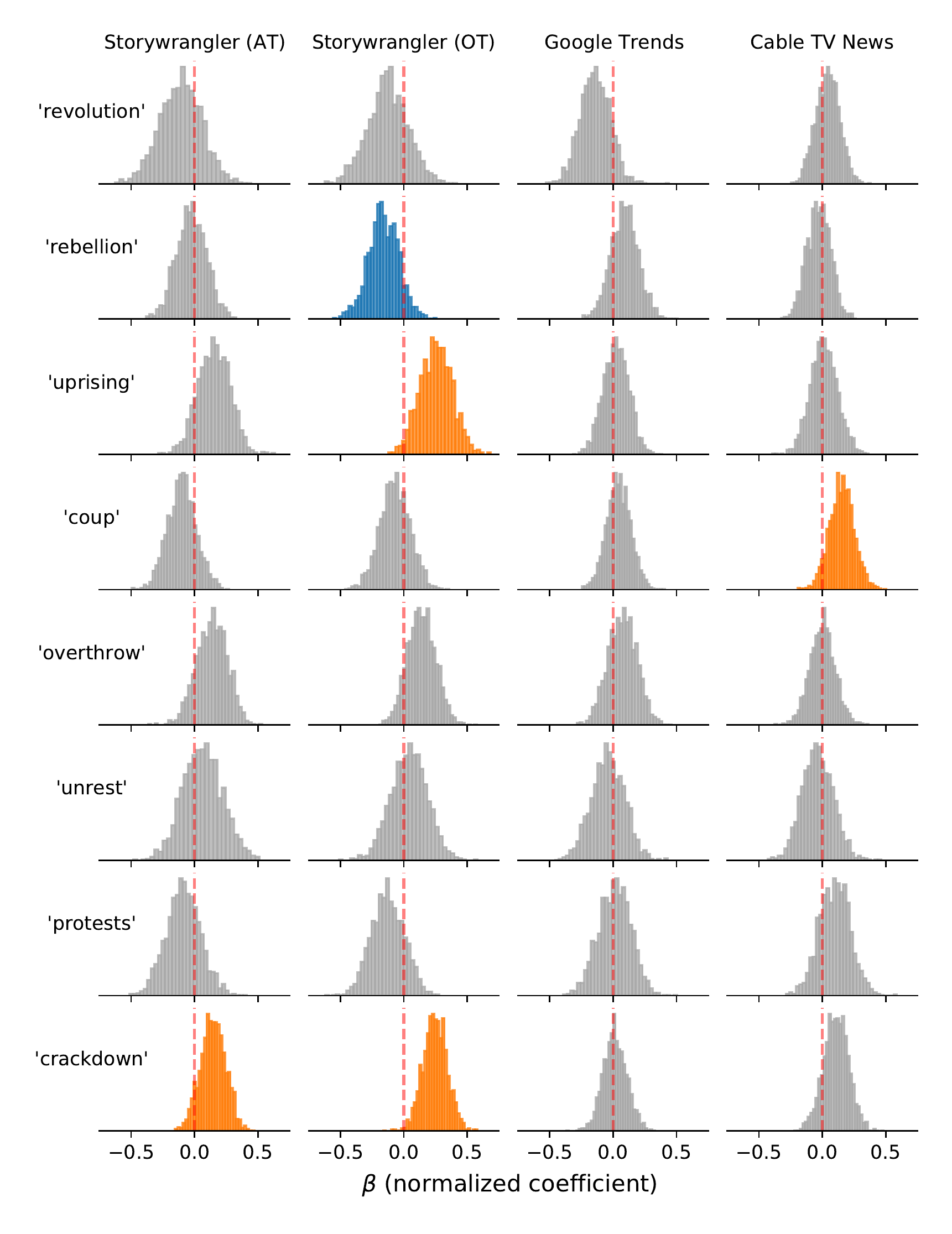}}
\end{figure*}

Twitter sentiment has already been shown to provide a useful signal in
monitoring public opinion~\cite{hong2011a,younus2011a}.
The aggregation process in which individual tweet documents are turned into popularity time series 
reduces the time and computation power required 
to use this data in models of political sentiment and public opinion. 
As a case study, we sought to predict values of a geopolitical risk (GPR) index 
using popularity of words that we heuristically associated with (inter)national unrest and popular discontent.
We conduct both an exploratory Bayesian analysis and a more rigorous frequentist analysis of the relationship between word time series and the GPR index. 

\subsection{Data description}\label{sec:risk-data}
The geopolitical risk index is developed by the U.S.\ Federal Reserve
(central bank)~\cite{caldara2017a}. We chose the words
``revolution'', ``rebellion'', ``uprising'', ``coup'', ``overthrow'',
``unrest'', ``protests'', and ``crackdown'' to include as predictors. 
We chose these words arbitrarily due to denotations and connotations of conflict associated with their common English meaning.
Since we could not reject the null hypothesis that at least one of the
logarithm of normalized frequency time series associated with these words contained a unit
root ($\textnormal{ADF(``overthrow'')} = -1.61,\ p = 0.474$), we computed
the difference of the logarithm of normalized frequency time series and used these
observations as features.  
We could also not reject the null hypothesis
that the geopolitical risk time series contained a unit root
($\textnormal{ADF} = -0.65$, $p = 0.858$) and therefore sought to predict
the log difference of the GPR.  
Because GPR is computed at monthly frequency, 
we resample normalized word frequencies to monthly normalized frequency by
taking the average of the lagged month's values.  
For example, 
the normalized frequency of the word ``crackdown'' sampled at month level timestamped
at 2010-03-31 is taken to be the average daily normalized frequency of the word ``crackdown'' 
from 2010-03-01 through 2010-03-31.

We compute the monthly normalized frequencies for each word in all tweets (AT), 
and originally authored tweets (OT), separately, 
for the last decade starting from 2010 to the end of 2019.
We also collect monthly $n$-gram usage time series for the same set of words 
from two other data sources---namely,
Google trends~\cite{choi2012a}, 
and the Stanford cable TV news analyzer~\cite{hong2020a} 
to examine the utility of the predictors derived from Storywrangler 
compared to existing data sources.
We conduct out of sample analyses on word frequency and GPR data from 2020-01-01 through 2020-05-01, 
which is the last day on which the GPR data was publicly accessible.

\subsection{Exploratory analysis}\label{sec:bayes-gpr}
We fit a linear model for each data source 
that hypothesized a linear relationship between the log difference in
normalized word frequencies for each of the words listed 
in the previous paragraph and the log difference in GPR. 
The likelihood function of this model took the form 
\begin{equation}
p(y|\beta, \sigma) = 
\prod_{t=0}^{T - 1} \textnormal{Normal}(y_{t + 1} | X_t\beta, \sigma^2).
\end{equation}
We denote $y \equiv (y_1,...,y_T)$ and $X = (X_0,...,X_{T - 1})$.
Each $y_t$ is the difference of log GPR measured on the first day of month $t$, while each $X_{t - 1}$ is the $p$-dimensional row vector of difference in log word frequencies averaged over the entirety of month $t - 1$. Thus, $y$ is a $T$-dimensional column vector and $X$ is a $T \times p$-dimensional matrix. 
Note that this model design respects temporal causality.
We regularize the model toward the null hypothesis of no relationship between $X$ and $y$ by placing a zero-mean normal prior on $\beta$, the $p$-dimensional column vector of coefficients, as $\beta \sim
\textnormal{MultivariateNormal}(0, I)$.  
We place a weakly informative prior
on the noise scale, $\sigma \sim \textnormal{LogNormal}(0, 1)$, 
as we are \textit{a priori} unsure of what noise level the data would exhibit.

We sample from the model using the No U-Turn Sampler (NUTS) algorithm~\cite{hoffman2014a} 
for 500 warmup iterations and 2000 iterations of sampling.  
There is strong evidence to suggest that the sampler
converged since the maximum Gelman-Rubin statistic~\cite{gelman1992a} 
for all priors was less than $1.01$ ($\max \hat R = 1.0009$).

We compute centered 80\% credible intervals for each of the model
coefficients $\beta_k$, $k = 0, ..., p$.
(A centered $Q\%$ credible
interval for the univariate random variable $Y \sim p(y)$ is an
interval $(a, b)$ such that
$\frac{1}{2}(1 - \frac{Q}{100}) = \int_{-\infty}^a dy \ p(y) = \int_b^{\infty}dy\ p(y)$.
For example, a centered 80\% credible interval is an interval such
that $0.1 = \int_{-\infty}^a dy\ p(y) = \int_b^\infty dy\ p(y)$.)
We termed a relationship significant if the 80\% credible interval did
not contain zero.  

In Fig.~\ref{fig:storywrangler.geopolitical_risk_linear_model}, 
we display the empirical distributions of the $\beta$ coefficient of each word for each model. 
The sign of the coefficient differs between the words, 
with positive significant associations highlighted in orange 
and negative significant associations highlighted in blue.

Using the predictors derived from Storywrangler, 
we recognize that the log difference in normalized usage frequency of ``rebellion''
is negatively associated with future log difference of GPR in originally authored tweets (OT). 
Similarly, ``uprising'' is positively associated with the percent change of GPR 
in organic tweets (OT), but not statistically significant in all tweets (AT).
The word ``crackdown'' is positively associated with GPR 
in both organic tweets (OT), and all tweets (AT).

We speculate that increases in the usage of ``crackdown’’ 
may imply that a popular revolt is already in the process of being crushed, 
a realization of increased geopolitical uncertainty.
Conversely, usage of ``uprising'' could be driven by 
a growing collective attention of a newborn social of movement 
with increased tension and uncertainty,
while usage of `rebellion' might be referring to past events~\cite{chenoweth2014a}.
Although our results using all tweets and organic tweets are fairly similar, 
future work can further investigate how social amplifications via retweets 
can influence our perception of geopolitical risk uncertainty. 

Furthermore, 
the log difference in normalized usage frequency of ``revolution'' and ``rebellion''
in Google trends search data~\cite{choi2012a} 
are associated with future log difference of the GPR index level but not statistically significant.
We believe that is likely associated with an increased number of searches for an ongoing revolt.    
We observe a similar signal for data derived 
from the Stanford cable TV news analyzer~\cite{hong2020a}, 
whereby growing mentions of ``coup'' 
in news outlets can be linked to higher levels of uncertainty in GPR. 

Although we see some overlap across the data streams examined here, 
Storywrangler provides a unique signal derived from everyday conversations on Twitter 
that is sometimes similarly portrayed across platforms, 
but often orthogonal and complementary to existing data sources.  
We foresee stronger potential for future work 
cross-referencing Storywrangler with other data repositories  
to enrich current data sources (e.g., search data and news).

\subsection{Qualitative comparison}

We conduct an additional analysis to differentiate between 
intra-GPR dynamics and 
potential predictive power of difference in log word frequency on log difference in GPR.
Our notation is identical to that used in Appendix~\ref{sec:bayes-gpr}.

We first fit a null autoregressive model, 
denoted by $M_0$, 
that assessed the effect of lagged GPR values on the current GPR value. 
We fit an autoregressive model of order $p = 4$ to the difference in log GPR data 
from 2010-01-01 to 2019-12-01, 
where $p = 4$ was chosen by minimizing Akaike information criterion (AIC).
This model takes the form
\begin{equation}\label{eq:ar}
y_t = \theta_0 + \sum_{p=1}^4 \theta_p y_{t - p} + u_t,
\end{equation}
with $u_t \sim \textnormal{Normal}(0, \sigma^2)$.
Here, 
$y_t$ is the difference of log GPR measured on the first day of month $t$. 
We fit the model using conditional MLE. 
Each of $\theta_p$ 
for $p \in \{1, 2, 3\}$ 
were significantly different from zero 
at the $p = 0.05$ confidence level (mean $\pm$ sd)
(
$\theta_1 = -0.4028 \pm 0.093$, $p < 0.001$; 
$\theta_2 = -0.3858 \pm 0.098$, $p < 0.001$;
$\theta_3 = -0.2650 \pm 0.098$, $p = 0.007$
).
The model is stable because all roots of the polynomial 
$P(z) = \sum_{p=1}^4 \theta_p z^{-p}$ are outside of the unit circle.
The model exhibited AIC = $-2.349$.
For out-of-sample comparison with other models, 
we computed the mean square forecast error (MSFE), defined as 
\begin{equation}
    \MSFE(M) = \frac{1}{t_2 - t_1}\sum_{t = t_1}^{t_2} (y_t - \hat y_t)^2,
\end{equation}
for $t_1, t_2 \in \mathbb N$ and $t_2 - t_1 >0$, 
where $\hat y$ are the out-of-sample predictions 
by model $M$ and $y$ are the true out-of-sample data. 
The null model $M_0$ had $\MSFE(M_0) = 0.6734$.

We then assessed the additional predictive power of each dataset, 
using an autoregressive with exogenous regressors (AR-X) model. 
For this model we used the same number of lags as the null AR model, 
$p = 4$. 
This model takes the form 
\begin{equation}\label{eq:ar-x}
    y_t = \theta_0 + \sum_{p=1}^4 \theta_p y_{t - p} + X_{t-1}\beta +  u_t,
\end{equation}
where $y_t$ has the same interpretation as in 
Eq.~\ref{eq:ar}, 
$X_{t-1}$ is the $N\times p$-dimensional of exogenous regressors 
(Storywrangler OT, Storywrangler AT, cable news, or Google trends) 
aggregated over month $t - 1$,
as described in Appendix~\ref{sec:risk-data}, 
and $\beta$ is the $p$-dimensional vector of regression coefficients. 
We fit each model using conditional MLE. 
Each estimated model was stable 
as indicated by the roots of the polynomial 
$P(z) = \sum_{p=1}^4 \theta_p z^{-p}$ 
being outside of the unit circle for each model.
No regression coefficient of any exogenous regressor for any of 
$\{$Storywrangler AT, Google trends, News$\}$ 
was significant at the Bonferroni corrected 
$p_c = p / N = 0.05 / 4 = 0.0125$ level. 
However, 
one exogenous 
regressor---the difference of log frequency for the word 
``crackdown''--- was significant 
at the $p_c = 0.0125$ level 
$\beta_{\text{crackdown}} = 0.0712 \pm 0.028,\ p_c = 0.010$, 
indicating that an increase in the average difference in log frequency
of usage of the word ``crackdown'' 
in month $t - 1$ 
is significantly associated with an increase 
in the difference of log GPR in month $t$, 
even after controlling for autoregressive effects. 
We display summaries of conditional MLE inference results 
of the null AR model 
and AR-X model for each data source 
in Tables~\ref{tab:gpr-ar-null}---\ref{tab:gpr-arx-news}.

We then compare the MSFE results of each AR-X model 
to the null AR model and display the results 
in Table~S6. 
The AR-X model using Storywrangler OT has the lowest MSFE (0.578), 
followed by Storywrangler AT (0.650).
However, 
AIC for each model 
(measured on training model) 
are not significantly different 
and do not differ significantly from AIC for the null AR model.

\begin{table}[!tph]
\centering
\caption{MSFE results of null AR and AR-X model with each dataset.}
\begin{tabularx}{\columnwidth}{l|C|C|C|C|C}
& \multicolumn{2}{c|}{\textbf{Storywrangler}}   &  \textbf{Cable}       &  \textbf{Google} & \textbf{Null} \\ 
& \textbf{OT}  & \textbf{AT}                    & \textbf{News}   & \textbf{Trends}  & $M_0$ \\
\hline
MSEF~ & 0.578 & 0.650 & 0.659 & 0.679 & 0.673 \\
$R^2$ & 0.288 & 0.226 & 0.233 & 0.212 & 0.186  \\
\end{tabularx}
\label{tab:gpr-msfe}
\end{table} 

To assess significance of associations between lagged difference
in log word frequency from each dataset and difference in log GPR 
without taking autoregressive behavior of GPR into account, 
we also estimate ordinary least squares models (OLS) models 
of the form 
\begin{equation}
    y_t = X_{t - 1}\beta + u_t,
\end{equation}
where all terms in the equation have their same meaning as in the previous paragraphs. 
We fit these models using exact least squares.
No regression coefficient in any design matrix was significant 
at the $p_c = 0.0125$ level except 
for ``crackdown'' in the Storywrangler OT dataset 
($\beta_{\text{crackdown}} = 0.0776 \pm 0.030,\ p = 0.011$).
We summarize results of model fits 
in Tables~\ref{tab:gpr-ols-twitter-ot}---\ref{tab:gpr-ols-news}.

The results of this analysis are mixed 
and warrant in-depth further investigation. 
We arbitrarily chose the set of words to analyze; 
this is not appropriate for an analysis 
of any rigor greater than a simple case study such as this. 
Rather, further analysis should choose 
1-, 
2-, 
and 3-grams 
to analyze using a more principled analysis, 
such as a subject matter expert interview that we now describe.

A group of acknowledged experts in geopolitical risk 
(e.g., military strategic planners, 
emerging markets funds managers, 
diplomats, 
and senior humanitarian aid workers) 
should be systematically interviewed and asked 
to provide $n$-grams that they would 
(1) use, or expect other experts to use, 
to describe both impactful and non-impactful events in developing countries, 
(2) expect to hear non-experts use to describe these events. 
The Storywrangler OT and AT datasets, 
as well as the Google trends, News, 
and similar datasets, 
can then be queried for these words and an association analysis performed. 
This methodology will help to eliminate the selection bias 
that we have probably introduced via our word selection.

\begin{table*}[!h]
\begin{tabularx}{\columnwidth}{lRRR}
\textbf{Model}         &         AutoReg(4)        & \textbf{AIC} &           -2.349    \\
\textbf{Method}        &      Conditional MLE & \textbf{BIC} &           -2.205          \\
\textbf{Data} &          GPR Only   &      \textbf{HQIC} &           -2.290            \\
\end{tabularx}
\begin{tabularx}{\columnwidth}{lCcCCCC}
\hline
                   & \textbf{coef} & \textbf{std err} & \textbf{z} & \textbf{P$> |$z$|$} & v\textbf{[0.025} & \textbf{0.975]}  \\
\hline
\textbf{intercept} &       0.0188  &        0.028     &     0.682  &         0.495        &       -0.035    &        0.073     \\
\textbf{GPR.L1}    &      -0.4028  &        0.093     &    -4.341  &         0.000        &       -0.585    &       -0.221     \\
\textbf{GPR.L2}    &      -0.3858  &        0.098     &    -3.926  &         0.000        &       -0.578    &       -0.193     \\
\textbf{GPR.L3}    &      -0.2650  &        0.098     &    -2.709  &         0.007        &       -0.457    &       -0.073     \\
\textbf{GPR.L4}    &      -0.1702  &        0.094     &    -1.815  &         0.070        &       -0.354    &        0.014     \\
\end{tabularx}
\begin{tabularx}{\columnwidth}{lCCCC}
\hline
& \textbf{          Real} & \textbf{       Imaginary} & \textbf{       Modulus} & \textbf{      Frequency}  \\
\hline
\textbf{AR.1} &                0.5127     &                -1.3379j     &                1.4328     &               -0.1918      \\
\textbf{AR.2} &                0.5127     &                +1.3379j     &                1.4328     &                0.1918       \\
\textbf{AR.3} &               -1.2913     &                -1.0930j     &                1.6918     &               -0.3882       \\
\textbf{AR.4} &               -1.2913     &                +1.0930j     &                1.6918     &               0.3882       \\
\end{tabularx}
\caption{Summary of conditional MLE inference results, 
fitting null AR model to GPR data only.
\label{tab:gpr-ar-null}}
\end{table*}

\begin{table*}[!h]
\begin{tabularx}{\columnwidth}{lRRR}
\textbf{Model}         &        AutoReg-X     & \textbf{AIC                } &           -2.332           \\
\textbf{Method}        &      Conditional MLE         & \textbf{BIC               } &           -1.996           \\
\textbf{Data} &          GPR + Storywrangler OT              & \textbf{HQIC               } &           -2.195           \\
\hline
\end{tabularx}
\begin{tabularx}{\columnwidth}{lCcCCCC}
                    & \textbf{coef} & \textbf{std err} & \textbf{z} & \textbf{P$> |$z$|$} & \textbf{[0.025} & \textbf{0.975]}  \\
\hline
\textbf{intercept}  &       0.0208  &        0.026     &     0.799  &         0.424        &       -0.030    &        0.072     \\
\textbf{GPR.L1}     &      -0.2999  &        0.092     &    -3.253  &         0.001        &       -0.481    &       -0.119     \\
\textbf{GPR.L2}     &      -0.3872  &        0.093     &    -4.155  &         0.000        &       -0.570    &       -0.205     \\
\textbf{GPR.L3}     &      -0.2371  &        0.095     &    -2.501  &         0.012        &       -0.423    &       -0.051     \\
\textbf{GPR.L4}     &      -0.2056  &        0.091     &    -2.249  &         0.025        &       -0.385    &       -0.026     \\
\hline
\textbf{revolution} &      -0.0461  &        0.042     &    -1.097  &         0.273        &       -0.128    &        0.036     \\
\textbf{rebellion}  &      -0.0315  &        0.032     &    -0.973  &         0.331        &       -0.095    &        0.032     \\
\textbf{uprising}   &       0.0628  &        0.035     &     1.782  &         0.075        &       -0.006    &        0.132     \\
\textbf{coup}       &      -0.0296  &        0.030     &    -0.996  &         0.319        &       -0.088    &        0.029     \\
\textbf{overthrow}  &       0.0426  &        0.031     &     1.369  &         0.171        &       -0.018    &        0.104     \\
\textbf{unrest}     &       0.0363  &        0.045     &     0.802  &         0.423        &       -0.052    &        0.125     \\
\textbf{protests}   &      -0.0486  &        0.040     &    -1.212  &         0.226        &       -0.127    &        0.030     \\
\textbf{crackdown}  &       0.0712  &        0.028     &     2.587  &         0.010        &        0.017    &        0.125     \\
\hline
\end{tabularx}
\begin{tabularx}{\columnwidth}{lCCCC}
              & \textbf{          Real} & \textbf{       Imaginary} & \textbf{       Modulus} & \textbf{      Frequency}  \\
\hline
\textbf{AR.1} &                0.5643     &                -1.2589j     &                1.3796     &               -0.1829       \\
\textbf{AR.2} &                0.5643     &                +1.2589j     &                1.3796     &                0.1829       \\
\textbf{AR.3} &               -1.1411     &                -1.1198j     &                1.5988     &               -0.3765       \\
\textbf{AR.4} &               -1.1411     &                +1.1198j     &                1.5988     &                0.3765      
\end{tabularx}
\caption{Summary of conditional MLE inference results, 
fitting AR-X model to GPR data with Storywrangler OT. 
\label{tab:gpr-arx-twitter-ot}}
\end{table*}

\begin{table*}[!h]
\begin{tabularx}{\columnwidth}{lRRR}
\textbf{Model}         &        AutoReg-X    & \textbf{AIC                } &           -2.269           \\
\textbf{Method}        &      Conditional MLE        & \textbf{BIC               } &           -1.933           \\
\textbf{Data} &          GPR + Storywrangler AT           & \textbf{HQIC               } &           -2.133           \\
\end{tabularx}
\begin{tabularx}{\columnwidth}{lCcCCCC}
\hline
                    & \textbf{coef} & \textbf{std err} & \textbf{z} & \textbf{P$> |$z$|$} & \textbf{[0.025} & \textbf{0.975]}  \\
\hline
\textbf{intercept}  &       0.0199  &        0.027     &     0.743  &         0.458        &       -0.033    &        0.072     \\
\textbf{GPR.L1}     &      -0.3365  &        0.094     &    -3.575  &         0.000        &       -0.521    &       -0.152     \\
\textbf{GPR.L2}     &      -0.3954  &        0.098     &    -4.030  &         0.000        &       -0.588    &       -0.203     \\
\textbf{GPR.L3}     &      -0.2547  &        0.096     &    -2.641  &         0.008        &       -0.444    &       -0.066     \\
\textbf{GPR.L4}     &      -0.1954  &        0.092     &    -2.114  &         0.034        &       -0.377    &       -0.014     \\
\hline
\textbf{revolution} &      -0.0366  &        0.045     &    -0.809  &         0.418        &       -0.125    &        0.052     \\
\textbf{rebellion}  &      -0.0085  &        0.035     &    -0.244  &         0.807        &       -0.077    &        0.060     \\
\textbf{uprising}   &       0.0381  &        0.037     &     1.022  &         0.307        &       -0.035    &        0.111     \\
\textbf{coup}       &      -0.0417  &        0.032     &    -1.304  &         0.192        &       -0.104    &        0.021     \\
\textbf{overthrow}  &       0.0431  &        0.036     &     1.197  &         0.231        &       -0.027    &        0.114     \\
\textbf{unrest}     &       0.0347  &        0.049     &     0.701  &         0.483        &       -0.062    &        0.132     \\
\textbf{protests}   &      -0.0382  &        0.041     &    -0.932  &         0.351        &       -0.118    &        0.042     \\
\textbf{crackdown}  &       0.0408  &        0.030     &     1.382  &         0.167        &       -0.017    &        0.099     \\
\end{tabularx}
\begin{tabularx}{\columnwidth}{lCCCC}
\hline
              & \textbf{          Real} & \textbf{       Imaginary} & \textbf{       Modulus} & \textbf{      Frequency}  \\
\hline
\textbf{AR.1} &                0.5379     &                -1.2792j     &                1.3877     &               -0.1866       \\
\textbf{AR.2} &                0.5379     &                +1.2792j     &                1.3877     &                0.1866       \\
\textbf{AR.3} &               -1.1896     &                -1.1147j     &                1.6302     &               -0.3802       \\
\textbf{AR.4} &               -1.1896     &                +1.1147j     &                1.6302     &                0.3802      
\end{tabularx}
\caption{Summary of conditional MLE inference results, 
fitting AR-X model to GPR data with Storywrangler AT. 
\label{tab:gpr-arx-twitter-at}}
\end{table*}

\begin{table*}[!h]
\begin{tabularx}{\columnwidth}{lRRR}
\textbf{Model}         &        AutoReg-X     & \textbf{AIC                } &           -2.236           \\
\textbf{Method}        &      Conditional MLE         & \textbf{BIC               } &           -1.900           \\
\textbf{Data} &          GPR + Google Trends            & \textbf{HQIC               } &           -2.099           \\
\end{tabularx}
\begin{tabularx}{\columnwidth}{lCcCCCC}
\hline
                    & \textbf{coef} & \textbf{std err} & \textbf{z} & \textbf{P$> |$z$|$} & \textbf{[0.025} & \textbf{0.975]}  \\
\hline
\textbf{intercept}  &       0.0192  &        0.027     &     0.705  &         0.481        &       -0.034    &        0.073     \\
\textbf{GPR.L1}     &      -0.4184  &        0.094     &    -4.442  &         0.000        &       -0.603    &       -0.234     \\
\textbf{GPR.L2}     &      -0.3944  &        0.101     &    -3.890  &         0.000        &       -0.593    &       -0.196     \\
\textbf{GPR.L3}     &      -0.2607  &        0.097     &    -2.694  &         0.007        &       -0.450    &       -0.071     \\
\textbf{GPR.L4}     &      -0.1809  &        0.094     &    -1.925  &         0.054        &       -0.365    &        0.003     \\
\hline
\textbf{revolution} &      -0.0262  &        0.034     &    -0.774  &         0.439        &       -0.092    &        0.040     \\
\textbf{rebellion}  &       0.0212  &        0.032     &     0.670  &         0.503        &       -0.041    &        0.083     \\
\textbf{uprising}   &       0.0265  &        0.027     &     0.987  &         0.324        &       -0.026    &        0.079     \\
\textbf{coup}       &       0.0198  &        0.029     &     0.685  &         0.493        &       -0.037    &        0.076     \\
\textbf{overthrow}  &       0.0067  &        0.033     &     0.204  &         0.838        &       -0.058    &        0.071     \\
\textbf{unrest}     &    7.62e-5  &        0.038     &     0.002  &         0.998        &       -0.074    &        0.074     \\
\textbf{protests}   &      -0.0025  &        0.035     &    -0.070  &         0.944        &       -0.072    &        0.067     \\
\textbf{crackdown}  &       0.0172  &        0.028     &     0.609  &         0.543        &       -0.038    &        0.072     
\end{tabularx}
\begin{tabularx}{\columnwidth}{lCCCC}
\hline
              & \textbf{          Real} & \textbf{       Imaginary} & \textbf{       Modulus} & \textbf{      Frequency}  \\
\hline
\textbf{AR.1} &                0.5177     &                -1.3376j     &                1.4342     &               -0.1912       \\
\textbf{AR.2} &                0.5177     &                +1.3376j     &                1.4342     &                0.1912       \\
\textbf{AR.3} &               -1.2383     &                -1.0741j     &                1.6392     &               -0.3863       \\
\textbf{AR.4} &               -1.2383     &                +1.0741j     &                1.6392     &                0.3863      
\end{tabularx}
\caption{Summary of conditional MLE inference results,
fitting AR-X model to GPR data with Google trends. 
\label{tab:gpr-arx-trends}}
\end{table*}

\begin{table*}[!h]
\begin{tabularx}{\columnwidth}{lRRR}
\textbf{Model}         &        AutoReg-X     & \textbf{AIC                } &           -2.241           \\
\textbf{Method}        &      Conditional MLE          & \textbf{BIC               } &           -1.905           \\
\textbf{Data} &          GPR + Cable News             & \textbf{HQIC               } &           -2.104           \\
\end{tabularx}
\begin{tabularx}{\columnwidth}{lCcCCCC}
\hline
                    & \textbf{coef} & \textbf{std err} & \textbf{z} & \textbf{P$> |$z$|$} & \textbf{[0.025} & \textbf{0.975]}  \\
\hline
\textbf{intercept}  &       0.0196  &        0.027     &     0.720  &         0.471        &       -0.034    &        0.073     \\
\textbf{GPR.L1}     &      -0.3702  &        0.095     &    -3.880  &         0.000        &       -0.557    &       -0.183     \\
\textbf{GPR.L2}     &      -0.4010  &        0.100     &    -3.994  &         0.000        &       -0.598    &       -0.204     \\
\textbf{GPR.L3}     &      -0.2528  &        0.100     &    -2.518  &         0.012        &       -0.450    &       -0.056     \\
\textbf{GPR.L4}     &      -0.2095  &        0.100     &    -2.104  &         0.035        &       -0.405    &       -0.014     \\
\hline
\textbf{revolution} &       0.0099  &        0.030     &     0.328  &         0.743        &       -0.049    &        0.069     \\
\textbf{rebellion}  &      -0.0159  &        0.029     &    -0.539  &         0.590        &       -0.074    &        0.042     \\
\textbf{uprising}   &      -0.0279  &        0.036     &    -0.778  &         0.437        &       -0.098    &        0.042     \\
\textbf{coup}       &       0.0367  &        0.032     &     1.155  &         0.248        &       -0.026    &        0.099     \\
\textbf{overthrow}  &       0.0235  &        0.030     &     0.779  &         0.436        &       -0.036    &        0.083     \\
\textbf{unrest}     &       0.0174  &        0.037     &     0.465  &         0.642        &       -0.056    &        0.091     \\
\textbf{protests}   &       0.0059  &        0.038     &     0.157  &         0.875        &       -0.068    &        0.079     \\
\textbf{crackdown}  &       0.0317  &        0.030     &     1.043  &         0.297        &       -0.028    &        0.091     
\end{tabularx}
\begin{tabularx}{\columnwidth}{lCCCC}
\hline
              & \textbf{          Real} & \textbf{       Imaginary} & \textbf{       Modulus} & \textbf{      Frequency}  \\
\hline
\textbf{AR.1} &                0.5425     &                -1.2819j     &                1.3920     &               -0.1863       \\
\textbf{AR.2} &                0.5425     &                +1.2819j     &                1.3920     &                0.1863       \\
\textbf{AR.3} &               -1.1461     &                -1.0726j     &                1.5697     &               -0.3803       \\
\textbf{AR.4} &               -1.1461     &                +1.0726j     &                1.5697     &                0.3803      
\end{tabularx}
\caption{Summary of conditional MLE inference results,
fitting AR-X model to GPR data with cable news.
\label{tab:gpr-arx-news}}
\end{table*}

\begin{table*}[!tp]
\begin{tabularx}{\columnwidth}{lClC}
\textbf{Model}            &       OLS & \textbf{Method}           &  Least Squares  \\
\textbf{R-squared         } &     0.126        & \textbf{Adj. R-squared    } &     0.062   \\
\textbf{Log-Likelihood    } &   -25.449   & \textbf{Durbin-Watson     } &    2.328   \\
\textbf{AIC               } &     68.90      & \textbf{BIC               } &     93.83   \\
\textbf{F-statistic       } &     1.973 & \textbf{Prob (F-statistic)} &   0.0565    \\
\textbf{Omnibus}       &  6.186 & \textbf{Prob(Omnibus)} &  0.045  \\
\textbf{Jarque-Bera (JB)} &    6.251 & \textbf{Prob(JB)          } &   0.0439  \\
\textbf{Kurtosis}      &  3.826  & \textbf{Skew}          &  0.384 \\
\end{tabularx}
\begin{tabularx}{\columnwidth}{lCcCCCC}
\hline
                    & \textbf{coef} & \textbf{std err} & \textbf{t} & \textbf{P$> |$t$|$} & \textbf{[0.025} & \textbf{0.975]}  \\
\hline
\textbf{const}      &       0.0064  &        0.029     &     0.223  &         0.824        &       -0.051    &        0.063     \\
\textbf{revolution} &      -0.0443  &        0.045     &    -0.989  &         0.325        &       -0.133    &        0.044     \\
\textbf{rebellion}  &      -0.0295  &        0.035     &    -0.834  &         0.406        &       -0.100    &        0.041     \\
\textbf{uprising}   &       0.0800  &        0.038     &     2.093  &         0.039        &        0.004    &        0.156     \\
\textbf{coup}       &      -0.0260  &        0.033     &    -0.781  &         0.437        &       -0.092    &        0.040     \\
\textbf{overthrow}  &       0.0469  &        0.035     &     1.354  &         0.179        &       -0.022    &        0.116     \\
\textbf{unrest}     &       0.0006  &        0.046     &     0.013  &         0.989        &       -0.091    &        0.092     \\
\textbf{protests}   &      -0.0394  &        0.042     &    -0.931  &         0.354        &       -0.123    &        0.045     \\
\textbf{crackdown}  &       0.0776  &        0.030     &     2.578  &         0.011        &        0.018    &        0.137 
\end{tabularx}
\caption{OLS summary: GPR and Storywrangler OT.
\label{tab:gpr-ols-twitter-ot}}
\end{table*}

\begin{table*}[!h]
\begin{tabularx}{\columnwidth}{lClC}
\textbf{Model}              &       OLS        & \textbf{Method}           &  Least Squares   \\
\textbf{R-squared         } &     0.071      & \textbf{Adj. R-squared    } &     0.003   \\
\textbf{Log-Likelihood    } &   -29.094      & \textbf{Durbin-Watson     } &    2.345    \\
\textbf{AIC               } &     76.19      & \textbf{BIC               } &     101.1   \\
\textbf{F-statistic       } &     1.039      & \textbf{Prob (F-statistic)} &    0.412   \\
\textbf{Omnibus}       &  8.949 &  \textbf{Prob(Omnibus)} &  0.011   \\
\textbf{Jarque-Bera (JB)  } &    9.355 & \textbf{Prob(JB)          } &  0.00930  \\
\textbf{Kurtosis}      &  3.885 & \textbf{Skew}          &  0.529  \\
\end{tabularx}
\begin{tabularx}{\columnwidth}{lCcCCCC}
\hline
                    & \textbf{coef} & \textbf{std err} & \textbf{t} & \textbf{P$> |$t$|$} & \textbf{[0.025} & \textbf{0.975]}  \\
\hline
\textbf{const}      &       0.0058  &        0.030     &     0.195  &         0.846        &       -0.053    &        0.065     \\
\textbf{revolution} &      -0.0457  &        0.050     &    -0.916  &         0.362        &       -0.145    &        0.053     \\
\textbf{rebellion}  &       0.0004  &        0.038     &     0.011  &         0.991        &       -0.075    &        0.076     \\
\textbf{uprising}   &       0.0600  &        0.041     &     1.473  &         0.144        &       -0.021    &        0.141     \\
\textbf{coup}       &      -0.0377  &        0.035     &    -1.077  &         0.284        &       -0.107    &        0.032     \\
\textbf{overthrow}  &       0.0598  &        0.040     &     1.486  &         0.140        &       -0.020    &        0.139     \\
\textbf{unrest}     &      -0.0043  &        0.052     &    -0.084  &         0.933        &       -0.107    &        0.098     \\
\textbf{protests}   &      -0.0294  &        0.043     &    -0.680  &         0.498        &       -0.115    &        0.056     \\
\textbf{crackdown}  &       0.0423  &        0.032     &     1.316  &         0.191        &       -0.021    &        0.106     \\
\end{tabularx}
\caption{OLS summary: GPR and Storywrangler AT.
\label{tab:gpr-ols-twitter-at}}
\end{table*}

\begin{table*}[!h]
\begin{tabularx}{\columnwidth}{lClC}
\textbf{Model}            &       OLS        & \textbf{Method}           &  Least Squares   \\
\textbf{R-squared         } &     0.013        & \textbf{Adj. R-squared    } &    -0.059   \\
\textbf{Log-Likelihood    } &   -32.636     & \textbf{Durbin-Watson     } &    2.479   \\
\textbf{AIC               } &     83.27       & \textbf{BIC               } &     108.2   \\
\textbf{F-statistic       } &    0.1843 & \textbf{Prob (F-statistic)} &    0.993    \\
\textbf{Omnibus}       & 12.770 & \textbf{Prob(Omnibus)} &  0.002  \\
\textbf{Jarque-Bera (JB)  } &   15.141 & \textbf{Prob(JB)          } & 0.000515  \\
\textbf{Kurtosis}      &  4.195 & \textbf{Skew}          &  0.643  \\
\end{tabularx}
\begin{tabularx}{\columnwidth}{lCcCCCC}
\hline
                    & \textbf{coef} & \textbf{std err} & \textbf{t} & \textbf{P$> |$t$|$} & \textbf{[0.025} & \textbf{0.975]}  \\
\hline
\textbf{const}      &       0.0058  &        0.031     &     0.188  &         0.851        &       -0.055    &        0.066     \\
\textbf{revolution} &      -0.0204  &        0.038     &    -0.537  &         0.592        &       -0.096    &        0.055     \\
\textbf{rebellion}  &       0.0206  &        0.036     &     0.575  &         0.567        &       -0.050    &        0.091     \\
\textbf{uprising}   &       0.0118  &        0.030     &     0.394  &         0.694        &       -0.048    &        0.071     \\
\textbf{coup}       &       0.0161  &        0.033     &     0.493  &         0.623        &       -0.049    &        0.081     \\
\textbf{overthrow}  &       0.0010  &        0.035     &     0.028  &         0.978        &       -0.069    &        0.071     \\
\textbf{unrest}     &       0.0079  &        0.041     &     0.191  &         0.849        &       -0.074    &        0.090     \\
\textbf{protests}   &      -0.0040  &        0.040     &    -0.100  &         0.920        &       -0.082    &        0.074     \\
\textbf{crackdown}  &       0.0192  &        0.032     &     0.598  &         0.551        &       -0.045    &        0.083     \\
\end{tabularx}
\caption{OLS summary: GPR and Google trends.
\label{tab:gpr-ols-trends}}
\end{table*}

\begin{table*}[!h]
\begin{tabularx}{\columnwidth}{lClC}
\textbf{Model}            &       OLS       &   \textbf{Method}           &  Least Squares \\
\textbf{R-squared         } &     0.039        & \textbf{Adj. R-squared    } &    -0.032   \\
\textbf{Log-Likelihood    } &   -31.092      &   \textbf{Durbin-Watson     } &    2.462 \\
\textbf{AIC               } &     80.18      & \textbf{BIC               } &     105.1   \\
\textbf{F-statistic       } &    0.5504 & \textbf{Prob (F-statistic)} &    0.816    \\
\textbf{Omnibus}       & 14.135 &  \textbf{Prob(Omnibus)} &  0.001 \\
\textbf{Jarque-Bera (JB)  } &   17.910 & \textbf{Prob(JB)          } & 0.000129  \\
\textbf{Kurtosis}      &  4.366 & \textbf{Skew}          &  0.666  \\
\end{tabularx}
\begin{tabularx}{\columnwidth}{lCcCCCC}
\hline
                    & \textbf{coef} & \textbf{std err} & \textbf{t} & \textbf{P$> |$t$|$} & \textbf{[0.025} & \textbf{0.975]}  \\
\hline
\textbf{const}      &       0.0068  &        0.030     &     0.226  &         0.821        &       -0.053    &        0.067     \\
\textbf{revolution} &       0.0217  &        0.033     &     0.659  &         0.511        &       -0.044    &        0.087     \\
\textbf{rebellion}  &      -0.0156  &        0.032     &    -0.486  &         0.628        &       -0.079    &        0.048     \\
\textbf{uprising}   &      -0.0187  &        0.035     &    -0.530  &         0.597        &       -0.089    &        0.051     \\
\textbf{coup}       &       0.0460  &        0.033     &     1.395  &         0.166        &       -0.019    &        0.111     \\
\textbf{overthrow}  &       0.0082  &        0.033     &     0.250  &         0.803        &       -0.057    &        0.074     \\
\textbf{unrest}     &       0.0104  &        0.041     &     0.252  &         0.801        &       -0.071    &        0.092     \\
\textbf{protests}   &       0.0210  &        0.041     &     0.518  &         0.606        &       -0.059    &        0.101     \\
\textbf{crackdown}  &       0.0326  &        0.033     &     0.978  &         0.330        &       -0.033    &        0.099     \\
\end{tabularx}
\caption{OLS summary: GPR and cable news.
\label{tab:gpr-ols-news}}
\end{table*}

\end{document}